\documentclass[twocolumn,twocolappendix]{aastex63}

\usepackage{xspace}
\usepackage{mathtools}
\usepackage{float}
\usepackage{amsmath}	
\usepackage{multirow}
\usepackage{enumitem}

\newcommand{\Omegam}{\xspace{\ensuremath{\Omega_{\mathrm{m}}}}\xspace}

\newcommand{\lcdm}{\xspace{\ensuremath{\Lambda\mathrm{CDM}}}\xspace}
\newcommand{\hmpc}{\xspace{$h^{-1}\mathrm{Mpc}$}\xspace}
\newcommand{\hkpc}{\xspace{$h^{-1}\mathrm{kpc}$}\xspace}
\newcommand{\hMsun}{\xspace{$h^{-1}\mathrm{M}_{\odot}$}\xspace}

\newcommand{\cemcee}{\xspace{\textsc{emcee\_in\_c}}\xspace}
\newcommand{\emcee}{\xspace{\textsc{emcee}}\xspace}

\newcommand{\mmin}{\xspace{$M_\mathrm{min}$}\xspace}
\newcommand{\logmmin}{\xspace{$\mathrm{log}{M_\mathrm{min}}$}\xspace}
\newcommand{\siglogm}{\xspace{$\sigma_{\mathrm{log} M}$}\xspace}
\newcommand{\logmzero}{\xspace{$\mathrm{log}{M_0}$}\xspace}
\newcommand{\logmone}{\xspace{$\mathrm{log}{M_1}$}\xspace}

\newcommand{\ngal}{\xspace{$n_{\mathrm{gal}}$}\xspace}
\newcommand{\wprp}{\xspace{$w_\mathrm{p}(r_\mathrm{p})$}\xspace}
\newcommand{\zxi}{\xspace{$\xi(s)$}\xspace}
\newcommand{\gmf}{\xspace{$n(N)$}\xspace}
\newcommand{\sigN}{\xspace{$\sigma_v(N)$}\xspace}
\newcommand{\mcf}{\xspace{$\mathrm{mcf}(s)$}\xspace}
\newcommand{\cic}{\xspace{$P_N(R)$}\xspace}
\newcommand{\vpf}{\xspace{$\mathrm{VPF}(R)$}\xspace}
\newcommand{\spf}{\xspace{$\mathrm{SPF}(R)$}\xspace}

\newcommand{\Linit}{\xspace{$\mathcal{L}_\mathrm{init}$}\xspace}
\newcommand{\siginit}{\xspace{$\sigma_\mathrm{par,init}$}\xspace}
\newcommand{\sigk}{\xspace{$\sigma_{\mathrm{par,}K}$}\xspace}
\newcommand{\wptwo}{\xspace{$w_\mathrm{p}(r_\mathrm{p} \sim 0.3 \ h^{-1}\mathrm{Mpc})$}\xspace}

\usepackage{color}
\usepackage[normalem]{ulem} 

\received{XXX}
\revised{XXX}
\accepted{XXX}
\submitjournal{ApJ}

\shorttitle{Accurate Galaxy Clustering II}
\shortauthors{Szewciw et al.}
\graphicspath{{./}{figures/}}

\begin{document}

\title{Toward Accurate Modeling of Galaxy Clustering on Small Scales: Constraining the Galaxy-Halo Connection with Optimal Statistics}

\correspondingauthor{Adam O Szewciw}
\email{adam.o.szewciw@vanderbilt.edu}

\author{Adam O. Szewciw}
\affiliation{Vanderbilt University \\
2201 West End Ave \\
Nashville, TN, 37235 USA}

\author[0000-0002-4392-8920]{Gillian D. Beltz-Mohrmann}
\affiliation{Vanderbilt University \\
2201 West End Ave \\
Nashville, TN, 37235 USA}

\author[0000-0002-1814-2002]{Andreas A. Berlind}
\affiliation{Vanderbilt University \\
2201 West End Ave \\
Nashville, TN, 37235 USA}

\author[0000-0002-4845-1228]{Manodeep Sinha}
\affiliation{SA 118, Center for Astrophysics \& Supercomputing \\
Swinburne University of Technology,  \\
1 Alfred St., Hawthorn, VIC 3122, Australia}
\affiliation{ARC Centre of Excellence for All Sky Astrophysics in 3 Dimensions
  (ASTRO 3D)}
\affiliation{Vanderbilt University \\
2201 West End Ave \\
Nashville, TN, 37235 USA}

\begin{abstract}
Applying halo models to analyze the small-scale clustering of galaxies is a proven method for characterizing the connection between galaxies and their host halos.
Such works are often plagued by systematic errors or are limited to clustering statistics which can be predicted analytically.
In this work, we employ a numerical mock-based modeling procedure to examine the clustering of SDSS DR7 galaxies.
We apply a standard halo occupation distribution (HOD) model to dark-matter-only simulations with a \lcdm cosmology.
To constrain the theoretical models, we utilize a combination of galaxy number density and selected scales of the projected correlation function, redshift-space correlation function, group multiplicity function, average group velocity dispersion, mark correlation function, and counts-in-cells statistics.
We design an algorithm to choose an optimal combination of measurements that yields tight and accurate constraints on our model parameters.
Compared to previous work using fewer clustering statistics, we find significant improvement in the constraints on all parameters of our halo model for two different luminosity-threshold galaxy samples. 
Most interestingly, we obtain unprecedented high-precision constraints on the scatter in the relationship between galaxy luminosity and halo mass.
However, our best-fit model results in significant tension ($\gtrsim 4 \sigma$) for both samples, indicating the need to add second-order features to the standard HOD model. 
To guarantee the robustness of these results, we perform an extensive analysis of the systematic and statistical errors in our modeling procedure, including a first-of-its-kind study of the sensitivity of our constraints to changes in the halo mass function due to baryonic physics.
\end{abstract}

\keywords{Large-scale structure of the universe (902) --- Galaxy dark matter halos (1880) --- Galaxy groups (597) --- Clustering (1908) --- Redshift surveys (1378)}

\section{Introduction} \label{sec:intro}

One of the great strengths of \lcdm is in its accurate predictions of the large-scale density field of the Universe today. 
It is common to quantify this density field through statistical measurements of the clustering of galaxies. 
\lcdm is highly successful at predicting galaxy clustering on large physical scales where galaxies are simple tracers of the underlying matter density field. 
On small scales ($\lesssim10$ \hmpc), however, the spatial distribution of galaxies is affected not only by our cosmological model but also by all of the poorly understood physics of galaxy formation and evolution. 

We can bypass our ignorance of galaxy physics via the use of so called ``halo models," which are motivated by our understanding that galaxies form and reside in gravitationally bound, virialized regions of dark matter known as halos \citep[e.g.,][]{Neyman1952,Peebles1974,McClelland1977,Scherrer1991,Jing1998,Benson2000,Ma2000,Peacock2000,Seljak2000,Scoccimarro2001,Sheth2001,White2001,Cooray2002}. 
These models assume that the clustering of galaxies can be fully described by (i) the clustering of their host halos and (ii) the way in which galaxies occupy these halos. 
The first piece can be accurately predicted by \lcdm via cosmological N-body simulations of dark matter only (i.e., gravity only). 
The second piece can be described by a halo model, which provides a statistical description of how galaxies occupy halos (i.e., the galaxy-halo connection). 

A key ingredient of the halo model is the Halo Occupation Distribution (HOD), which specifies via a few parameters the probability that a halo of mass $M$ contains $N$ galaxies (above some luminosity threshold) and provides a prescription for the distribution of galaxies within a halo \citep{Berlind2002,Berlind2003}. 
The standard form of the HOD \citep{Zheng2005} contains at most five free parameters that specify the mean occupation number of galaxies and assumes that galaxies trace the dark matter inside halos.
Constraining these parameters when fitting to observational data provides a useful empirical measurement against which we can test competing theories of galaxy formation and evolution.

Several studies which use this procedure produce fits which, if taken at face value, would result in ruling out the \lcdm+ HOD model \citep[e.g.,][]{Zehavi2011}. 
The errors used in these studies are typically derived via the jackknife method, which has been shown to produce biased results \citep{Norberg2009}. 
Meanwhile, spectroscopic surveys such as the Sloan Digital Sky Survey \citep[SDSS;][]{York2000} are providing increasingly high precision measurements of galaxy clustering at a wide range of physical scales. 
If we wish to take advantage of the information present at small scales to constrain the galaxy-halo connection, it is imperative that we carefully understand and minimize the uncertainty in our modeling procedure.

Recently, \citet{Sinha2018} (S18, hereafter) developed a numerical mock-based modeling procedure that significantly improved the accuracy of HOD modeling.
They compared the clustering of SDSS galaxies to a \lcdm+ HOD model, measuring the projected correlation function, group multiplicity function, and galaxy number density. 
Carefully controlling for systematic errors allowed them to interpret the goodness of fit of their model.
They found that their best-fit HOD model was unable to jointly fit the clustering statistics, revealing significant tension between SDSS galaxies and their \lcdm+ HOD model.
Because this tension did not exist when they considered only measurements of the projected correlation function (as is done in many studies), S18 demonstrated the value of adding additional statistics in small-scale clustering analyses.

Motivated by these findings, in this work we extend the procedure used in S18 in order to maximize the return from spectroscopic surveys.
We do so with two main goals in mind: to tighten the constraints on our adopted five-parameter HOD model and to increase the tension found in S18.
To accomplish these goals, we include in our analysis the same clustering statistics used in S18 (galaxy number density, projected correlation function, and group multiplicity function) as well as four additional clustering statistics: redshift-space correlation function, group velocity dispersion, mark correlation function, and counts-in-cells statistics.
The fully numerical modeling procedure built by S18 allows us to accurately model all of these clustering statistics without having to rely solely on analytic predictions (which would limit us to two-point statistics) as is done in most studies.

This work can be seen as the second in a series of several works attempting to model the clustering of SDSS galaxies to constrain the galaxy-halo connection.
Ultimately, our goal is to expand our HOD model to include features like assembly bias \citep{Gao2005, Wechsler2006, Croton2007, Pujol2014, Hearin2016, Pujol2017, Artale2018, Salcedo2018, Xu2018, Zehavi2018, Bose2019, Contreras2019, Padilla2019, Wang2019, Zentner2019, Salcedo2020, Hadzhiyska2020, Hadzhiyska2021b, Montero2021}, spatial bias \citep{Watson2012, Piscionere2015}, and velocity bias \citep{Guo2015a, Guo2015b}. 
However, in order to constrain these extra features, it is crucial that we first incorporate clustering statistics into our modeling procedure that are sensitive to these biases. 
Thus, in this work, as we add clustering statistics to heighten the tension found in S18, we do so keeping in mind the ability of each statistic to constrain extended features of the HOD model.

The projected correlation function has been utilized in many galaxy clustering studies \citep[e.g.,][]{Zheng2007,Zehavi2011,Leauthaud2012,Coupon2015} and could potentially be used to constrain assembly bias in future work \citep{Zentner2014,Zentner2019,Vakili2019}.
A few studies have used the redshift-space correlation function \citep[e.g.,][]{Tinker2006a,Parejko2013,Guo2015b, Beltz-Mohrmann2020}, which is sensitive to the velocity assumptions in our HOD model and thus could be used to investigate velocity bias.
Even fewer studies have incorporated other clustering statistics, like the mark correlation function, which is sensitive to assembly bias \citep[e.g.,][]{Zu2018}; the group multiplicity function, which is sensitive to both high-mass occupation \citep[e.g.,][]{Zheng2007a,Sinha2018} and spatial bias \citep[e.g.,][]{Beltz-Mohrmann2020}; the group velocity dispersion, which must be sensitive to velocity bias; or counts-in-cells statistics, which are sensitive to assembly bias \citep[e.g.,][]{Tinker2006,McCullagh2017, Walsh2019,Wang2019,Beltz-Mohrmann2020}.

Although the HOD model that we use in this work does not include exotic features, these new clustering statistics can still provide significant constraining power for our current model.
In this work we use an optimal combination of these statistics on various scales to extract more clustering information from the SDSS survey than in all previous studies.
Additionally, we greatly improve the accuracy of our modeling procedure compared to S18 and perform an extensive analysis of both the statistical and systematic errors present in our procedure, further pushing galaxy clustering analysis into the accurate regime.
These improvements will allow us to tighten our constraints on the model and more robustly characterize the tension found in S18.

We discuss our observational data in Section~\ref{sec:obs}, our clustering statistics in Section~\ref{sec:stats}, our simulations and mock galaxy catalogs in Section~\ref{sec:tools}, and our modeling procedure in Section~\ref{sec:model}.
In Section~\ref{sec:opt} we discuss our selection process for the measurements with the most constraining power. 
We discuss our results in Section~\ref{sec:results} and provide conclusions in Section~\ref{sec:conc}.
Finally, in the appendices we provide a detailed analysis of the accuracy of several components of our modeling procedure, including our fiber collision correction (Appendix~\ref{app:fib}), our model estimation (Appendix~\ref{app:model_error}) and covariance matrix (Appendix~\ref{app:cov_error}), and the ability of our procedure to recover the correct HOD parameters of mock galaxy catalogs (Appendix~\ref{app:mock_chain}).

\section{Observational Dataset} \label{sec:obs}

In this work, we make use of the seventh data release \citep[DR7;][]{Abazajian2009} of the Sloan Digital Sky Survey \citep[SDSS;][]{York2000}. 
Specifically, we utilize the large scale structure samples from the NYU Value Added Galaxy Catalog \citep[NYU-VAGC;][]{Blanton2005}.
As in S18, we use a parent sample of over 500,000 galaxies covering the northern footprint of SDSS (see S18 for more details).
The absolute magnitudes of these galaxies have been k-corrected to rest-frame magnitudes at redshift $z=0.1$ \citep{Blanton2003b} but, unlike in S18, have not been corrected for passive luminosity evolution.
We find that, when comparing volume-limited samples constructed with and without the passive luminosity correction, the number density of the sample without the correction is closer to being constant in redshift.
Therefore, we do not include this correction.

\begin{deluxetable}{cccccc}
\tablenum{1}
\tablecaption{SDSS Volume-limited Sample Parameters\label{tab:sdss}}
\tablewidth{0pt}
\tablehead{
\colhead{$M_r^\mathrm{lim}$} & \colhead{$z_\mathrm{min}$} & \colhead{$z_\mathrm{max}$} & \colhead{$z_\mathrm{median}$} &
\colhead{$V_\mathrm{eff}$} & \colhead{$n_g$} \\
\colhead{} & \colhead{} & \colhead{} & \colhead{} &
\colhead{$(h^{-3}\mathrm{Mpc}^3)$} & \colhead{$(h^3\mathrm{Mpc}^{-3})$}
}
\startdata
$-19$ & 0.02 & 0.07 & 0.0562 & 6,087,119 & 0.01463 \\
$-21$ & 0.02 & 0.158 & 0.1285 & 67,174,396 & 0.00123
\enddata
\tablecomments{The columns list (from left to right): the absolute magnitude threshold of each sample at $z = 0.1$; the minimum, maximum, and median redshifts, respectively; the effective volume of each sample (corrected for incompleteness); and the galaxy number density of each sample.}
\vspace{-3mm}
\end{deluxetable}

From this parent sample, we construct two volume-limited subsamples, each complete down to a specified r-band absolute magnitude threshold. 
To construct these samples, we first choose redshift limits, $z_\mathrm{min}$ and $z_\mathrm{max}$.
For both samples, we adopt a $z_\mathrm{min}$ of 0.02 so that the redshift-space distortions are small compared to cosmological redshifts.
We choose a value of $z_\mathrm{max}$ such that the redshifted spectra of all galaxies with our limiting absolute magnitude would still fall within the redshift survey’s apparent magnitude and surface brightness limits if placed at $z_\mathrm{max}$.
Our low-luminosity sample is complete down to an $r$-band absolute magnitude of $-19$ in the redshift range $0.02 - 0.07$, while our high-luminosity sample is complete down to an $r$-band absolute magnitude of $-21$ in the redshift range $0.02 - 0.158$.
We refer to our two samples as ``$-19$" and ``$-21$" henceforth.
The redshift limits, median redshift, effective volume, and number density of these two samples are listed in Table~\ref{tab:sdss}. 
These values are slightly different than those used in S18.

Throughout this paper, we adopt a flat \lcdm cosmological model with \Omegam = 0.302 and $h=1$.
The co-moving distances of SDSS galaxies are determined based on this assumption and have units of \hmpc.
Similarly, reported absolute magnitudes are actually $M_r + 5\mathrm{log}h$\footnote{Throughout this paper, $\mathrm{log}$ refers to $\mathrm{log}_{10}$.}.

\section{Clustering Statistics} \label{sec:stats}

Many studies which use clustering statistics\footnote{We refer to a type of measurement (e.g., \wprp) as a ``clustering statistic" and a measurement in a specific bin (e.g., \wptwo) as an ``observable."} to constrain the galaxy-halo connection \citep[e.g.,][]{Zehavi2011} focus on analytic predictions of these measurements.
Such studies are limited to employing only those clustering statistics for which analytic predictions are possible.
Given the difficulty in writing down the analytic form of an arbitrary clustering statistic, as well as the need to consider the impacts of survey geometry and incompleteness, this limitation unnecessarily excludes many clustering statistics with potentially useful information for the galaxy-halo connection. 
One of the key advantages of employing a numerical modeling procedure is that clustering statistics can be measured in \textit{identical} ways on both the SDSS galaxies \textit{and} on mock galaxy catalogs generated according to our model.
Such a procedure thus allows us to make apples-to-apples comparisons between the data and our theoretical model, allowing us to employ any arbitrary clustering statistic with the knowledge that any systematic errors (e.g., edge effects) are reflected the same way in both the model and the data.

\citet{Berlind2002}, and more recently S18 and \citet{Hadzhiyska2021a}, demonstrate that different clustering statistics are sensitive to different combinations of HOD parameters, and that analyses involving several different galaxy clustering statistics have the most power to constrain galaxy bias within an HOD framework.
In their analysis, S18 employ the galaxy number density, the projected correlation function, and the group multiplicity function. 
With the aim of increasing the precision of the S18 results, we utilize four additional clustering statistics in this work: the redshift-space correlation function, the average group velocity dispersion, the mark correlation function, and counts in cells statistics.
We choose these new clustering statistics for their potential ability to constrain models beyond the standard HOD (i.e., those that include assembly bias, spatial bias, and velocity bias), which we intend to explore in future work.
All of our clustering statistics (other than galaxy number density) are described below.

\subsection{The projected correlation function: \wprp}
\label{subsec:wp}
In general, a correlation function is the excess number of galaxy pairs above that which is expected for a random distribution of points, as a function of pair separation.
Widely used in studies of galaxy clustering \citep[e.g.,][S18]{Zehavi2011}, the projected correlation function aims to remove the effect of redshift-space distortions by first counting pairs of galaxies in bins of their line-of-sight and projected components, $\pi$ and $r_\mathrm{p}$, and then integrating over $\pi$:
\begin{equation}
w_\mathrm{p}(r_\mathrm{p}) = 2 \int_{0}^{\pi_\mathrm{max}}\xi(r_\mathrm{p},\pi)d\pi.
\end{equation}

We use the \citet{Zehavi2002} definitions of $r_\mathrm{p}$ and $\pi$, and we count pairs of galaxies in 14 evenly spaced logarithmic bins of projected separation $r_\mathrm{p}$ between $0.1$ and $20$ \hmpc. 
We then integrate out to $\pi_\mathrm{max}$ of $40$ and $80$ \hmpc for the $-19$ and $-21$ samples, respectively. 
We calculate $\xi(r_\mathrm{p},\pi)$ with the Landy-Szalay estimator $(DD - 2DR + RR)/RR$ \citep{Landy1993}, where $DD$, $DR$, and $RR$ are the normalized numbers of data-data, data-random, and random-random pairs, respectively.
We use the publicly available \textsc{corrfunc} \citep{Corrfunc2019,Corrfunc2020} to compute our projected correlation function. 

\subsection{The redshift-space correlation function: \zxi}
\label{subsec:zxi}
The redshift-space correlation function \zxi is the excess number of galaxy pairs above that which is expected for a random distribution of points, as a function of redshift-space (i.e., not projected) pair separation $s$. 
Because the redshift-space distortions of galaxies depend on their velocities, measuring \zxi will allow us to investigate the assumption that galaxies trace the velocity distribution of dark matter within their halo.
To compute \zxi, we once again use \textsc{corrfunc} and the Landy-Szalay estimator, counting pairs in the same bins of separation as those used for \wprp (but not projected).

\subsection{The group multiplicity function: \gmf}
\label{subsec:gmf}
The group multiplicity function is the number density of galaxy groups as a function of the number of galaxies (``richness") in the group, $n(N)$ \citep[e.g.,][]{Berlind2002}.
We use the \citet{Berlind2006} friends-of-friends algorithm to identify groups: galaxies are linked together if their projected and line-of-sight separations are both less than a corresponding linking length. 
We adopt the \citet{Berlind2006} linking lengths of $b_{\bot}=0.14$ and $b_{\parallel}=0.75$, which are given in units of the mean inter-galaxy separation $n_\mathrm{g}^{-1/3}$, where $n_\mathrm{g}$ is the sample number density.
For the $-21$ sample, we measure groups in the following five (inclusive) bins of $N: (5-6), (7-9), (10-13), (14-19), (20-27).$
For the $-19$ sample, we measure groups in the following seven (inclusive) bins of $N: (5-6), (7-9), (10-13), (14-19), (20-32), (33-52), (53-84).$ 
These are the same bins as those used in S18, with the exception of the largest bin in S18 for each sample, which is excluded from this analysis.

\subsection{Average group velocity dispersion: \sigN}
\label{subsec:sigN}
We also measure the average velocity dispersion of galaxy groups as a function of richness, \sigN.
The velocity of each galaxy in the group is first computed by subtracting the velocity of the entire group (we only consider line-of-sight peculiar velocities).
For each galaxy group, we then calculate the group velocity dispersion.
Lastly we find the average group velocity dispersion of all groups within a particular richness bin (e.g., groups containing $10-13$ galaxies).
We use the same richness binning as in the group multiplicity function.
Like \zxi, measuring \sigN also probes the assumption that galaxies trace the velocities of the underlying dark matter particles within the halo.

\subsection{The mark correlation function: \mcf}
\label{subsec:mcf}

The mark correlation function, \mcf, is given by
\begin{equation}
\label{eq:mcf}
\mathrm{mcf}(s) = \frac{\mathrm{MM}(s)}{\mathrm{DD}(s)},
\end{equation}
where $\mathrm{DD}(s)$ is the number of galaxy-galaxy pairs as a function of redshift-space separation $s$, and $\mathrm{MM}(s)$ is the sum of a weighted pair-product of the same galaxy-galaxy pairs.
Both DD and MM are normalized by, respectively, the total number of galaxy-galaxy pairs and the sum of the product of weighted galaxy-galaxy pairs, at all separations.

To compute \mcf, we must first assign a weight (or ``mark") to each galaxy.
Motivated by \citet{Salcedo2018}, who found that secondary bias of halos can be explained by a halo's distance to a massive neighbor, we choose as our mark the distance from a galaxy to the nearest galaxy cluster.
If the galaxy is located in a cluster, then we take the distance to the cluster's center.
To choose the minimum number of galaxies to be considered a ``cluster," we take the best-fit HOD parameters from S18 and determine the average integer number of galaxies that would be placed in a halo of mass $10^{14} M_\odot$ according to Equations \ref{eq:ncen} and \ref{eq:nsat} (see Section~\ref{subsec:mocks}).
With this definition, to be considered a cluster, a galaxy-group must contain 15 and three galaxies for the $-19$ and $-21$ samples, respectively.
For our purposes, whether or not this definition accurately weights each galaxy by its distance to the nearest cluster is irrelevant.
Ultimately, we only care whether or not this statistic contains information that can be used to constrain our model, a question we explore in Section~\ref{sec:opt}.
We again make use of \textsc{corrfunc} to compute \mcf.

\subsection{Counts in cells: \cic}
\label{subsec:cic}
Counts-in-cells statistics measure the probability of finding a given number of galaxies within a randomly placed finite region (e.g., a sphere) as a function of region size (e.g., radius). 
One special case of this is the void probability function (VPF), which measures the probability of finding no galaxies in a random region of space.
Another variation of counts-in-cells is the ``singular probability function," (SPF) or the probability of finding exactly one galaxy in a randomly placed region.

\citet{Tinker2006} and more recently \citet{McCullagh2017} and \citet{Wang2019} found that, in contrast to the projected correlation function, the VPF is sensitive to environmental variations of the HOD, and thus could be used to rule out certain HOD models. 
We measure counts-in-cells (both the VPF and the SPF) in spheres of evenly spaced bins with radii of 2, 4, 6, 8, and 10 \hmpc.
We again make use of \textsc{corrfunc} to compute VPF and SPF.

\section{From Simulations to Mock Galaxy Catalogs} \label{sec:tools}

\subsection{Simulations and halo catalogs}
\label{subsec:sims}

In this work we make use of dark matter only (DMO) cosmological N-body simulations from the Large Suite of Dark Matter Simulations project \citep[LasDamas;][]{McBride2009}.
These simulations were run on the Texas Advanced Computing Center's Stampede supercomputer using the public code \textsc{gadget-2} \citep{Springel2005}. 
We generate power spectra and initial conditions with \textsc{cmbfast} \citep{Seljak1996,Zaldarriaga1998,Zaldarriaga2000} and \textsc{2lptic} \citep{Scoccimarro1998, Crocce2006}, respectively. 
All simulations were run with the following cosmological parameters \citep{Planck2014}: $\Omega_\mathrm{m}=0.302$, $\Omega_{\Lambda}=0.698$, $\Omega_\mathrm{b}=0.048$, $h=0.681$, $\sigma_8=0.828$, and $n_s=0.96$. 
In particular, we run two sets of simulations: one to mimic the SDSS $-19$ sample (which we call Consuelo), and another to mimic the SDSS $-21$ sample (which we call Carmen). 
Starting from $z=99$, we run Consuelo to $z=0.054$, and we run Carmen to $z=0.132$, which are roughly the median redshifts of the SDSS $-19$ and $-21$ samples, respectively\footnote{These values are the median redshifts of the SDSS samples in S18. 
Our median redshifts (Table~\ref{tab:sdss}) differ slightly due to changes in how we process SDSS (see Section~\ref{sec:obs}).}. 

\begin{deluxetable*}{ccccccccc}
\tablenum{2}
\tablecaption{Simulation Parameters\label{tab:simulations}}
\tablewidth{0pt}
\tablehead{
\colhead{Use} & \colhead{Sample} & \colhead{Simulation} & \colhead{Seeds} &
\colhead{$L_\mathrm{box}$} & \colhead{$N_\mathrm{part}$} & \colhead{$m_\mathrm{part}$} & \colhead{$\epsilon$} & \colhead{Number} \\
\colhead{} & \colhead{} & \colhead{} & \colhead{} &
\colhead{(\hmpc)} & \colhead{} & \colhead{(\hMsun)} & \colhead{(\hkpc)} & \colhead{}
}
\startdata
Covariance matrix & $-19$ & Consuelo & 4001 - 4100 & 420 & $1400^3$ & $2.26 \times 10^9$ & 8 & 100 \\
Covariance matrix & $-21$ & Carmen & 2001 - 2100 & 1000 & $1120^3$ & $5.97 \times 10^{10}$ & 25 & 100 \\
MCMC & $-19$ & ConsueloHD & 4002, 4022 & 420 & $2240^3$ & $5.53 \times 10^8$ & 5 & 2 \\
MCMC & $-21$ & CarmenHD & 2007, 2023 & 1000 & $2240^3$ & $7.46 \times 10^9$ & 12 & 2
\enddata
\tablecomments{The columns list (from left to right): what each simulation is used for, the absolute magnitude threshold of the corresponding SDSS sample, the name of the simulation, the seeds used, the boxsize, number of particles, mass resolution, force softening, and the number of simulations.}
\vspace{-5mm}
\end{deluxetable*}

For each sample, we run two high-resolution simulations and 100 low-resolution simulations.
We use the high-resolution simulations to estimate the model observables and the low-resolution simulations to construct a covariance matrix representing cosmic variance (see Section~\ref{sec:model}). 
Each of the low-resolution simulations differs in the phases of the density modes of the power spectrum, which is controlled by a seed supplied to \textsc{2lptic}.
The seeds used for the high-resolution simulations were chosen from the 100 low-resolution seeds to minimize the cosmic variance error in our model (see S18 for more details). 
The details of these simulations are given in Table~\ref{tab:simulations}. 

We identify halos using a spherical over-density \citep[SO;][]{Lacey1994} algorithm using the \textsc{rockstar} phase-space temporal halo finder \citep{Behroozi2013}. 
We find halos using the $M_\mathrm{vir}$ halo definition, where the halo density depends on both cosmology and redshift, as given by \citet{Bryan1998}. 

\subsection{Building mock galaxy catalogs}
\label{subsec:mocks}
We can construct mock galaxy catalogs by directly populating the dark matter halos from our simulations.
This population is performed via the Halo Occupation Distribution (HOD).
Specifically, the form of the HOD we use is the ``vanilla" model used by S18 and previously \citet{Zheng2007}.
In this form of the HOD, the population statistics of dark matter halos depend only on halo mass, with central and satellite galaxies treated separately \citep{Kravtsov2004,Zheng2005}.

For a halo of mass $M$, the average number of central galaxies is given by
\begin{equation}
\label{eq:ncen}
\langle N_\mathrm{cen} \rangle = \frac{1}{2}\bigg[1 + \mathrm{erf} \bigg(\frac{\mathrm{log} M - \mathrm{log}M_\mathrm{min}}{\sigma_{\mathrm{log} M}}\bigg)\bigg].
\end{equation}
Within the error function, there are only two parameters, \mmin and \siglogm, which control the population statistics for central galaxies. 
\mmin is the mass at which approximately half of the halos host a central galaxy, while \siglogm dictates how rapidly the central population goes to zero with decreasing mass.
When we consider whether or not to assign a central galaxy to a specific halo of mass $M$, we draw a random number $R$ from a uniform distribution on the interval $[0,1)$. 
If $R < \langle N_\mathrm{cen} \rangle$, then a galaxy is assigned.
This central galaxy is placed at the halo center and given the mean velocity of the halo. 

If we have placed a central galaxy in a halo of mass $M$, then the mean number of satellite galaxies is given by
\begin{equation}
\label{eq:nsat}
\langle N_\mathrm{sat} \rangle = \langle N_\mathrm{cen} \rangle \times \bigg(\frac{M - M_0}{M_1}\bigg)^\alpha.
\end{equation}
The exact number of satellite galaxies we place in a specific halo of mass $M$ is determined by drawing from a Poisson distribution with a mean $\langle N_\mathrm{sat} \rangle$.
While there is a dependence on $\langle N_\mathrm{cen} \rangle$ (and thus on \mmin and \siglogm), the satellite population statistics are primarily governed by three parameters: $M_1$, $\alpha$, and $M_0$.
More intuitively, halos of mass $M_1$ on average host one satellite galaxy, while $\alpha$ dictates how rapidly halo occupation increases with increasing halo mass.
The parameter $M_0$ technically sets the halo mass below which we do not find any satellite galaxies, but in practice this parameter has not been well-constrained in studies which use luminosity samples similar to our own \citep[e.g.,][S18]{Guo2015b}.
After we choose the exact number of satellite galaxies in a specific halo, each satellite is given the position and velocity of a randomly chosen dark matter particle in the halo.

Once we populate our dark matter halos with galaxies, we must build realistic mock galaxy catalogs that resemble our SDSS samples of interest.
To do this, we first need to transpose the mock galaxies from Cartesian to spherical coordinates by positioning an observer at the center of the box and converting the positions of the galaxies into RA, DEC, and co-moving distances.
Due to the smaller volume of the SDSS sample, we are able to carve out four independent mock galaxy catalogs from each simulation box.
We then need to incorporate the same systematic effects that plague our observational dataset, such as redshift-space distortions, sample geometry, and incompleteness. 

To introduce redshift-space distortions, we determine the line-of-sight peculiar velocities of the galaxies and calculate the redshift as $1+z = (1+z_\mathrm{cosmological})(1+z_\mathrm{doppler})$, where $z_\mathrm{cosmological}$ is the cosmological redshift and $z_\mathrm{doppler}$ is the redshift due to the radial peculiar velocity.
We then eliminate any mock galaxies outside the redshift limits or sky footprint of our SDSS sample.
This procedure ensures that any effect on measured clustering statistics due to sample geometry are present in both the mock galaxies used in our model and in the SDSS data.

\subsection{Correcting for fiber collisions}
\label{subsec:fiber}
In SDSS, spectroscopic fibers cannot be placed closer to each other than their own thickness ($55''$). 
This limitation results in approximately 7\% of targeted galaxies lacking a spectroscopically measured redshift due to their proximity to a neighboring galaxy.
Modeling these fiber collisions is difficult primarily because it requires a simulation with a large enough volume and a high enough resolution to reproduce the flux-limited footprint of SDSS.
Instead of incorporating fiber collisions into our model, we choose to make a correction to the SDSS data.
It should be noted that this is the one case where we necessarily diverge from our general modeling philosophy -- instead of incorporating this systematic error into our generative model, we attempt to remove it from the SDSS data.

To accomplish this task, we first apply the nearest neighbor correction to galaxies lacking a spectroscopic redshift \citep{Zehavi2002}.
We then estimate the error in measured clustering statistics from using \textit{only} the nearest neighbor correction and adjust our measurements accordingly.
Our full procedure is detailed in Appendix~\ref{app:fib}, but we provide a brief summary here.
We utilize the SDSS plate overlap regions to estimate the error in our clustering statistics that results from applying this nearest neighbor correction.
In overlap regions, we have spectroscopic redshifts for many galaxies that are within $55''$ of a neighboring galaxy.
These regions thus allow us to investigate the impact of applying the nearest neighbor correction.
Briefly, we find that the nearest neighbor correction alone is not sufficient for many of the clustering statistics we use.
Thus, we choose to adjust the values of our observables to account for this error.
We use the adjusted observables in our analysis of SDSS (Section~\ref{sec:results}).
We provide both the original nearest neighbor SDSS observables and the adjusted observables in the machine-readable format.

\section{Summary of Modeling Procedure} 
\label{sec:model}
Our main goal in this work is to constrain the galaxy-halo connection.
To achieve this end we implement a fully numerical modeling methodology, based on and adapted from S18.
We summarize the procedure here, as well as highlight a few key differences from S18.

To employ our numerical modeling procedure, we utilize the Texas Advanced Computing Center's Stampede2 supercomputer.
We use a Markov Chain Monte Carlo (MCMC) algorithm to explore the HOD parameter space.
In particular, we developed a C-implementation of the popular affine-invariant sampler \emcee \citep{EMCEE2013}, which we call \cemcee\footnote{https://github.com/aszewciw/emcee\_in\_c Our decision to use \cemcee was made to meet the software restrictions of the Stampede2 supercomputer. We have verified that \cemcee performs identically to \emcee.}.

An MCMC (or ``chain") involves probabilistic sampling of an unknown posterior distribution of parameter values given some assumed prior distribution.
We employ flat prior distributions on the same parameter ranges given in S18.
At each HOD point we sample, we must evaluate the likelihood that this HOD model could have generated a dataset with the same clustering as SDSS.
Given a K-dimensional vector \textbf{D} of observables measured on the SDSS dataset and a corresponding vector \textbf{M} of the same observables for an HOD model (``model observables" hereafter), this likelihood is given by
\begin{equation}
    \label{eq:likelihood}
    \mathcal{L}(\mathbf{D}|\mathbf{M}) = \frac{\exp(-\frac{1}{2}{(\mathbf{D} - \mathbf{M}) \mathbf{C}^{-1} (\mathbf{D} - \mathbf{M})^{T}})}{\sqrt{(2\pi)^K\mathrm{det}(\mathbf{C})}},
\end{equation}
where \textbf{C} is the K-dimensional covariance matrix of these observables representing cosmic variance (see Equation \ref{eq:covariance}).
Ignoring the factor of $-1/2$, the term within the exponential is $\chi^2$.

In the context of a numerical modeling methodology, ideally we would obtain the observable vector \textbf{M} by first populating a large number of high-resolution DMO simulations with galaxies according to our set of HOD parameters.
From these populated simulation boxes, we would then carve out SDSS-like mock galaxy catalogs, measure all observables, and find the average value of each observable across all mocks.
In practice, we need to populate halos in a large enough volume in order to yield a cosmic variance error in \textbf{M} that is sufficiently smaller than the uncertainty in \textbf{D} so as not to dominate the overall error budget.
As discussed in Section~\ref{subsec:sims}, S18 found that two boxes were sufficient to accomplish this task for an analysis using \ngal, \wprp, and \gmf.
We continue forward with this approach using the same boxes as S18 and investigate the cosmic variance error in \textbf{M} for the expanded set of clustering statistics we use.

\begin{deluxetable}{ccc}
\tablenum{3}
\tablecaption{Clustering Statistics Measurement Method\label{tab:method}}
\tablewidth{0pt}
\setlength{\tabcolsep}{18pt}
\tablehead{
\colhead{Statistic} & \colhead{$-19$ Method} & \colhead{$-21$ Method}}
\startdata
\ngal & B-02 & B-07 \\
\wprp & B-02 & B-07 \\
\zxi & B-02, B-22 & B-23 \\
\gmf & M-22 & M-07, M-23 \\
\sigN & M-02, M-22 & M-07, M-23 \\
\mcf & M-02, M-22 & M-07, M-23 \\
\cic & B-02, B-22 & M-07, M-23
\enddata
\tablecomments{Shown here are the methods for calculating the clustering statistics for the $-19$ and $-21$ samples. For each sample, there are two simulations: 4002 and 4022 for $-19$, and 2007 and 2023 for $-21$. From each box we can create 4 mocks. Therefore, each statistic was either calculated on one box (e.g., Box 4002, or B-02), 4 mocks (e.g., M-22), two boxes (e.g., B-02, B-22), or 8 mocks (e.g., M-02, M-22).}
\vspace{-5mm}
\end{deluxetable}

For each HOD point we explore in the MCMC, we directly populate the halos in these two boxes\footnote{To meet the memory requirements of the Stampede2 compute nodes, we use downsampled versions of these boxes in which we only keep a randomly selected 5\% of the dark matter particles in each halo. This still leaves ample particles on which to place satellites ($\sim500$ times more particles than satellites on average). Thus, downsampling results in no loss of accuracy.}.
We have a choice as to whether to measure each clustering statistic either on the full box(es) or on mock galaxy catalogs carved out from these boxes (see Section~\ref{subsec:mocks}).
The goal is to measure each statistic in a way that yields model observables closest to the ideal case described above (i.e., mean of many mocks).
Briefly, utilizing the two low-resolution counterparts to our high-resolution boxes, we compare different methods of measurement and choose the method for each clustering statistic that produces model observables closest to the mean measurement across 400 low-resolution mocks.
We describe our full procedure in Appendix~\ref{app:model_error}.
We summarize the results of this test in Table~\ref{tab:method}, where, for each sample, we indicate whether we choose to measure each clustering statistic on a box (B) or on mocks (M) and which box(es)/mock(s) are used (see table description for more details).

We take one last step when estimating the model observables in our chains.
For a given HOD, the process of populating dark matter halos with galaxies is stochastic (see Section~\ref{subsec:mocks}) and is controlled with a ``population seed."
For a fixed HOD, varying this population seed primarily affects the number of centrals/satellites assigned to each halo and can lead to significant differences in measured clustering statistics (see Appendix~\ref{app:model_error} and Figure~\ref{fig:model_error}).
Even when using a fixed population seed, a slight change in HOD can also lead to significant differences in measured clustering statistics.
Such differences do not arise from an actual difference in the likelihoods of these two different models but rather from the noise in how we are estimating the model observables.
To reduce this noise, for each HOD point in our chain, we populate our simulation boxes four times, using four fixed population seeds.
We then measure the observables (in the way given in Table~\ref{tab:method}) on each of these four resultant galaxy catalogs.
We use the average of these four measurements as our model observables.

Similar to the model observables \textbf{M}, ideally we would obtain the covariance matrix \textbf{C} using a large number of high-resolution mock galaxy catalogs.
Because of the high computational cost associated with running many high-resolution DMO simulations, we instead construct the covariance matrix from 400 mock galaxy catalogs built from 100 low-resolution boxes, all populated according to a set of fiducial HOD parameters (see Sections~\ref{sec:opt} and \ref{sec:results}).
The elements of the covariance matrix are given by
\begin{equation}
\label{eq:covariance}
C_{ij} = \frac{1}{N-1} \sum_{1}^{N}(y_i - \overline{y_i})(y_j - \overline{y_j}).
\end{equation}
Here, the sum is taken over the $N=400$ mocks.
The values $y_i$ and $y_j$ are the $i^{th}$ and $j^{th}$ observables measured on each mock and thus vary in the sum.
The values $\overline{y_i}$ and $\overline{y_j}$ are the mean values of the $i^{th}$ and $j^{th}$ observables, respectively, and thus are fixed in the sum.
Each diagonal element, $C_{ii}$, of the matrix is simply the variance among 400 mocks for observable $i$.
In this paper, we refer to $\sqrt{C_{ii}}$ as the ``cosmic variance uncertainty" of observable $i$.
For simplicity, we denote the cosmic variance uncertainty of an arbitrary observable as $\sigma_\mathrm{obs}$.

There are three important differences between the covariance matrices we use and those used in S18.
First, we use twice the number of mocks as S18, resulting in less noise in our covariance matrix.
Second, the low-resolution boxes we use to construct the matrix all have the same cosmology and halo definition as the high-resolution boxes we use to obtain our model observables.
This is not the case in many of the chains run in S18.
Third, instead of using the raw covariance matrix produced by these 400 mocks, S18 rescale the elements of the covariance matrix by the ratio of SDSS to mock measurements (see S18 for more details).
When evaluating $\mathcal{L}(D|M)$, however, the covariance matrix represents the variation in clustering statistics calculated on different mock realizations generated by the same set of HOD parameters.
Therefore, we instead simply use the raw covariance matrix in our likelihood calculation and do not rescale its elements.
While we view all three of these points as improvements to the modeling procedure, we do not quantify their impact.

In addition to these changes, we investigate sources of noise in our covariance matrix in Appendix~\ref{app:cov_error}.
We examine only the noise in the cosmic variance uncertainty of each observable, ignoring off-diagonal elements of the matrix.
To summarize, for the $-19$ sample, the major source of noise is due to the number of mocks we use, which gives an error of $\sim$3.5\% on all values of $\sigma_\mathrm{obs}$.
For a few observables, the largest source of noise is due to our choice to use a fiducial matrix instead of re-building the matrix for each set of HOD parameters we evaluate in the chain.
Still, this error is at worst only 10\%.
For the $-21$ sample, the largest source of noise comes from using low-resolution mocks instead of high-resolution mocks to build our matrices.
In the worst case, we find that we overestimate the errors by $\sim$15\% for several small scales of \wprp and \zxi.
We do not make any corrections to our matrix to account for this discrepancy, believing that overestimating our errors will lead to more conservative posterior results.

While the tests in Appendix~\ref{app:cov_error} quantify the error in the elements of the covariance matrix, it is a much more difficult task to quantify the joint impact of these errors.
In particular, we wish to know how these errors affect our constraints on SDSS.
In general, using more observables will produce tighter constraints, but this improvement occurs at the cost of more noise in the covariance matrix and thus in the resulting constraints.
We therefore seek to obtain a set of observables that provide both tight and reliable constrains, a task we address in the following section.

\section{Optimization Algorithm} \label{sec:opt}

From our set of $N$ observables, we seek to find the subset of size $K\leq N$ that gives the tightest constraints on our HOD parameters, while still being robust with respect to the systematic errors in our modeling procedure.
Because we do not know what an acceptable value for $K$ is, exploring all combinations quickly becomes an impossible task.
With $N = 63$, as in our $-21$ sample, $N$ choose $K$ for $K: 1 \rightarrow N$ represents $\sim 10^{19}$ unique combinations!
Therefore, we design and implement an algorithm that is an approximation of this task.

The algorithm we implement orders our $N$ observables by joint constraining power.
When choosing the $K^{th}$ observable, we pick the one that, when combined with our $K-1$ previously chosen observables, produces the tightest constraints on our HOD parameters.
Once the order is created, we analyze the reliability of our constraints for different values of $K$.
This analysis leads to a choice of $K$ observables which we utilize in an MCMC to constrain the HOD of our SDSS volume-limited samples.
In the following subsections, we describe our algorithm, its raw results, and our choice of observables.

\subsection{Algorithm design}
\label{subsec:alg}

The first step in our algorithm is to set up a grid of HOD parameters which we will use to explore the constraining power of different combinations of observables.
Because some of our parameters are strongly correlated (e.g., \logmmin and \siglogm), a uniformly-spaced grid contains a large number of unrealistic HOD parameter combinations.
Therefore, to set up a grid we perform the following steps:

\begin{enumerate}[noitemsep]
    \item Choose a fiducial set of HOD parameters.
    \item Using these parameters, create a high-resolution fiducial mock galaxy catalog.
    \item Measure \ngal and $w_\mathrm{p}(r_\mathrm{p} \sim 0.3 \ h^{-1}\mathrm{Mpc})$ on this mock.
    \item Run an MCMC with this mock's observables as the ``data." For each HOD point:
    \begin{itemize}[noitemsep]
        \item Calculate and record \textit{all} $N$ model observables. (See Section~\ref{sec:model} for a description of how model observables are calculated.)
        \item Evaluate likelihood \Linit using \textit{only} \ngal and $w_\mathrm{p}(r_\mathrm{p} \sim 0.3 \ h^{-1}\mathrm{Mpc})$.
    \end{itemize}
\end{enumerate}

This procedure produces an initial non-uniform grid of fairly reasonable HOD points which could have generated our fiducial mock.
We choose to perform this analysis on a mock galaxy catalog for two reasons.
First, we wish to know whether a chain run with a specific set of observables is able to recover the true HOD, which is not something we can determine with SDSS.
Second, we wish to choose observables based on their universal constraining power.
By using a mock galaxy catalog, we avoid choosing observables that are over-fit to SDSS.
We are, however, potentially over-fitting to this particular mock galaxy catalog.
Therefore, we perform steps 2-4 for four different mock galaxy catalogs.
These mocks have the same HOD and only differ due to cosmic variance.
Thus, we construct \textit{four} different grids which we use to explore the constraining power of different combinations of observables.

For a given mock/grid, we quantify the constraining power of \ngal and \wptwo by measuring the standard deviation of HOD values of each parameter.
We call this initial constraint \siginit.
We wish to examine how different combinations of observables can improve upon the initial constraints.
To do so, we utilize the fact that the value of \Linit at a particular point in HOD-space is proportional to the density of points in the grid at that location.
Because we measure all N observables when constructing the grid, for each HOD point we can compute a likelihood $\mathcal{L}_K$ using any arbitrary set of $K$ observables and the appropriate $K$ by $K$ covariance matrix.
We can then assign each point in the grid a weight $\mathcal{L}_K$/\Linit. 
The weight assigned to each HOD point is proportional to the density of points we would have at this location if we had we run a chain on this mock using this set of $K$ observables.
Computing a weighted standard deviation \sigk for each parameter gives an estimate of the constraints from said chain.
This process, known as ``importance sampling," thus provides a way of estimating the constraining power of an arbitrary combination of $K$ observables.

We use importance sampling to order our $N$ observables by their cumulative constraining power.
In short, we choose observables, one at a time, always choosing the observable that, when combined with the previously chosen observables, best constrains all parameters of interest.
To choose the $K^{th}$ observable, we perform the following steps:

\begin{enumerate}[noitemsep]
    \item From the $N-(K-1)$ observables not yet chosen, pick a ``trial observable."
    \item In one of the four grids, using this trial observable and the $K-1$ previously chosen observables, compute a likelihood $\mathcal{L}_K$ for each point.
    \item Weight each point in the grid by $\mathcal{L}_K$/\Linit.
    \item Compute a weighted standard deviation, \sigk, for each HOD parameter.
    \item Add in quadrature the fractional reduction in the constraints on each parameter achieved by including this observable: $M_\mathrm{grid}=\sum_{\mathrm{par}}\bigg(\frac{\sigma_{\mathrm{par},K}}{\sigma_{\mathrm{par},K-1}}\bigg)^2$.
    \item Repeat steps 2 - 5 for each of the grids. Add the values of $M_\mathrm{grid}$ to get the metric $M$ for this trial observable.
    \item Repeat steps 1 - 6 for each of the $N-(K-1)$ possible trial observables.
    \item Choose the trial observable with the lowest value of $M$.
\end{enumerate}

Once the $K^{th}$ observable is chosen, we move on to choose the $(K+1)^{th}$ until all $N$ observables have been ordered.
The metric $M$ always compares the improvement in constraints when going from $K-1$ to $K$ observables.
This metric rewards observables which constrain parameters that have not already been well-constrained by the first $K-1$ observables.
Additionally, by minimizing the sum of $M_\mathrm{grid}$ across multiple grids, the choices we make are more robust to the peculiarities of a specific mock due to cosmic variance.

In principle, this procedure could be used to order all observables from $K=1$ to $K=N$.
In practice, however, we make the decision to start with \ngal and \wptwo and only choose observables from $K=3$ to $K=N$.
In general, importance sampling works well when the target distribution is close to or contained within the initial distribution.
This criteria is met when choosing the third observable but is not always true when we attempt to choose the first or second.
Our choice to build the initial grid using \ngal and \wptwo thus affects all successive choices of observables.
We assume that \ngal will be included in any clustering analysis, given the low computational cost and high information content.
Thus, including it in the construction of this grid seems reasonable.
We first attempted to create our grid using only \ngal, but this grid contained HOD points very unlike SDSS, particularly in the satellite parameters.
In preliminary analysis, we found that \wptwo is particularly good at constraining the satellite parameters.
Thus, we include it to acquire a more reasonable starting grid of HODs.

Importance sampling works if the grid contains a sufficient number of points in the region of the target distribution.
As we go to higher values of $K$, the target distributions occupy a smaller region in HOD space, and the density of our grid can become an issue.
Ideally, each time we choose a new observable, we would construct a new grid by running a new chain with the already chosen observables.
This denser grid would then be used to choose the next observable.
Such a procedure would minimize the error in importance sampling but has a high computational cost.
We make the decision to build a denser grid whenever the importance sampling procedure produces noisy posterior distributions, which we determine by visual inspection.
We note at which number of observables we make the decision to construct denser grids in the following section.

\begin{figure*}
    \centering
    \includegraphics[width=6in]{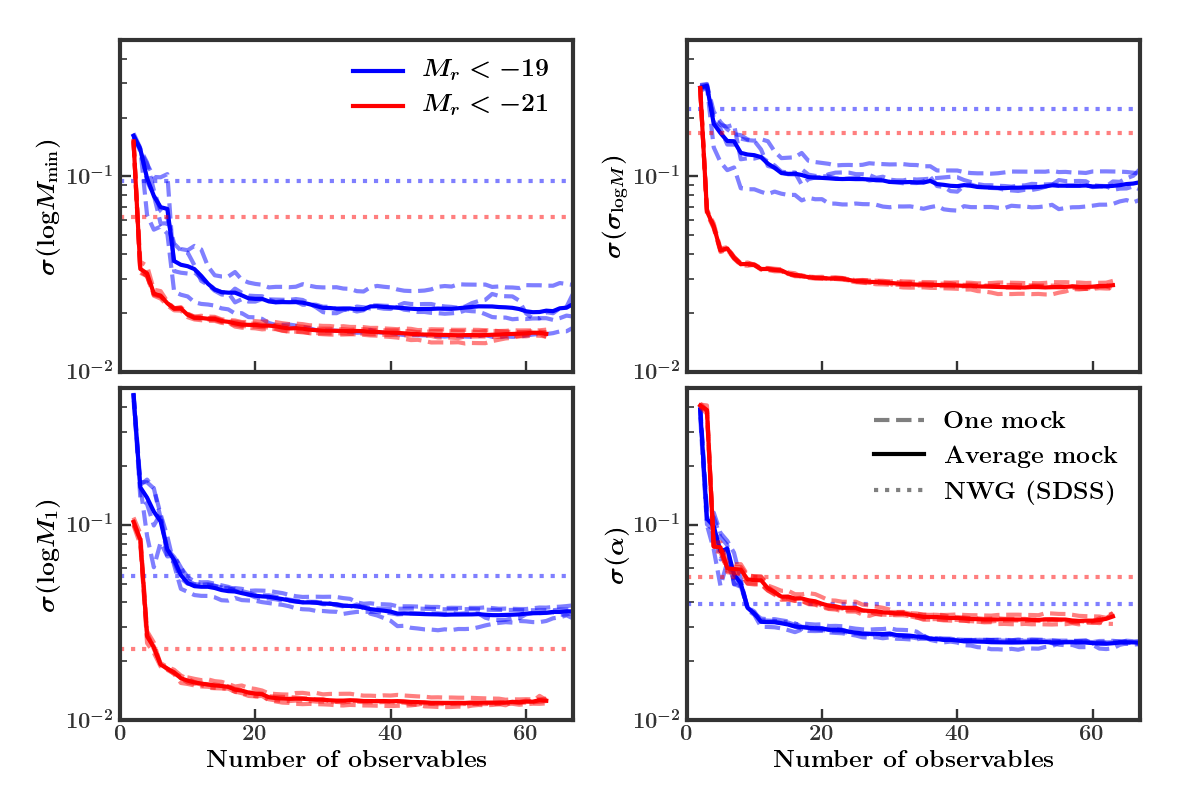}
    \caption{Projected constraints (1-$\sigma$) on four of our HOD parameters as we increase the number of observables. The constraints for the $-19$ and $-21$ mocks are shown in blue and red, respectively. The order of the observables is determined by our algorithm and is shown in Table~\ref{tab:order}. Each dashed line gives the projected constraints from one of our grids while the solid line shows the average of all four grids. For comparison, the constraints on our SDSS samples from a chain run using \ngal, \wprp, and \gmf (as was done in S18) are shown with dotted horizontal lines.}
    \label{fig:orderings}
\end{figure*}

\subsection{Ordering of observables}
\label{subsec:order}
We apply the algorithm outlined in the previous section to order the observables according to their potential constraining power.
We perform this procedure (separately, for each sample) on mocks constructed using HODs appropriate for the $-19$ and $-21$ SDSS volume-limited samples.
The fiducial HOD parameters we use in building the grids are given in Table~\ref{tab:fid}.
These parameters are the best-fit HOD values from a chain run on SDSS\footnote{These values are the best-fit parameters for a chain run using the unadjusted SDSS observables (see Appendix~\ref{app:fib}). They are slightly different than the best-fit values reported in Table~\ref{tab:bestfit}.} using only \ngal, \wprp, and \gmf and thus constitute reasonable parameter values for SDSS.
For each sample, we construct both the mocks and the covariance matrix from these HODs.

\begin{deluxetable}{cccccc}
\tablenum{4}
\tablecaption{Fiducial HOD Parameters\label{tab:fid}}
\tablewidth{0pt}
\tablehead{
\colhead{$M_r^\mathrm{lim}$} & \colhead{\logmmin} & \colhead{\siglogm} & \colhead{\logmzero} & \colhead{\logmone} & \colhead{$\alpha$}
}
\startdata
$-19$ & 11.54 & 0.22 & 12.01 & 12.74 & 0.92 \\
$-21$ & 12.72 & 0.46 & 7.87 & 13.95 & 1.17
\enddata
\tablecomments{Unless otherwise noted, we use these fiducial HOD parameters to construct the covariance matrices and mocks we use throughout this paper.}
\vspace{-5mm}
\end{deluxetable}

For the $-21$ mocks, the value of \logmzero is too low to have any impact on the mathematical form of our HOD.
For the $-19$ mocks, however, \logmzero is large enough to have a significant impact on halo occupation.
Therefore, we decide to simultaneously optimize using all five parameters for the $-19$ mocks but ignore \logmzero when optimizing the $-21$ mocks (i.e., in our calculation of $M_\mathrm{grid}$).

In Figure~\ref{fig:orderings}, we present the results of our algorithm for four of the five HOD parameters (excluding \logmzero).
The dashed/solid lines show, for our $-19$ (blue) and $-21$ (red) mocks, the estimated posterior constraints (weighted standard deviations) on each parameter as we increase the number of observables in the order chosen according to our algorithm.
Each dashed curve gives the estimated constraint obtained from one of the grids (mocks), while the solid curve gives the average constraint.
The horizontal dotted lines show the constraints achieved on our SDSS volume-limited sample when we only use \ngal, \wprp, and \gmf (NWG, hereafter), as was done in S18 (see Section~\ref{sec:results}).

\begin{figure}
    \centering
    \includegraphics[width=3in]{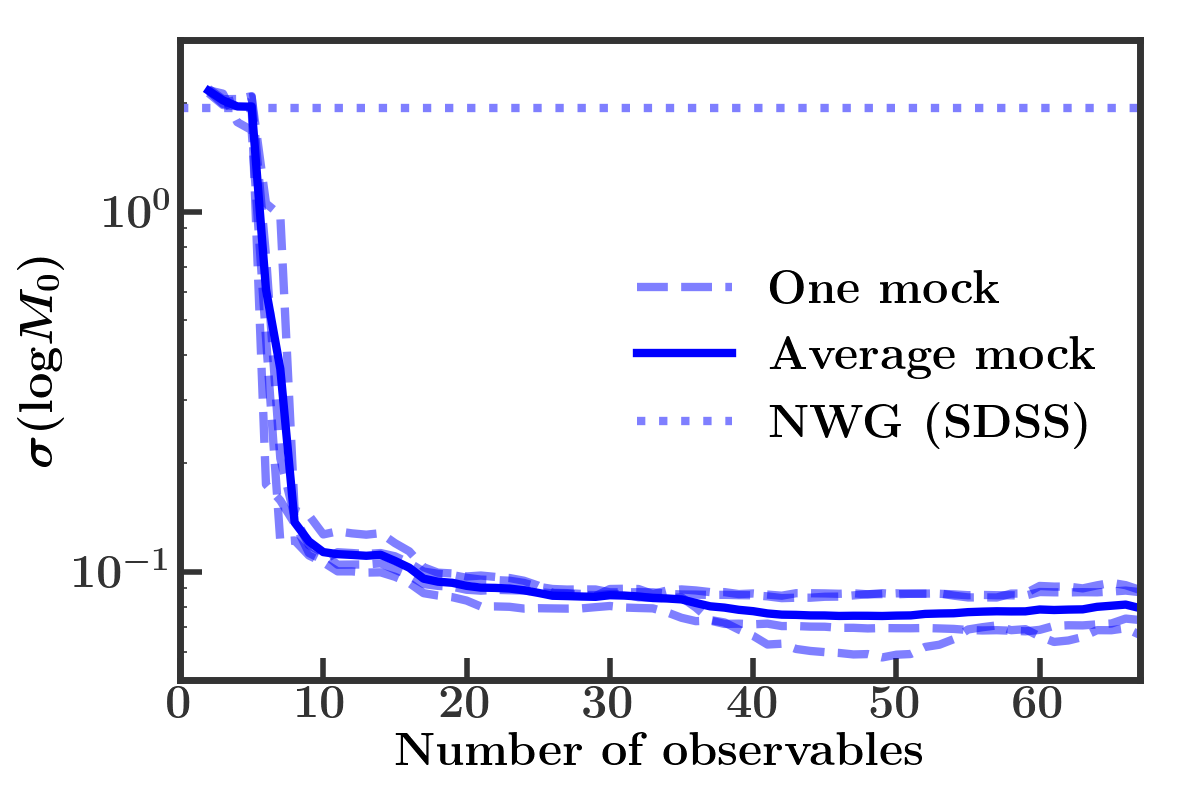}
    \caption{Projected constraints on \logmzero for the $-19$ mocks. The lines and observable order are the same as in Figure~\ref{fig:orderings}. We are unable to constrain \logmzero for the $-21$ mocks and therefore do not include it.}
    \label{fig:mzero}
\end{figure}

In Figure~\ref{fig:orderings}, we can see that our average projected constraints are tighter than the NWG constraints for all parameters by the time we reach $\sim 10$ observables.
Thus, we need only half as many observables as are used in the NWG chain to obtain the same constraining power.
For many of the parameters (e.g., $-21$ central parameters), we are outperforming the NWG constraints with even fewer ($\sim 5$) observables.
These results highlight the strong constraining power that comes from combining various scales of multiple clustering statistics. 

Comparing the two samples, we project tighter constraints on three of the parameters for the $-21$ mocks compared to the $-19$ mocks, with $\alpha$ as the only parameter better constrained for $-19$.
As can be seen by the horizontal dotted lines, this trend is also true when constraining SDSS using only \ngal, \wprp, and \gmf.
Additionally, the projected constraints are all quite similar among the four mocks for $-21$, while there is a greater dispersion among the $-19$ mocks, particularly for \logmmin and \siglogm.
Because the fiducial value of \siglogm in the $-19$ mocks (0.22) is closer to zero than in the $-21$ mocks (0.46), the posterior contours are often cut off by our flat prior lower bound (\siglogm cannot be negative).
This results in the shapes of the contours having more variety between the four mock realizations, affecting both \logmmin and \siglogm due to their correlation.

Remarkably, we are also able to obtain tight constraints on \logmzero for the $-19$ mocks using a collection of $\lesssim 10$ observables.
We show this result in Figure~\ref{fig:mzero}, which has the same format as Figure~\ref{fig:orderings} but excludes the $-21$ results.
When using only \ngal, \wprp, and \gmf, \logmzero is almost entirely unconstrained.
With the availability of the clustering statistics we introduce in this paper, our algorithm very quickly selects observables which constrain \logmzero.

\begin{deluxetable}{ccccc}
\tablenum{5}
\tablecaption{Observable Order\label{tab:order}}
\tablewidth{\columnwidth}
\tablehead{
\colhead{$M_r^\mathrm{lim}$} & \colhead{Number} & \colhead{Clustering Statistic} & \colhead{Bin}
}
\startdata
& 1 & \ngal &  \\ 
& 2 & \wprp & 2 \\ 
& 3 & \wprp & 4 \\ 
& 4 & \vpf & 3  \\ 
& 5 & \wprp & 8 \\ 
& 6 & \zxi & 1 \\ 
& 7 & \gmf & 3 \\ 
$-19$ & 8 & \zxi & 5  \\ 
& 9 & \gmf & 2 \\ 
& 10 & \gmf & 4 \\ 
& 11 & \gmf & 1 \\ 
& 12 & \spf & 4 \\ 
& 13 & \zxi & 13 \\ 
& 14 & \mcf & 14 \\ 
& 15 & \zxi & 6 \\
\hline
& 1 & \ngal &  \\ 
& 2 & \wprp & 2 \\ 
& 3 & \zxi & 8 \\ 
& 4 & \wprp & 4 \\ 
& 5 & \mcf & 9 \\ 
& 6 & \wprp & 1 \\ 
& 7 & \zxi & 9 \\ 
$-21$ & 8 & \mcf & 7 \\ 
& 9 & \zxi & 4 \\ 
& 10 & \zxi & 7 \\ 
& 11 & \mcf & 10 \\ 
& 12 & \zxi & 1 \\ 
& 13 & \wprp & 14 \\ 
& 14 & \gmf & 1 \\ 
& 15 & \spf & 4
\enddata
\tablecomments{The type of clustering statistic and the bin number (1-indexing) for the first 15 observables chosen (in order) for each sample. Note that the first two observables were fixed. We include the full order in the machine-readable format.}
\vspace{-5mm}
\end{deluxetable}

\begin{figure*}
    \centering
    \includegraphics[width=6in]{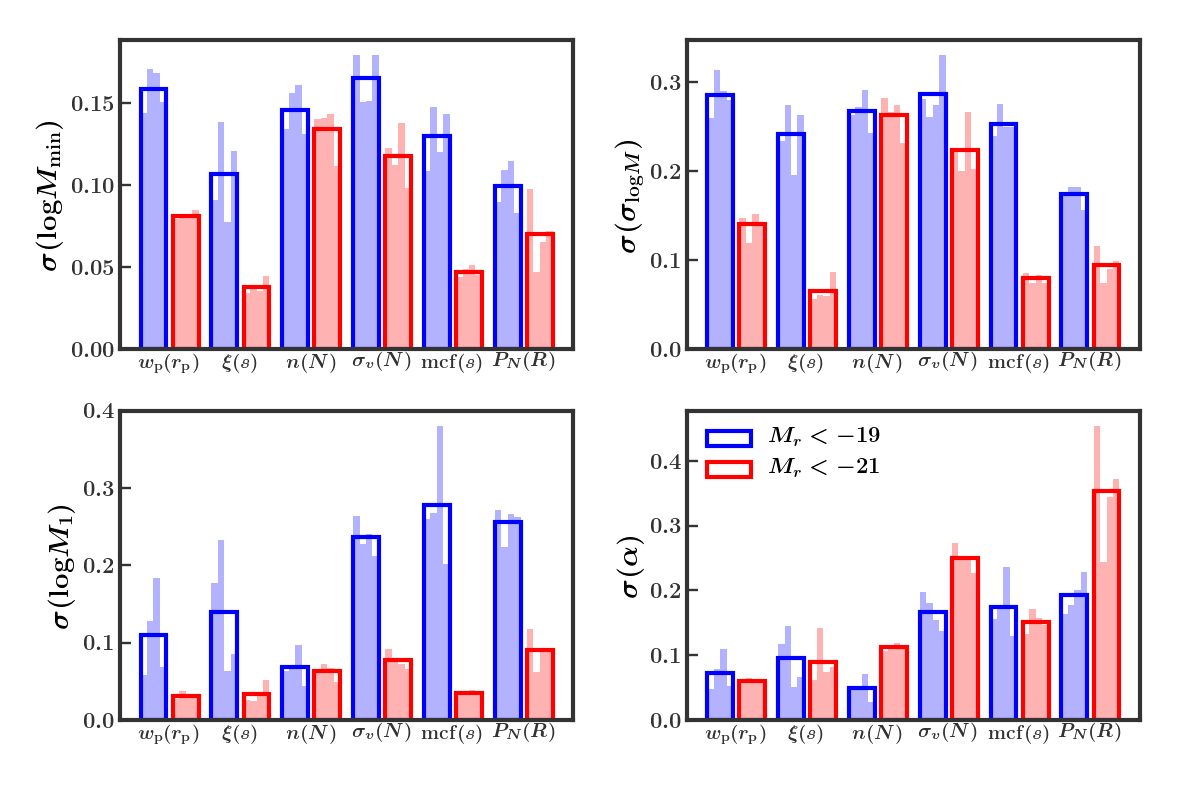}
    \caption{Projected constraining power of each individual clustering statistic (+\ngal). The height of each smaller vertical bar shows the projected constraints on one mock, while the larger open bar shows the average constraint across all four mocks. The vertical axes and colors are the same as in Figure~\ref{fig:orderings}.}
    \label{fig:ind_stat}
\end{figure*}

At the end of Section~\ref{subsec:alg}, we describe how we construct denser grids whenever we deem it necessary.
For both samples, we choose observables $3-5$ using our $K=2$ grid (constructed with just \ngal and \wptwo).
After choosing observable 5, the posterior contours arising from importance sampling appear to be noisy.
We thus reconstruct denser grids for both samples using the first five observables ($K=5$ grids).
For $-21$, we use these $K=5$ grids to choose observables $6-63$.
For $-19$, constraining \logmzero notably has a significant impact on the allowable region of \logmmin and \logmone, which are both anti-correlated with \logmzero.
Once we begin to obtain tight constraints on \logmzero, the density of our grid is again no longer sufficient.
Therefore, for $-19$, we only use the $K=5$ grid to choose observables $6-8$ and switch to an even denser grid at $K=8$.
We then use this $K=8$ grid to choose observables for all larger values of $K$.

In Table~\ref{tab:order}, we show the order in which the first 15 observables are chosen for each sample.
We indicate the clustering statistic and the bin number, where Bin~1 is the smallest bin of the corresponding clustering statistic.
We include the full order in the machine-readable format.
Several small and intermediate ($\lesssim$ 1 \hmpc) scales of \wprp and \zxi are chosen early for both samples.
While these scales are highly correlated and thus provide similar information, their joint constraining power is apparently strong enough to justify their inclusion.
For $-19$, among the first 15 observables are four scales of \gmf, while only one scale is found among the first 15 choices for $-21$.
This statistic is sensitive to the satellite population and therefore may be more relevant for $-19$, which has a higher satellite fraction.
On the other hand, for $-21$, we find four scales of \mcf among the first 15 choices, while finding only one scale for $-19$.

While commenting on the precise reasons for the choice of each observable is beyond the scope of this paper, we wish to get a general picture of the ability of each individual clustering statistic to constrain each HOD parameter.
To estimate the results of running a chain with only one clustering statistic (e.g., \mcf), we perform a task very similar to the one described in Section~\ref{subsec:alg}.
Using the sparsest ($K=2$) grids, we weight every point in our grid by $\mathcal{L}_\mathrm{stat}/\mathcal{L}_\mathrm{init}$, where $\mathcal{L}_\mathrm{stat}$ is the likelihood calculated using all scales of one clustering statistic and $\mathcal{L}_\mathrm{init}$ is the likelihood calculated using only \ngal and \wptwo.
We then importance sample the grid to estimate the constraints on each HOD parameter that would be obtained from running a chain with this clustering statistic.

We show the results of this test in Figure~\ref{fig:ind_stat}.
Like in Figure~\ref{fig:orderings}, each panel shows the estimated constraints on a different HOD parameter, with constraints for $-19$ and $-21$ shown in blue and red, respectively. 
The height of each smaller, filled bar shows the constraints on an individual mock when using the clustering statistic\footnote{For each statistic, we include \ngal in the calculation of $\mathcal{L}_\mathrm{stat}$. Thus, these are the estimated constraints for each clustering statistic plus \ngal.} indicated on the horizontal axis, while the empty vertical bar shows the average constraint across all four mocks.
We choose to treat \vpf and \spf as one statistic (\cic) in this figure.

For the central parameters, \logmmin and \siglogm, it is notable that no one clustering statistic does a particularly good job of constraining the $-19$ mocks.
Contrast this with the $-21$ mocks, where \zxi and \mcf (as well as \wprp and \cic to a lesser extent) each alone provide very tight constraints.
In Figure~\ref{fig:orderings}, however, we see that the achievable constraints on \logmmin for $-19$ and $-21$ are quite similar.
This highlights the advantage of using a variety of clustering statistics measured at different scales.

For both samples, we see that \wprp and \zxi provide excellent constraints on the satellite parameters, \logmone and $\alpha$.
We can thus see why so many scales of these observables are chosen early for both $-19$ and $-21$.
For \logmone, it is notable that \mcf again provides extremely tight constraints for $-21$ but not for $-19$, further illustrating why so many \mcf scales are chosen early.
Similarly for $\alpha$, \gmf provides better constraints for $-19$ than for $-21$, illustrating why more scales are chosen for that sample.

We should caution that the results presented in Figure~\ref{fig:ind_stat} are \textit{estimates}.
As we have already mentioned, importance sampling works well if the grid has a sufficient density of points in the neighborhood of the target distribution.
Because we are using our sparse $K=2$ grids, this condition is not generally met for the results shown in Figure~\ref{fig:ind_stat}.
This fact is reflected by the large variation from mock to mock for a given statistic (e.g., \zxi for $-19$).
We only show this figure to provide some qualitative insight into the order of observables presented in Table~\ref{tab:order}.

Looking again at Figures~\ref{fig:orderings} and \ref{fig:mzero}, we can see that by $\sim$ 20 observables, the constraints on virtually all parameters begin to asymptote.
There is still, however, a slight improvement in the constraints beyond 20 observables.
From these figures alone, however, we do not have a handle on the noise present in each of these estimates of our constraints.
It is thus difficult to gauge how much information we gain as we increase the number of observables.
We address this point in the following section.

\subsection{Choosing an optimal set}
\label{subsec:choose_obs}

In Figure~\ref{fig:orderings}, we see that using more observables always improves the constraints on our HOD parameters.
Using more observables also adds more elements to the covariance matrix.
All elements of the covariance matrix contain some noise because we calculate them using a finite number of mocks.
In principle this noise would disappear if we had an infinite number of mocks, but in practice we have only 400.
Because calculating a likelihood involves an inversion of the covariance matrix (see Equation~\ref{eq:likelihood}), any noise in the matrix is propagated into the likelihood.
Using more observables introduces more noise into the calculation of a likelihood and thus into the posterior results of an MCMC.
Therefore, as we increase the number of observables, there is a trade-off between the increase in constraining power and the added noise in these constraints.

\begin{figure*}
    \centering
    \includegraphics[width=6in]{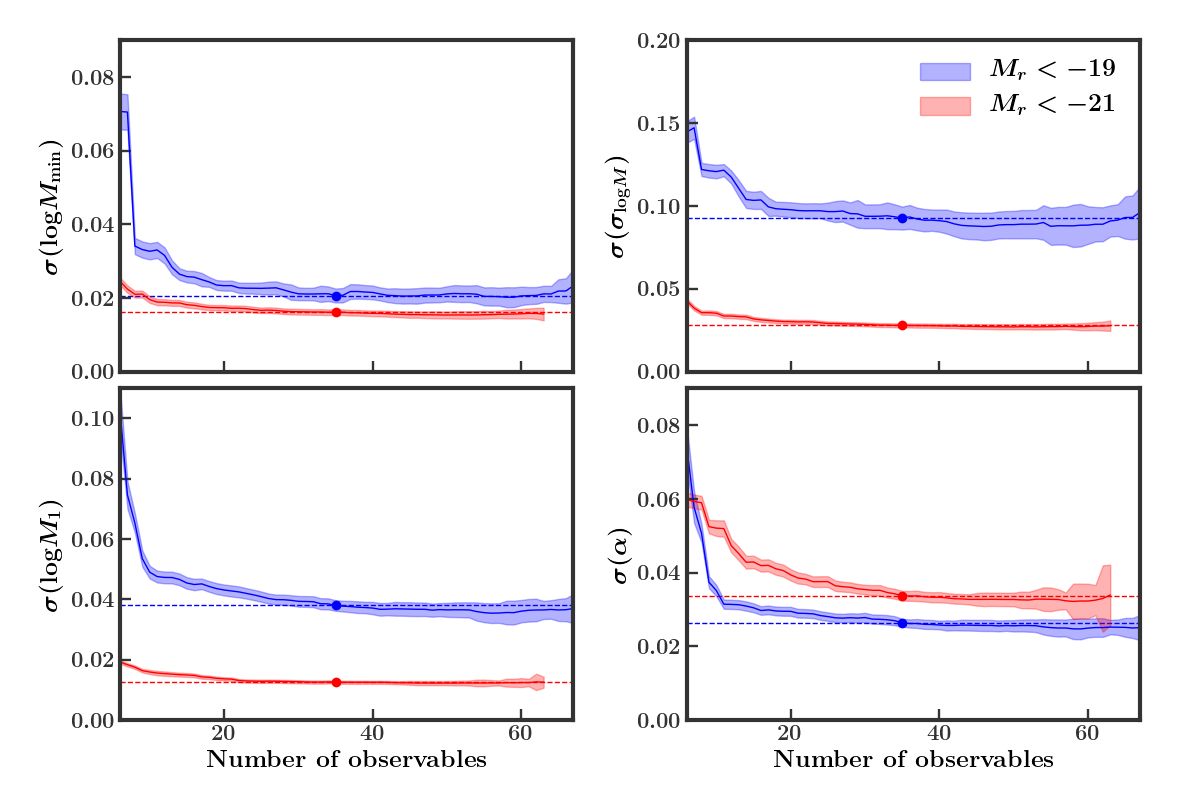}
    \caption{Our method for choosing how many observables to use. Each solid line shows the average mock constraint (1-$\sigma$) from Figure~\ref{fig:orderings}, and the shaded region shows our estimate of the uncertainty (inner 68\%) in our constraints. The dot and dashed line indicate the number of observables we decide to use and the projected constraints for each parameter. We only show results here for $\geq 6$ observables. (See text for more details.)}
    \label{fig:posterior_error}
\end{figure*}

We wish to know, given a particular set of $K$ observables, how our constraints would change if we were to estimate the covariance matrix using a different set of 400 mocks.
We design a test to estimate this ``error" in our constraints for different values of $K$.
We show the results of this test in Figure~\ref{fig:posterior_error}.
As in previous figures, the $-19$ and $-21$ samples are plotted in blue and red, respectively.
The dark solid lines show the grid-averaged estimated constraint (i.e., the same lines as are in Figure~\ref{fig:orderings}), while the shaded region shows our estimate of the error in these constraints.
To estimate this error, we perform the following steps for each of our grids and for each ordered subset of $K$ observables:

\begin{enumerate}[noitemsep]
    \item Treating the $K$ by $K$ covariance matrix as a multivariate Gaussian, randomly sample 400 points from this distribution.
    \item From these 400 points, construct a new ``resampled" covariance matrix.
    \item Compute a new likelihood $\mathcal{L}_K$ for each point in the grid\footnote{We use the most dense available grid for each value of $K$.}. In this calculation, use the resampled $K$-dimensional covariance matrix.
    \item Importance sample the grid, weighting each point by $\mathcal{L}_K/\mathcal{L}_\mathrm{init}$ (see Section~\ref{subsec:alg} for description of $\mathcal{L}_\mathrm{init}$).
    \item Re-calculate the weighted posterior standard deviations, $\sigma_\mathrm{par,resamp}$, of each HOD parameter.
    \item Repeat steps $1-5$ 100 times, yielding 100 values of $\sigma_\mathrm{par,resamp}$ for each HOD parameter.
    \item For each parameter, compute an error $\delta(\sigma_\mathrm{par})$ by taking the standard deviation of these 100 values of $\sigma_\mathrm{par,resamp}$.
\end{enumerate}

Performing these steps for each of our four grids yields four estimates of $\delta(\sigma_\mathrm{par})$ for each HOD parameter as we increase $K$.
For each value of $K$, we choose the largest value of $\delta(\sigma_\mathrm{par})$ among these four estimates as our error\footnote{We also explore taking the average error among the four grids. The choice makes very little difference so we opt for the more conservative approach (i.e., larger errors).}.

As expected, we see in Figure~\ref{fig:posterior_error} that $\delta(\sigma_\mathrm{par})$ increases as we increase the number of observables.
We wish to choose a value of $K$ where both the constraint ($\sigma_\mathrm{par}$) and the error on the constraint ($\delta(\sigma_\mathrm{par})$) are small.
To accomplish this task, we employ the ``one standard-error rule": we choose the lowest value of $K$ such that the constraint at this value is within one standard error of the constraint at all higher values of $K$.
We have to simultaneously meet this condition for all parameters of interest to us, which pushes us to higher values of $K$ than are necessary for some parameters.
Coincidentally, for both samples, this happens to be the same number: $K=36$.
We mark this value of $K$ in Figure~\ref{fig:posterior_error} with a dot.
We also indicate the estimated constraining power for $K=36$ with a horizontal dashed line.
Although not shown here, \logmzero is also taken into account when choosing $K$ for the $-19$ sample.

In choosing a value of $K$, we have focused solely on the constraining power achievable by using a particular combination of observables.
The other critical requirement of our set of observables is that, when used in an MCMC, the true HOD parameters used to construct our mock catalogs should lie somewhere inside of our posterior contours.
Where exactly these parameters lie depends on the cosmic variance of the dataset on which a chain is run and thus will differ for each of our four mocks.
Testing whether any bias exists for a given set of observables, therefore, is a far greater challenge.
Instead, we choose to just visually inspect whether or not the true HOD parameters lie within the 95\% posterior probability region for each of our HOD parameters.
We find that for our $K=36$ observables, this condition is met for all mocks in both samples.
In Appendix~\ref{app:mock_chain}, we go one step beyond importance sampling and run a full MCMC on one mock from each sample, using our $K=36$ observables.
We again find that the true HOD parameters are recovered for both samples.

\begin{figure*}
    \centering
    \includegraphics[width=6in]{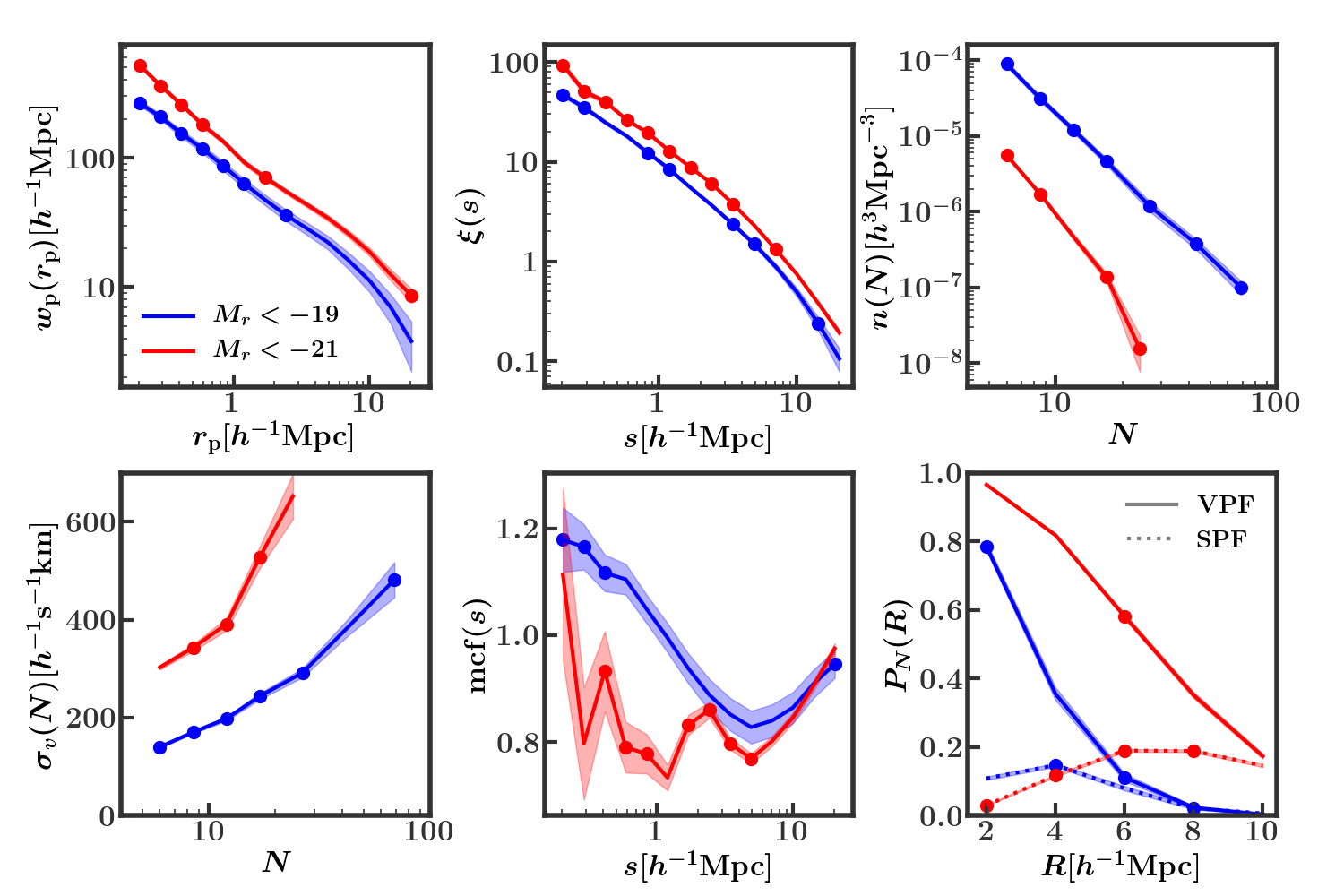}
    \caption{We show here the SDSS measurements of the observables we use in our analysis, with each panel showing measurements for a different clustering statistic. For the $-19$ (blue) and $-21$ (red) samples, the solid lines give the SDSS measurements on all scales we consider using of each clustering statistic. The points mark the scales we ultimately choose to use in our analysis (see Section~\ref{subsec:choose_obs}). We also use galaxy number density, which is not shown here. For illustrative purposes, we show the cosmic variance uncertainty from our 400 mocks with the shaded region centered on the SDSS measurements.}
    \label{fig:sdss_obs}
\end{figure*}

We highlight the variety of observables chosen by our algorithm in Figure~\ref{fig:sdss_obs}.
We show measurements of all scales of each clustering statistic (excluding \ngal) for our two volume-limited SDSS samples.
The observables we ultimately choose are indicated with solid points. 
We also provide a shaded region to indicate the size of the cosmic variance uncertainty used in our analysis, which represents the variance associated with our process of generating SDSS-like mocks.

\begin{figure*}
    \centering
    \includegraphics[width=6.5in]{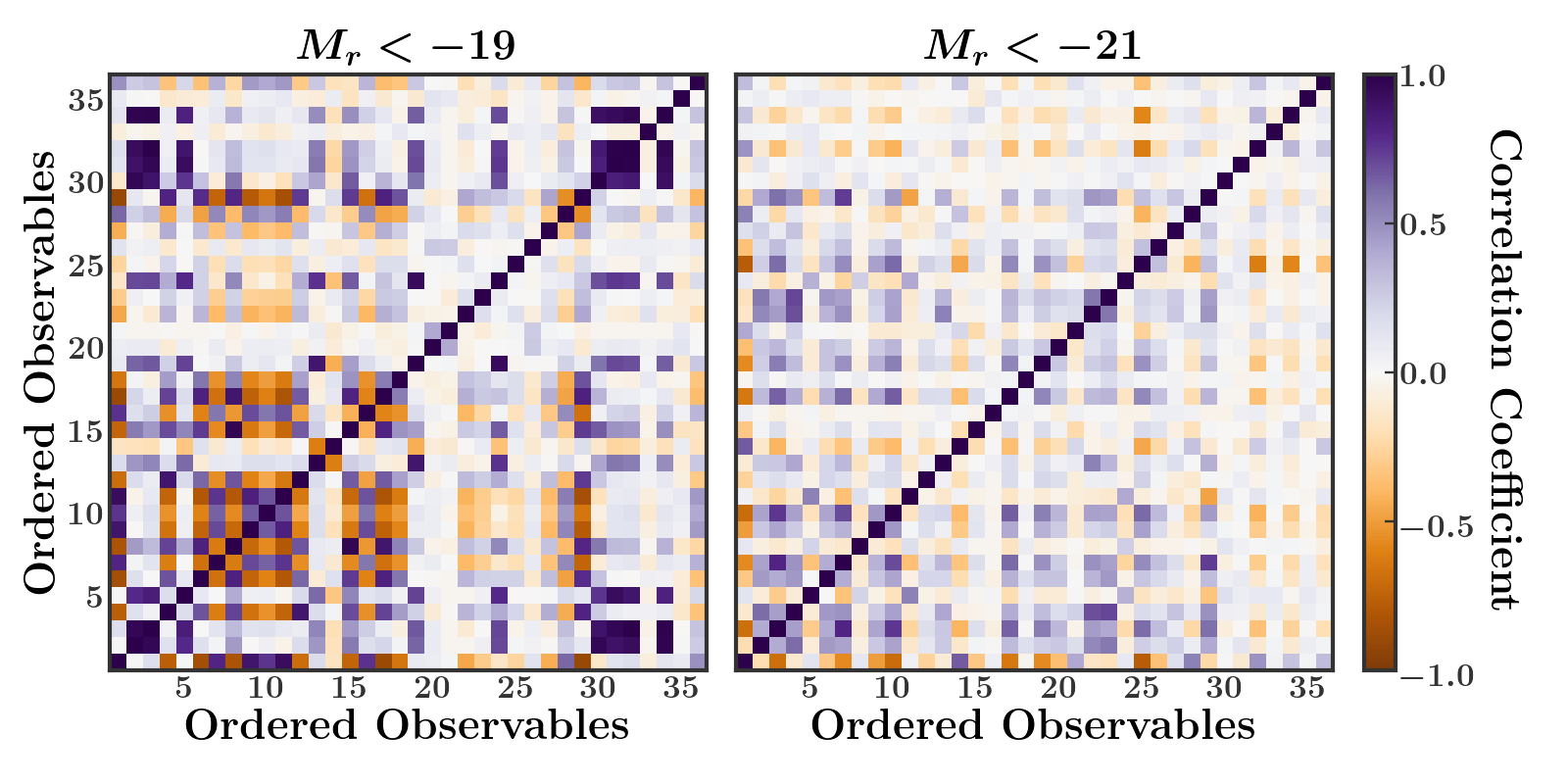}
    \caption{Correlation matrices for our $-19$ and $-21$ samples, built from 400 mock galaxy catalogs constructed using the fiducial parameters listed in Table~\ref{tab:fid}. Observables are placed in the order in which they are chosen by our algorithm (see Section~\ref{subsec:choose_obs}). We show only the first 36 observables (i.e., those which we use in our analysis).}
    \label{fig:matrices}
\end{figure*}

The degree of correlation between any two observables $i$ and $j$ is given by the correlation coefficient,
\begin{equation}
R_{ij} = \frac{C_{ij}}{\sqrt{C_{ii}C_{jj}}},
\end{equation}
where $C_{ij}$ is an element of the covariance matrix as defined in Equation~\ref{eq:covariance}.
Figure~\ref{fig:matrices} shows the correlation matrices for the $-19$ and $-21$ samples.
On both axes the observables are placed in the order in which they are chosen according to our algorithm (listed in Table~\ref{tab:order}).
We only show the 36 observables we use in our analysis, excluding all others.
The shading of a particular cell represents the degree of correlation between an observable on the x-axis and the corresponding observable on the y-axis.

The most remarkable feature of Figure~\ref{fig:matrices} is that both matrices are extremely diagonal, particularly when compared to those of S18.
By using our algorithm, we generally avoid choosing highly correlated observables which contain redundant information.
Looking deeper, the observables for the $-19$ sample exhibit a higher degree of correlation than the corresponding observables for the $-21$ sample.
This is true in general and not just for the observables chosen here.
Still, we can see that several highly correlated observables are chosen early for both samples, particularly for $-19$.
While this is not what we anticipated, it is clear that these correlated observables contain enough joint constraining power to be selected by our algorithm.

In this section, we have established a framework for selecting optimal observables that can be used to constrain the clustering of SDSS galaxies for a given HOD model. 
While this set of observables is specifically optimized to the HOD model employed in this work, this procedure could be repeated if we make any changes to our HOD model (e.g., adding assembly bias) in the future.
In the next section, we use our chosen observables to constrain and test our standard HOD model against our two SDSS volume-limited samples.

\section{Results} 
\label{sec:results}

\subsection{SDSS results}
\label{subsec:sdss}

With a set of 36 optimal observables chosen to produce tight and reliable constraints, we run an MCMC to constrain the HOD for each of our two volume-limited SDSS samples.
We refer to these chains as the ``OPT" chains henceforth.
We wish to compare the results of these OPT chains to those of S18.
Because we have altered the SDSS dataset (Sections~\ref{sec:obs} and \ref{subsec:fiber}) and details of the modeling procedure (Section~\ref{sec:model}), a direct comparison is not appropriate.
Therefore, we also run chains using only \ngal, \wprp, and \gmf, as was done in S18, and refer to these chains as ``NWG."
For each sample, we use the same HOD to construct the covariance matrix for both the OPT and NWG chains.
The parameters we use to construct this matrix are given in Table~\ref{tab:fid}.

All chains were run on the Texas Advanced Computing Center's Stampede2 supercomputer using 272 KNL CPUs spread across four compute-nodes.
Each chain was run using 542 walkers for $\sim$1000 iterations and thus involved $\sim$500,000 total HOD model evaluations.
For the OPT chains, each parameter evaluation took approximately five CPU-minutes, and thus a full chain took $\sim$45,000 CPU-hours (or $\sim$650 node-hours).
We determine that a chain has converged when the probability distribution of each parameter stabilizes, which occurs after $\sim$200 iterations.

\begin{figure*}
    \centering
    \includegraphics[width=5.9in]{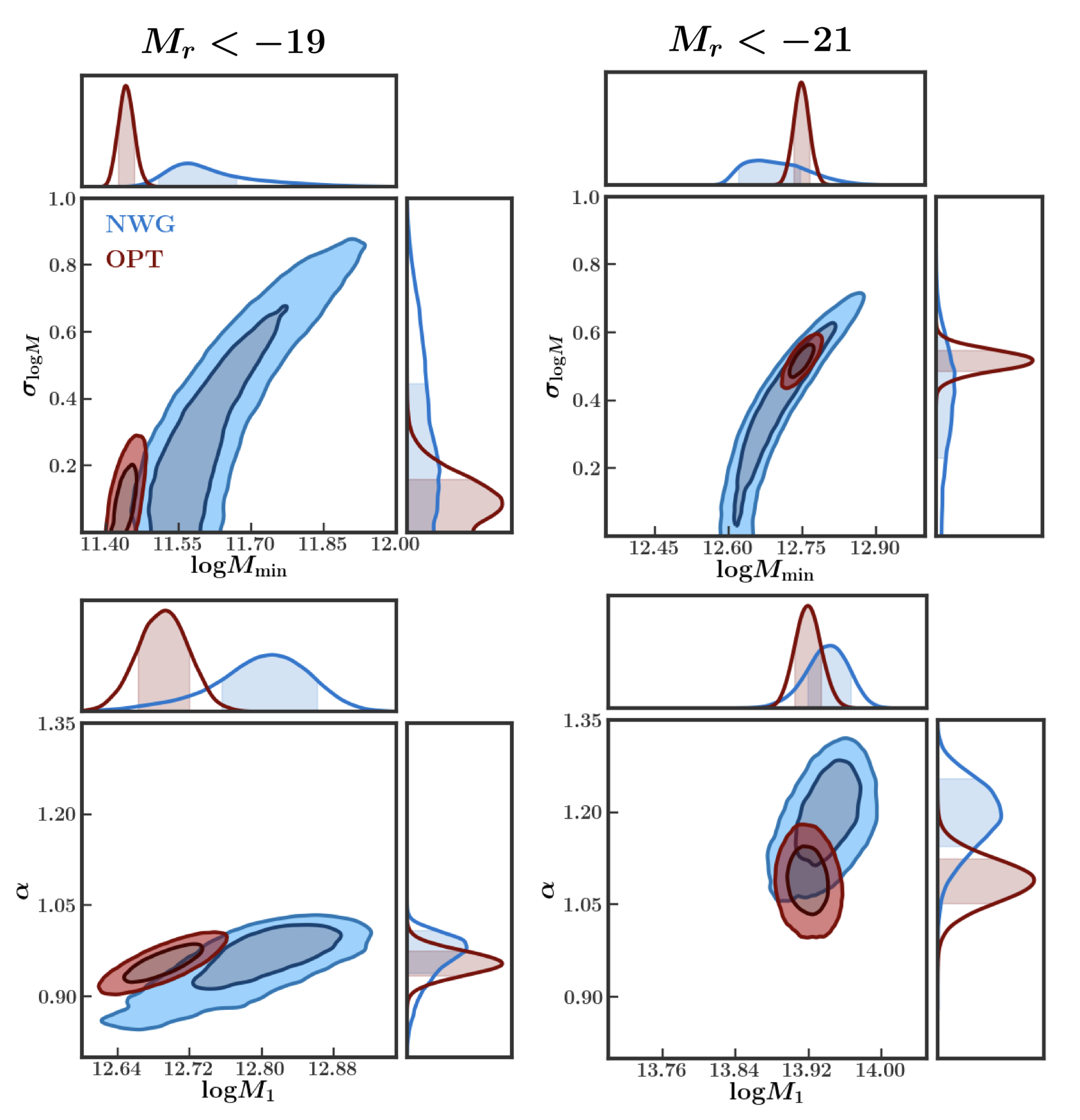}
    \caption{Results of our chains run on SDSS volume-limited samples. The left and right panels show results for the $-19$ and $-21$ samples, respectively. The blue contours are obtained from running a chain only using \ngal, \wprp, and \gmf (NWG). The red contours are the results when using the set of observables we choose in Section~\ref{subsec:choose_obs} (OPT).}
    \label{fig:sdss_chain}
\end{figure*}

The joint parameter constraints resulting from these chains are shown in Figure~\ref{fig:sdss_chain}.
The left and right panels show the constraints on the $-19$ and $-21$ samples, respectively.
The top panels show \siglogm vs. \logmmin while the bottom panels show $\alpha$ vs. \logmone.
The blue contours show the constraints achieved in the NWG chains, while the red contours show our results when running the OPT chains.
The shading shows the 68 and 95\% probability regions.

It is clear from Figure~\ref{fig:sdss_chain} that we achieve significantly tighter constraints on the joint distributions of all HOD parameters in the OPT chains compared to the NWG chains.
With these tighter constraints, we demonstrate the power of combining an assortment of clustering statistics measured at various physical scales.
As a result, we can now detect a significant difference in the values of \siglogm between the two samples, with the $-21$ sample having a higher value than the $-19$ sample.
This result is to be expected, given that the value of \siglogm can roughly be interpreted as the scatter in halo mass at fixed luminosity, which is generally greater for more luminous galaxies \citep{Behroozi2010}.
However, several previous HOD modeling works that rely on clustering \citep[e.g.,][]{Sinha2018, Zentner2019}, have not successfully detected this distinction in \siglogm between different luminosity samples.
Moreover, the few studies that have found that \siglogm generally increases with luminosity \citep{Zehavi2011, Guo2015b} did not have tight enough constraints to detect this difference to the level of significance that we achieve in this work.

\begin{figure}
    \centering
    \includegraphics[width=3in]{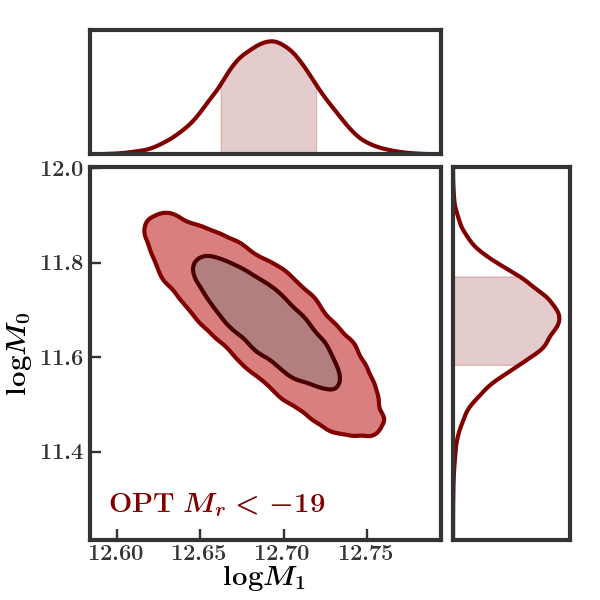}
    \caption{Constraints on \logmzero and \logmone for the $-19$ sample, using the set of observables we choose in Section~\ref{subsec:choose_obs}. Unlike the S18 results, we are able to obtain very tight constraints on \logmzero, which has a significant anti-correlation with \logmone.}
    \label{fig:sdss_m0}
\end{figure}

In the NWG chains (and the chains in S18), \logmzero is entirely unconstrained for both samples.
Using our chosen set of observables, however, we are able to obtain tight constraints on \logmzero for the $-19$ sample.
In Figure~\ref{fig:sdss_m0}, we show the joint constraints on \logmzero and \logmone for the $-19$ OPT chain.
We exclude the results for the NWG chain, where the contours extend all the way down to the lower prior bound on \logmzero at a value of 6.
This parameter is anti-correlated with both \logmone and \logmmin (not shown).
Our ability to constrain \logmzero, thus, affects the allowable values of these two parameters.

In Table~\ref{tab:constraints} we present the marginalized constraints on each parameter for the NWG and OPT chains.
For each sample, we give the median parameter values as well as the upper and lower limits corresponding to the 84 and 16 percentile parameter values.
To summarize the level of improvement, we divide the range of the inner 68\% of parameter values for the NWG chain by the corresponding range for the OPT chain.
We report this result in Table~\ref{tab:constraints} in the column labeled ``Constraint Ratio."
This ratio gives the factor by which we shrink the marginalized constraints when using our set of chosen observables, compared to using just the S18 observables.
As was evident in Figure~\ref{fig:sdss_chain}, we see improvement in the constraints on every parameter.
For both samples, we can see a particularly strong improvement in the constraints on the central parameters, \logmmin and \siglogm, and a weaker improvement for the satellite parameters, \logmone and $\alpha$.
The greatest improvement is for \logmzero in the $-19$ sample, again reflecting the fact that this parameter is unconstrained when using only \ngal, \wprp, and \gmf.

\begin{deluxetable*}{ccccc}
\tablenum{6}
\tablecaption{SDSS Constraints\label{tab:constraints}}
\tablewidth{0pt}
\tablehead{
\colhead{$M_r^\mathrm{lim}$} & \colhead{HOD Parameter} & NWG Constraint & OPT Constraint & \colhead{Constraint Ratio}
}
\startdata
 $-19$ & \logmmin & $11.597^{+0.124}_{-0.055}$ & $11.442^{+0.016}_{-0.015}$ & 5.774 \\
 & \siglogm & $0.289^{+0.293}_{-0.192}$ & $0.106^{+0.074}_{-0.065}$ & 3.489 \\
 & \logmzero & $10.385^{+1.519}_{-2.935}$ & $11.674^{+0.089}_{-0.094}$ & 24.339 \\
 & \logmone & $12.803^{+0.046}_{-0.058}$ & $12.691^{+0.028}_{-0.029}$ & 1.825 \\
 & $\alpha$ & $0.969^{+0.028}_{-0.047}$ & $0.954^{+0.019}_{-0.019}$ & 1.974 \\
$-21$ & \logmmin & $12.694^{+0.071}_{-0.058}$ & $12.748^{+0.015}_{-0.015}$ & 4.300 \\
 & \siglogm & $0.391^{+0.150}_{-0.201}$ & $0.517^{+0.029}_{-0.029}$ & 6.052 \\
 & \logmzero & $9.220^{+2.136}_{-2.183}$ & $9.015^{+2.017}_{-2.036}$ & 1.066 \\
 & \logmone & $13.941^{+0.021}_{-0.024}$ & $13.919^{+0.014}_{-0.014}$ & 1.607 \\
 & $\alpha$ & $1.195^{+0.051}_{-0.057}$ & $1.088^{+0.031}_{-0.033}$ & 1.688
\enddata
\tablecomments{Marginalized constraints on SDSS for each chain shown in Figure~\ref{fig:sdss_chain}. We present the median parameter values along with upper and lower limits corresponding to the 84 and 16 percentiles respectively. We also provide the ratio of constraints (inner 68 percentile range) of the NWG chain to the OPT chain. These numbers indicate the factor by which we have improved our constraints.}
\vspace{-12mm}
\end{deluxetable*}

\begin{deluxetable*}{cccccccccc}
\tablenum{7}
\tablecaption{SDSS Best-fit\label{tab:bestfit}}
\tablewidth{0pt}
\tablehead{
\colhead{$M_r^\mathrm{lim}$} & \colhead{Chain} & \colhead{\logmmin} & \colhead{\siglogm} & \colhead{\logmzero} & \colhead{\logmone} & \colhead{$\alpha$} & \colhead{$\chi^2$} & \colhead{d.o.f.} & \colhead{p-value}
}
\startdata
$-19$ & NWG & 11.552 & 0.229 & 12.107 & 12.707 & 0.905 & 18.083 & 17 & 0.384 \\
& OPT & 11.445 & 0.099 & 11.651 & 12.703 & 0.958 & 77.770 & 31 & $6.8\cdot10^{-6}$ \\
$-21$ & NWG & 12.691 & 0.377 & 12.075 & 13.938 & 1.191 & 20.577 & 15 & 0.151 \\
& OPT & 12.728 & 0.467 & 9.015 & 13.929 & 1.112 & 72.539 & 31 & $3.5\cdot10^{-5}$
\enddata
\tablecomments{Best-fit HOD parameters from all chains shown in Figure~\ref{fig:sdss_chain}. We also indicate the goodness of fit of each parameter combination with a $\chi^2$, the number degrees of freedom (d.o.f), and a p-value.}
\vspace{-9mm}
\end{deluxetable*}

In addition to tightening the constraints on our HOD parameters, we have also significantly heightened the tension between our model and SDSS.
In Table~\ref{tab:bestfit}, for each chain, we provide the best-fit HOD parameters and their associated $\chi^2$, degrees of freedom, and p-values.
For the NWG chains, the p-values of the best-fit points suggest that our model is sufficient to describe the clustering of SDSS\footnote{Our result is contrary to that of S18, who found 2.3 $\sigma$ tension for the $-21$ sample. This discrepancy could be due to any one of several improvements we make to the modeling procedure (Section~\ref{sec:model}) or processing of SDSS (Sections~\ref{sec:obs} and \ref{subsec:fiber}).}.
For the OPT chains, however, the p-values indicate that our model is unable to accurately match the clustering of SDSS. 
We find significant tension of 4.5 and 4.1 $\sigma$ for the $-19$ and $-21$ samples, respectively.

\begin{figure*}
    \centering
    \includegraphics[width=6in]{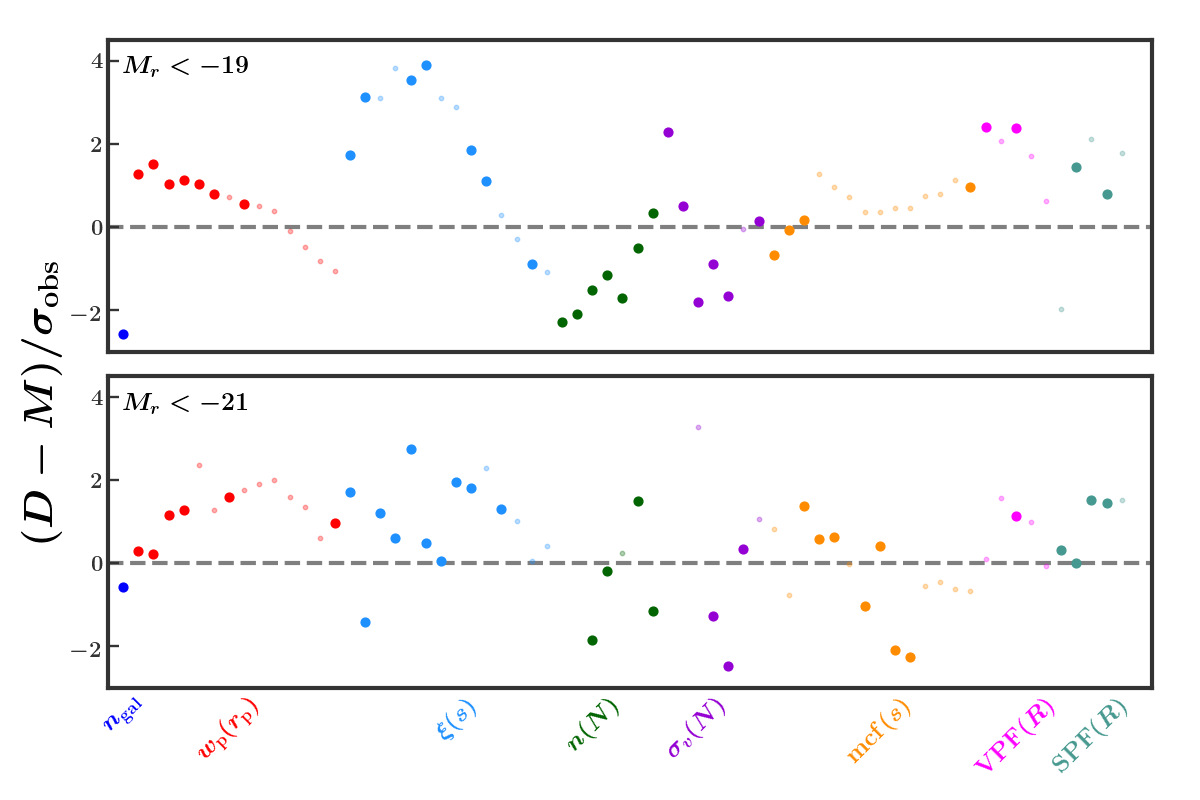}
    \caption{Residual measurements of our $-19$ (top) and $-21$ (bottom) SDSS volume-limited samples ($D$) compared to the best-fit HOD model ($M$) from our OPT chains. Each clustering statistic is given a different color and is labeled on the x-axis. We show the value $\chi_\mathrm{obs} = (D-M)/\sigma_\mathrm{obs}$ for all observables. The larger, darker points indicate those observables we actually use in our chains. Deviations from zero thus indicate that, ignoring any correlations, an observable has a strong individual contribution to the overall value of $\chi^2$.}
    \label{fig:sdss_resid}
\end{figure*}

To explore this tension further, we wish to identify which observables are contributing the most to the high value of $\chi^2$ for each sample.
In Figure~\ref{fig:sdss_resid}, for each observable, we show the value $\chi_\mathrm{obs}=(D-M)/\sigma_\mathrm{obs}$, where $D$ is the measurement on SDSS, $M$ is the best-fit model measurement, and $\sigma_\mathrm{obs}$ is the cosmic variance uncertainty.
For each clustering statistic, indicated with a color-coded label, we show $\chi_\mathrm{obs}$ values for all observables.
We indicate with larger, darker points those observables that are actually used in our analysis and thus contribute to the overall $\chi^2$ value.
Because many of the observables are correlated, the overall $\chi^2$ value is not just the sum of squares of these values, but we can still make some general conclusions about which observables are contributing the most to our high $\chi^2$ values.
For the $-19$ sample, notably both \ngal and small scales ($\lesssim 1\ h^{-1}\mathrm{Mpc}$) of \zxi are quite poorly fit. 
With the exceptions of \wprp and \mcf, some scales of all other clustering statistics seem to contribute significantly to the tension as well.
For $-21$, while \ngal is well-fit, all other clustering statistics appear to be poorly fit on at least one scale.
Overall, it is clear that the tension arises due to the inability of our five-parameter model to jointly fit all of these observables.

These results suggest that extensions to the HOD may be necessary in order to match the clustering of SDSS.
If our model requires additional features (e.g., assembly bias, velocity bias) which are not included in our five-parameter HOD, then our posterior results may have significant systematic errors \citep{Zentner2014}.
In particular, such errors may be exacerbated in the case of the OPT chains where we include clustering statistics which are sensitive to effects beyond those included in our form of the HOD.
Attempting to fit to the measurements of clustering statistics on SDSS with an inadequate model may cause a bias in our posterior results.
Indeed, looking again at Figure~\ref{fig:sdss_chain}, this effect may explain the offsets we observe between several of the parameter constraints of the NWG and OPT chains.
Without including these additional features in our HOD, however, we cannot verify this assertion.

We also consider the impact of our decision to adjust our observational data in order to account for the errors due to our use of the nearest neighbor correction (Section~\ref{subsec:fiber}).
Particularly, we wish to know how this treatment affects our posterior results and general conclusions about the success of our five-parameter HOD model.
Therefore, using the OPT observables, we also run chains on the unadjusted datasets for each sample and compare the results to our chains on the adjusted datasets.
For $-19$, we find shifts in the median parameter values of $\lesssim$1$\sigma$ and $\lesssim$2$\sigma$ for central and satellite parameters, respectively.
Additionally the p-value of the best-fit HOD point is even lower for the unadjusted chain.
For $-21$, we find the opposite trends: shifts in the median parameter values of $\lesssim$2$\sigma$ and $\lesssim$1$\sigma$ for central and satellite parameters, respectively.
The p-value of the best-fit HOD point is slightly larger, but this change is only enough to decrease the tension from 4.1 to 3.5 $\sigma$.
Thus, our treatment of fiber collisions has little impact on the general conclusions we draw about the ability of our five-parameter HOD to match the clustering of SDSS.

Before continuing, we address one last final detail concerning our results.
When running an MCMC, we are assuming that our fiducial matrix is not too dissimilar from the matrices we would construct from each HOD point in our chain.
The HOD parameters we use to construct our fiducial matrix, however, are far from the locations of our posterior (2 $\sigma$) contours for both the $-19$ and $-21$ OPT chains.
We consider performing an iterative procedure in which we reconstruct the covariance matrix from the best-fit HOD point of the OPT chain and then re-run the OPT chain with this new matrix.
This procedure could be repeated until some sort of convergence criteria is met.
To gauge whether or not this method is necessary, in Appendix~\ref{app:mock_chain}, we test whether two chains run on the same mock galaxy catalog, but differing in the HOD used to construct the matrix, produce similar posterior results.
We find that the results are largely similar and thus do not perform any iteration for our SDSS results.

\subsection{Modifying the halo mass function}
\label{subsec:hmf}

\begin{figure*}
    \centering
    \includegraphics[width=5.9in]{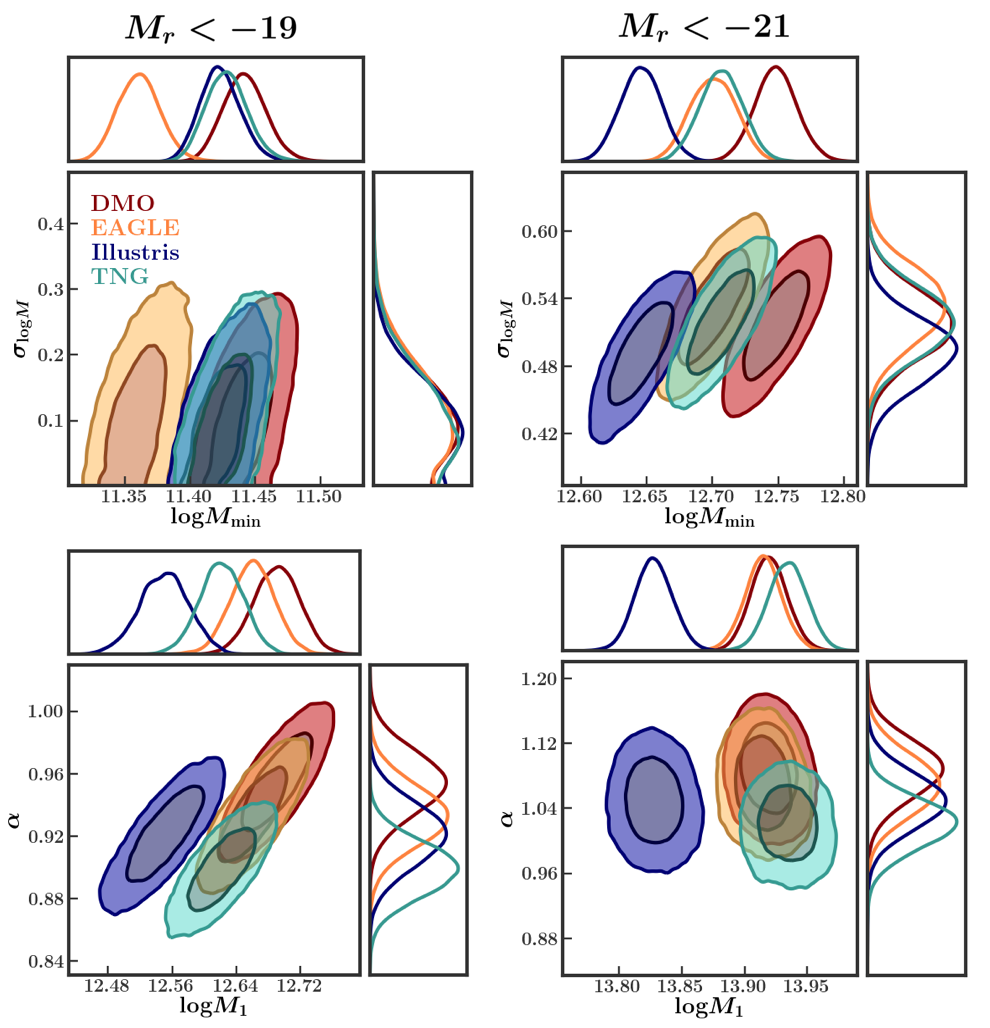}
    \caption{Results of chains run on SDSS using halo catalogs with masses modified according to the prescriptions of BM21. The shifts in \logmmin and \logmone can be understood by considering the global number density, while the shifts in \siglogm and $\alpha$ are in general more complex (see main text for more details).}
    \label{fig:halomass}
\end{figure*}

The use of the HOD framework as we have applied it is predicated on an assumption that our dark matter only simulations produce the correct halo mass function (HMF).
Hydrodynamic effects, however, can alter the masses of dark matter halos \citep[e.g.,][]{IntroducingIllustris, Schaller2015, Springel2018, Beltz-Mohrmann2020, Beltz-Mohrmann2021}. 
By not including baryonic physics, our simulations potentially have incorrect HMFs compared to that of the real Universe.

\citet{Beltz-Mohrmann2021} (BM21 hereafter) investigate the differences in HMFs between matched DMO and hydrodynamic simulation boxes in EAGLE, Illustris, and IllustrisTNG (TNG hereafter).
For halos at $z=0$, they find that in each simulation, stellar feedback generally reduces the masses of low mass halos ($\lesssim 10^{11}$ \hMsun), while AGN feedback generally reduces the masses of high mass halos (between $10^{12}$ and $10^{13}$ \hMsun) compared to their DMO counterparts. 
However, the exact effect that feedback has on the halo masses differs dramatically from one simulation to the next.

Based on each of these simulations, BM21 identify formulae that can be used to ``correct" the masses of halos from a DMO simulation in order to reproduce the HMF from a given hydrodynamic simulation. 
They provide these corrections for several different redshifts and halo definitions.
In this work, we apply the $z=0$, $M_\mathrm{vir}$ corrections to modify the masses of the halos in our DMO simulations.
We then re-run our OPT SDSS chains three times, using halos modified according to the prescriptions for EAGLE, Illustris, and TNG.
In each of these chains we use the same covariance matrix as in the previous section with parameters listed in Table~\ref{tab:fid}.

We show the results of these chains in Figure~\ref{fig:halomass}, with the same general format as Figure~\ref{fig:sdss_chain}.
In each panel, the EAGLE results are shown in orange, Illustris in blue, and TNG in green. 
``DMO" (red) refers to the chains run using the original halos (replicated from Figure~\ref{fig:sdss_chain}).
For both samples, we see that some of the HMF modifications produce significant shifts in the median values of the HOD parameters.
For the $-21$ sample, we see significant shifts in \logmmin and \logmone and smaller shifts in \siglogm and $\alpha$.
For the $-19$ sample, we see significant shifts in all parameters except \siglogm.

In the $-21$ sample, we can see large decreases in the values of \logmmin and \logmone in the Illustris chain. 
These shifts are due to the fact that the Illustris correction reduces the masses of all halos above $10^{12} h^{-1} M_\odot$ by $\gtrsim$15\%.
The EAGLE and TNG corrections only slightly reduce the masses of halos between $10^{12}$ and $10^{13} h^{-1} M_\odot$ but have little to no impact on the highest mass halos. 
This explains the slight decrease in \logmmin and the lack of change in \logmone for these chains.

In the $-19$ sample, we can see a large decrease in the value of \logmmin in the EAGLE chain but little change in the Illustris and TNG chains.
This effect is because the EAGLE correction reduces the masses of halos between $10^{11}$ and $10^{12} h^{-1} M_\odot$ by up to 20\%, while the Illustris and TNG corrections have little impact in this regime.
Like in the $-21$ sample, the Illustris correction leads to the biggest change in the value of \logmone.
Unlike $-21$, however, we see some shifts in \logmone for EAGLE and TNG because we are in a regime in which the halo masses are slightly reduced by these corrections.
Perhaps the most significant result of Figure~\ref{fig:halomass} is that there is no shift in the value of \siglogm for the -19 sample and very little shift for the -21 sample.
Thus, our claim in Section~\ref{subsec:sdss} concerning the detectable difference in \siglogm between the two samples holds even when considering the impact of baryons on the HMF.

While the changes in \logmmin and \logmone are fairly straightforward, the small changes in $\alpha$ in both samples (and in \siglogm in the $-21$ sample) are more difficult to intuit.
This demonstrates the fact that the halo mass corrections in BM21 are not simple parameter shifts but complex functions of mass that impact all of our HOD parameters.

Regardless of the reasons for the observed shifts, with the increased constraining power afforded by our numerical approach, the differences in recovered HOD parameter values are generally not robust with respect to the uncertainty in the HMF due to baryonic physics.
We certainly do not believe that these three hydrodynamic simulations capture the full uncertainty in baryonic physics or that any of them necessarily result in more accurate HMFs than are produced by DMO simulations.
This exercise is simply meant to emphasize that our HOD parameter constraints are subject to baryonic effects.

Despite the observed changes in the posterior contours, the tension that we find between SDSS and our best-fit HOD model remains largely unchanged for both samples.
In the -19 sample, the tension that we find between SDSS and our best-fit HOD model applied to DMO simulations (4.5 $\sigma$) remains the same after the Illustris correction, and increases slightly (to 4.6 $\sigma$) after the EAGLE and TNG corrections. 
In the -21 sample, the original tension (4.1 $\sigma$) remains the same after the Illustris correction, increases slightly (to 4.2 $\sigma$) after the EAGLE correction, and decreases slightly (to 3.9 $\sigma$) after the TNG correction.
In any case, we can conclude that the tension that we find between SDSS and our standard HOD model is not significantly reduced or exacerbated by the halo mass corrections we employ here.

In this section, we have only demonstrated the effects of correcting the mass function for the simple five-parameter HOD we use in this work.
As we look to constrain additional features of the galaxy-halo connection (e.g., assembly bias) in the future, we must continue to consider the impact of baryonic physics on our posterior results.

\section{Conclusions} \label{sec:conc}

In this work, we explore the ability of a variety of clustering statistics to jointly constrain the connection between galaxies and the dark matter halos in which they live.
We develop an algorithm to choose a set of ``optimal" observables based on estimates of their joint constraining power.
We choose these observables from a superset of multiple scales of the projected correlation function, the redshift-space correlation function, the group multiplicity function, the group velocity dispersion, the mark correlation function, and counts-in-cells statistics.
In particular, for two different volume-limited SDSS samples, we employ a Markov Chain Monte Carlo (MCMC) method to constrain a simple five-parameter halo occupation distribution model for connecting galaxies to dark matter halos.
We also explore how robust our results are with respect to the influence of baryonic physics on the halo mass function.
Our main results are summarized below:

\begin{itemize}[noitemsep]
\item Compared to a similar analysis using only \ngal, \wprp, and \gmf, when using our chosen set of observables, we are able to achieve tighter constraints on all HOD parameters for both the $-19$ and -$21$ SDSS volume-limited samples. Specifically, we tighten the constraints on the central parameters (\logmmin and \siglogm) by factors of $\sim$5.8 and $\sim$3.5 for the $-19$ sample and $\sim$4.3 and $\sim$6.1 for the $-21$ sample. We tighten the constraints on the satellite parameters (\logmone and  $\alpha$) by factors of $\sim$1.8 and $\sim$2.0 for the $-19$ sample and $\sim$1.6 and $\sim$1.7 for the $-21$ sample. The fifth parameter, \logmzero, is unconstrained when using \ngal, \wprp, and \gmf but is well-constrained for the $-19$ sample when using our chosen observables. (See Figures~\ref{fig:sdss_chain} and \ref{fig:sdss_m0}, as well as Table~\ref{tab:constraints}.)
\item Our tighter constraints have yielded high precision measurements of \siglogm for the $-19$ and $-21$ samples, an informative parameter which quantifies the scatter in halo mass at fixed luminosity and is related to the scatter in the stellar-to-halo mass relation (see Figure~\ref{fig:sdss_chain} and Table~\ref{tab:constraints}). In particular, we can clearly distinguish between the low and high scatter for the -19 and -21 samples, respectively.
\item For the $-19$ sample, we find a significant shift in the posterior distributions when using our chosen observables, compared to when we use only \ngal, \wprp, and \gmf. This shift may be because we incorporate in our modeling procedure new clustering statistics that depend on additional HOD features not included in our five-parameter model, such as assembly bias, velocity bias, etc. (see Figure~\ref{fig:sdss_chain}).
\item We find substantial tension between SDSS and our best-fit HOD model. This tension is slightly stronger for the $-19$ sample (4.5 $\sigma$) than for $-21$ (4.1 $\sigma$). This is in contrast to when we use only \ngal, \wprp, and \gmf to constrain the model, which exhibits no tension. Our inability to jointly match the clustering of SDSS suggests the need for a more flexible HOD model. (See Table~\ref{tab:bestfit}.)
\item We find that our joint and marginalized constraints are affected by the impact of baryonic physics on the masses of dark matter halos. The prescriptions of \citet{Beltz-Mohrmann2021} which we use to modify our halo masses, according to the hydrodynamic physics of EAGLE, Illustris, and IllustrisTNG, have varying degrees of impact on \logmmin, \logmone, and $\alpha$ for the two samples, though \siglogm appears relatively stable, particularly for -19 (See Figure~\ref{fig:halomass}.)
\end{itemize}

These results demonstrate the power of our numerical modeling methodology to constrain the galaxy-halo connection.
By directly populating dark matter simulations and carving out realistic mock galaxy catalogs, we are able to measure clustering statistics in the exact same way on both our SDSS dataset and our model, allowing us to employ less widely used clustering statistics.
Further, by choosing a combination of different physical scales for each clustering statistic, we avoid using observables which introduce more noise than information into our posterior results.
Using this numerical approach, we are also able to quantify and minimize the error associated with both our model observables (Appendix~\ref{app:model_error}) and our covariance matrix (Appendix~\ref{app:cov_error}).
We are also able to verify that an MCMC run on a mock galaxy catalog using our chosen set of observables is able to recover the true HOD parameters (Appendix~\ref{app:mock_chain}), suggesting that the remaining error in our methodology has little impact on our posterior results.

Potentially the largest source of error in this work is our treatment of fiber collisions (Appendix~\ref{app:fib}).
Lacking an accurate model of fiber collisions, we instead decide to adjust our SDSS measurements to account for the error arising from \textit{only} using nearest neighbor corrections.
While we believe that our estimation of this error and use of the resulting adjusted measurements are an improvement over nearest neighbor, this approach still may not yield accurate clustering statistics.
Unfortunately, there is no way to verify its accuracy without actually obtaining the redshift of every galaxy in SDSS.
To get some handle on the impact of our treatment of fiber collisions, however, we also run an MCMC using the unadjusted SDSS measurements. 
We find that the results have little impact on our general conclusions (see Section~\ref{subsec:sdss}).

Overall, compared to S18, we have succeeded in both achieving tighter constraints on the parameters of our HOD model and in heightening the tension between our model and SDSS.
The failure of our five-parameter HOD model to fit the clustering of SDSS for both of volume-limited samples we explore suggests that the assumptions of the standard HOD model may be incorrect, a result in line with several recent works.
For instance, halo occupation may depend on secondary properties of the halo, such as age or concentration, in addition to mass  \citep[i.e., there may be assembly bias; ][]{Zentner2014,Zentner2019,Vakili2019,Beltz-Mohrmann2020,Hadzhiyska2020}.
Additionally, the number of satellite galaxies in a halo of a given mass may not follow a Poisson distribution \citep{Boylan-Kolchin2010,Mao2015,Jimenez2019}.
Finally, central galaxies may not always reside at the center of the halo and move with the mean velocity of the halo (i.e., there might be central spatial and/or velocity bias), and satellite galaxies may not trace the spatial and velocity distribution of the dark matter \citep[i.e., there might be satellite spatial and/or velocity bias; ] []{vandenBosch2005,Watson2012,Piscionere2015,Guo2015a,Guo2015b,Beltz-Mohrmann2020}.
In the case of splashback galaxies, satellites may even exist well outside the radius of their halo \citep[e.g.,][]{Wetzel2010,Sinha2012}.

In the future, we intend to apply our updated modeling methodology (including our algorithm for choosing observables, treatment of fiber collision errors, and halo mass function corrections) to an expanded HOD model including these additional features.
In doing so, we hope to further our understanding of the galaxy-halo connection by exploring its more subtle aspects.
An additional possibility that could be responsible for some of the tension that we find is that the cosmological model that we assume in our analysis is incorrect. Fully exploring the cosmological parameter space with our mock-based numerical model is computationally challenging. Nevertheless this research direction is worth pursuing because the high precision that comes from multiple clustering statistics could yield interesting cosmological constraints that are independent from those on larger scales.

\acknowledgments
This project has been supported by the National Science Foundation (NSF) through Award (AST-1909631).
We thank Jeremy Tinker for helpful conversations during the course of this project and Erin Sheldon for sharing a template which we used to build \cemcee.

This research has made use of NASA's Astrophysics Data System; \textsc{matplotlib}, a Python library for publication quality graphics \citep{Hunter:2007}; \textsc{scipy} \citep{Virtanen_2020}; the \textsc{ipython} package \citep{PER-GRA:2007}; \textsc{astropy}, a community-developed core Python package for Astronomy \citep{2018AJ....156..123A, 2013A&A...558A..33A}; \textsc{numpy} \citep{harris2020array}; \textsc{pandas} \citep{McKinney_2010, McKinney_2011}, and \textsc{chainconsumer} \citep{Hinton2016}.

Funding for the SDSS and SDSS-II has been provided by the Alfred P. Sloan Foundation, the Participating Institutions, the National Science Foundation, the U.S. Department of Energy, the National Aeronautics and Space Administration, the Japanese Monbukagakusho, the Max Planck Society, and the Higher Education Funding Council for England. The SDSS Web Site is \url{http://www.sdss.org/}. 
The SDSS is managed by the Astrophysical Research Consortium for the Participating Institutions. 
The Participating Institutions are the American Museum of Natural History, Astrophysical Institute Potsdam, University of Basel, University of Cambridge, Case Western Reserve University, University of Chicago, Drexel University, Fermilab, the Institute for Advanced Study, the Japan Participation Group, Johns Hopkins University, the Joint Institute for Nuclear Astrophysics, the Kavli Institute for Particle Astrophysics and Cosmology, the Korean Scientist Group, the Chinese Academy of Sciences (LAMOST), Los Alamos National Laboratory, the Max-Planck-Institute for Astronomy (MPIA), the Max-Planck-Institute for Astrophysics (MPA), New Mexico State University, Ohio State University, University of Pittsburgh, University of Portsmouth, Princeton University, the United States Naval Observatory, and the University of Washington.

The mock catalogs used in this paper were produced by the LasDamas project (\url{http://lss.phy.vanderbilt.edu/lasdamas/}); we thank NSF XSEDE for providing the computational resources for LasDamas. 
Some of the computational facilities used in this project were provided by the Vanderbilt Advanced Computing Center for Research and Education (ACCRE). Parts of this research were conducted by the Australian Research Council Centre of Excellence for All Sky Astrophysics in 3 Dimensions (ASTRO 3D), through project number CE170100013.
Finally, in writing this paper we made extensive use of \url{thesaurus.com}.
These acknowledgements were compiled using the Astronomy Acknowledgement Generator (\url{http://astrofrog.github.io/acknowledgment-generator/}).

\pagebreak
\appendix

\section{Fiber Collisions} \label{app:fib}

In this section, we estimate the error in our clustering statistics that results from applying the nearest neighbor correction to SDSS.
To do so, we make use of the SDSS plate overlap regions.
In an overlap region, the same portion of sky is targeted with two (or more) spectroscopic plates.
Approximately 1/3 of the SDSS footprint consists of overlap regions, and thus we have spectra for $\sim 1/3$ of galaxies that would otherwise be lost due to fiber collisions.
We apply the nearest neighbor correction to galaxies in the overlap regions, effectively pretending some of these galaxies were missed due to fiber collisions. 
We then examine the resulting impact on our clustering statistics.

For each luminosity threshold, we first construct a baseline fiducial volume-limited sample as described in Section~\ref{sec:obs}.
We treat this fiducial sample as an ideal case, pretending that all of the redshifts were spectroscopically obtained.
In constructing this fiducial sample we must decide how to treat the galaxies in SDSS which lack spectroscopic redshifts.
We choose to assign to these galaxies the redshift of their nearest neighbor.
We are thus effectively pretending that the nearest neighbor redshifts are the ``correct," spectroscopic redshifts\footnote{We repeated this test instead dropping the galaxies without spectroscopic redshifts, and the results did not change.}.

Next, we construct modified samples in which we apply the nearest neighbor correction to a fraction of galaxies in the overlap regions.
We identify a galaxy as belonging to the overlap regions if both of the following conditions are met: (i) it is within $55''$ of another galaxy and (ii) spectroscopic redshifts were obtained for both galaxies\footnote{Strictly speaking, these galaxies might not all be in SDSS plate overlap regions. In the parent sample we use, redshifts for some galaxies are obtained from other surveys. This distinction, however, does not matter for purposes of the test we describe.}.
To construct our modified samples, we first identify friends-of-friends groups of galaxies, considering only those galaxies in the overlap regions.
Galaxies are linked together as part of a group if their angular separation is less than $55''$.

We next decide whether each galaxy in the group will either keep its own spectroscopic redshift or receive a nearest neighbor correction.
We call a galaxy which is selected at this stage to keep its own redshift a ``hit" and one which is not a ``non-hit.''
For each collision group, we maximize the number of galaxies which could have received spectroscopic fibers from only one plate.
If there exists more than one possible set of galaxies meeting this condition (e.g., a group only of only two galaxies), then we randomly choose one of these sets and designate the galaxies in this set as hits.
The non-hits still may or may not be assigned their spectroscopic redshifts, as we'll soon discuss.

We wish to investigate the impact of the nearest neighbor correction as we increase the fraction of galaxies affected by it.
We therefore only assign a nearest neighbor correction to a fraction of the non-hits.
We consider fractions of 1/4, 1/2, and 1.
For fractions of 1/4 and 1/2, we randomly re-designate said fraction of non-hits as ``misses."
The remaining non-hits are then re-designated as hits.
All hits keep their spectroscopic redshifts, while the misses are assigned the redshift of their nearest neighbor (which must be selected from among the hits).
For a fraction of 1, all non-hits are designated misses.
After assigning a galaxy a new redshift, we recompute this galaxy's k-correction using \textsc{kcorrect v4\_3} \citep{Blanton2007} and absolute magnitude.

For each fraction we consider (1/4, 1/2, 1), we build 100 modified parent samples.
Each parent sample is different due to the randomness associated with choosing hits and misses.
Finally, we construct one $-19$ and $-21$ modified volume-limited sample from each modified parent sample.
These samples differ from our fiducial cases because we have modified the redshifts and absolute magnitudes of many of the galaxies.
Some galaxies which previously made our sample cuts may be dropped in the new samples and vice versa.
We compute clustering statistics on each of these modified samples and compare the results to the fiducial case.

\begin{figure}
    \centering
    \includegraphics[width=\columnwidth]{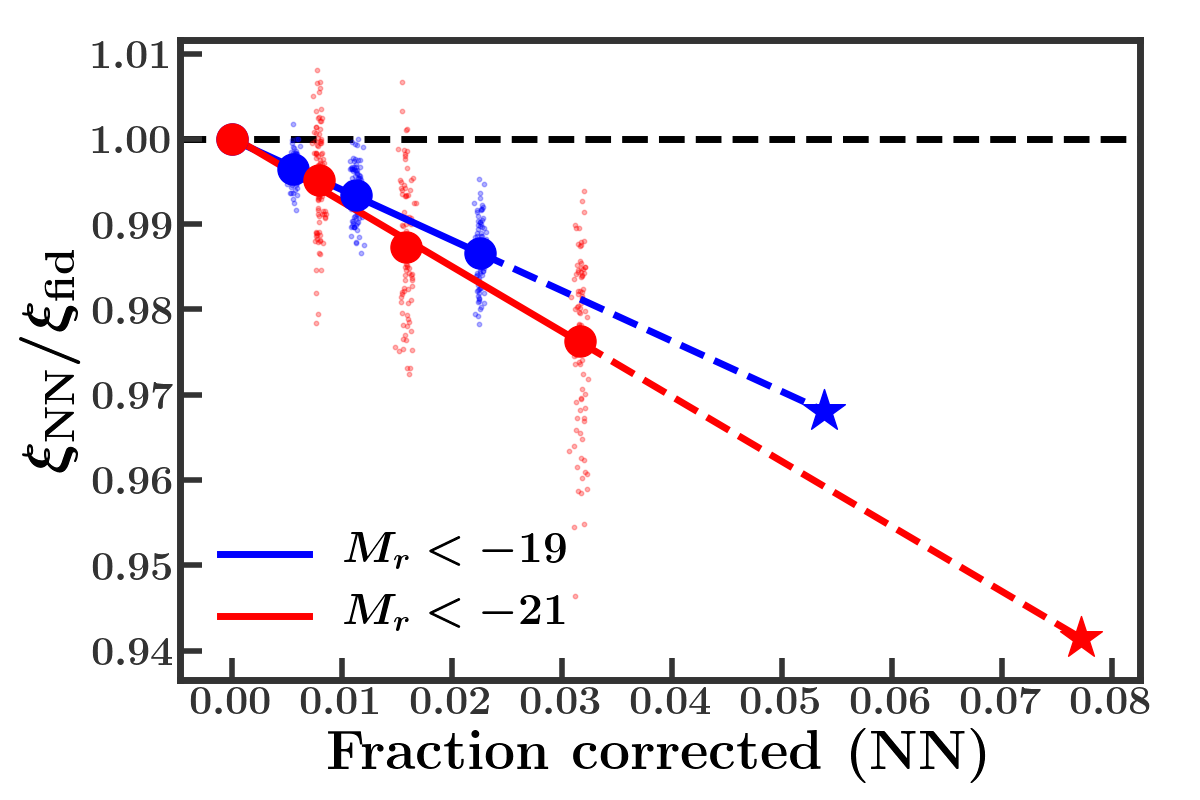}
    \caption{The estimated error in $\xi(s\sim 0.4\ h^{-1}\mathrm{Mpc})$ due to using the nearest neighbor correction to handle fiber collisions. The measurement $\xi_\mathrm{fid}$ is made on a fiducial SDSS sample to which nearest neighbor corrections have been applied. We treat this sample as if redshifts for all galaxies (including the ones which received the nearest neighbor correction) were spectroscopically obtained. The measurement $\xi_\mathrm{NN}$ is made on a modified version of the fiducial sample to which we additionally apply the nearest neighbor correction to galaxies in the SDSS plate overlap regions, reassigning their redshifts. We give $\xi_\mathrm{NN}/\xi_\mathrm{fid}$ as a function of the fraction of galaxies in the overall sample to which we have reassigned nearest neighbor redshifts. Thus, this fraction for the fiducial sample is zero. The small points represent different realizations of which galaxies receive the correction, while the large points give the mean for a fixed fraction. We fit a line to the means (solid line) and extrapolate (dashed line) to estimate the error for our actual SDSS catalogs (star). Results are shown for $-19$ and $-21$ in blue and red, respectively. We perform this same procedure for all of our observables, for both samples. (See text for more details.)}
    \label{fig:fiber}
\end{figure}

\begin{figure*}
    \centering
    \includegraphics[width=6in]{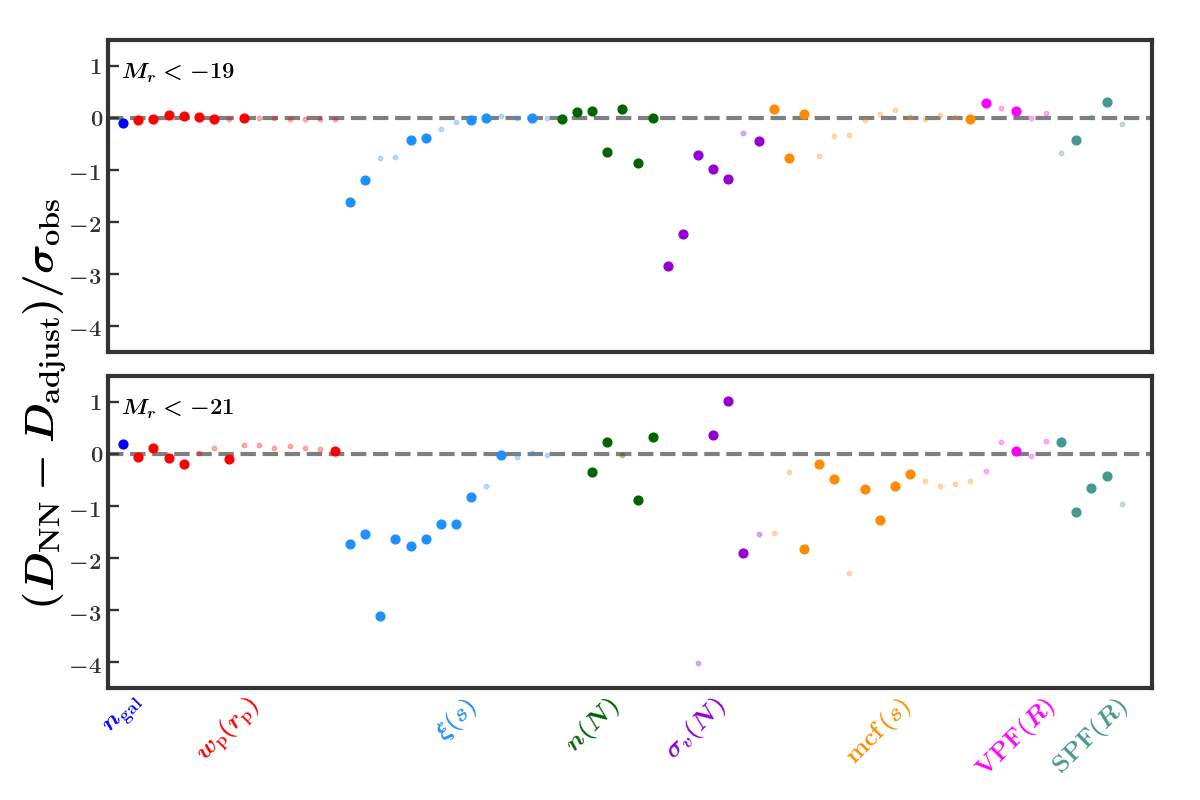}
    \caption{Comparison between the original SDSS measurements and the adjusted measurements which account for the error due to using the nearest neighbor correction to handle fiber collisions. We show results for both the -19 (top) and -21 (bottom) volume-limited samples. Each clustering statistic is indicated on the x-axis with a color-coded label. For each observable, the y-axis gives the difference between $D_\mathrm{NN}$, the unadjusted measurement on the SDSS catalog (which employs nearest neighbor corrections), and $D_\mathrm{adjust}$, the adjusted measurements we use in our analysis. We plot this difference in units of the cosmic variance uncertainty, $\sigma_\mathrm{obs}$. The y-axis thus can be viewed as what happens to the measurement of an observable when we apply nearest neighbor corrections in lieu of having the actual redshifts of all galaxies. The larger points mark the observables we use in our chains on SDSS (Section~\ref{subsec:sdss}).}
    \label{fig:fibers_all}
\end{figure*}

In Figure~\ref{fig:fiber}, we show the results of this test for our two volume-limited samples ($-19$ in blue and $-21$ in red) for one observable: \zxi on a scale of $\sim$ 0.4 \hmpc.
The x-axis is the fraction of galaxies in the total sample to which we reassign a nearest neighbor redshift.
For example, when we apply the nearest-neighbor correction to half of the non-hits, this results in a modified $-21$ sample in which approximately 1.7\% of the total galaxies receive the correction.
The y-axis is the measurement of \zxi on one sample divided by the measurement on the fiducial sample.
Each small point is the result for one of the modified samples, while the large points are the mean taken across samples of equal correction fraction.
Errors in the mean are plotted but are smaller than the point size.
We also show the point (0,1) in which no galaxies receive the correction (i.e., the fiducial case).

We can see in Figure~\ref{fig:fiber} that the error in $\xi(s \sim 0.4 \ h^{-1}\mathrm{Mpc})$ increases linearly as we increase the fraction of galaxies affected by fiber collisions.
We estimate this relationship by fitting a line to the means, including the point (0,1) when fitting.
This relationship is shown for both samples with the solid lines in Figure~\ref{fig:fiber}.
Our actual SDSS volume-limited samples, however, contain a higher fraction of galaxies affected by fiber collisions than we can approximate using overlap regions.
When building our modified samples, applying the nearest neighbor correction to 100 percent of the non-hits produces $-19$ and $-21$ samples in which approximately 2.3 and 3.2 percent of galaxies receive the correction, respectively.
In SDSS, however, approximately 5.4 and 7.7 percent of galaxies actually receive the correction for $-19$ and $-21$, respectively.
To estimate the error for our SDSS samples, we extrapolate the linear fit to the appropriate fraction.
We show this extrapolation with a dashed line in Figure~\ref{fig:fiber} and mark the corresponding value of $\xi/\xi_\mathrm{fid}$ with a star.
We call this value $c_\mathrm{adjust}$.

Remarkably, the relationship between the error in an observable and the fraction of galaxies affected by fiber collisions is linear for all the observables we consider.
We perform the linear fit and extrapolation described above to obtain a value of $c_\mathrm{adjust}$ for each observable.
This value is an estimate of the ratio of the measurement of an observable in our SDSS sample (given the fraction of galaxies which did not receive spectroscopic fibers and thus required the nearest neighbor correction) to the measurement we would make if we had spectroscopic redshifts for all galaxies.
Therefore, we can multiply the value of an observable measured on the fiducial SDSS sample by $1/c_\mathrm{adjust}$ to obtain an adjusted measurement.

In Figure~\ref{fig:fibers_all}, we show the effect that fiber collisions have on the clustering statistics we use in our analysis.
For each observable we take the difference between the measurement on SDSS before adjusting the observables ($D_\mathrm{NN}$) and the measurement after adjusting the observables ($D_\mathrm{adjust}$).
We divide this quantity by the cosmic variance uncertainty and plot the result in Figure~\ref{fig:fibers_all}.
Results are shown for $-19$ in the top panel and $-21$ in the bottom.
Each point corresponds to a specific observable, indicated on the x-axis.
We indicate the observables we use in our analysis with larger, darker points (see Sections~\ref{sec:opt} and \ref{sec:results}).
In general, the nearest neighbor correction works better for the $-19$ sample than it does for $-21$.
This may be due to the higher fraction of galaxies which received the correction in $-21$ compared to $-19$.
Although the correction does well for \ngal, \wprp, and \vpf, it is inadequate for the small scales of \zxi and several scales of \gmf, \sigN, \mcf, and \spf.

The results of this test indicate that a failure to adjust the clustering statistics measured on SDSS catalogs that use nearest neighbor corrections can result in significant errors ($\sim1-4\ \sigma_\mathrm{obs}$) for many of our observables.
Therefore, in this paper, we use the adjusted measurements for all observables.
We provide both the original nearest neighbor measurements and the adjusted measurements in the machine-readable format.
We do not claim that using these adjusted measurements is a complete solution to the issue of fiber collisions.
It is simply the best solution we could establish, given the clustering statistics we wish to use in this paper.
The assumptions of this test certainly warrant more exploration in future work, but our present goals are not to establish a complete method for dealing with fiber collisions.
In any case, in Section~\ref{subsec:sdss} we perform our analysis on the SDSS datasets with and without the adjustments.
While there are some shifts in the posterior results, our general conclusions about the goodness of fit of our five-parameter HOD model do not change.

\section{Accuracy of Model Observables} \label{app:model_error}

In this work, we expand the methodology of S18 using four additional clustering statistics: the redshift-space correlation function \zxi, the mean group velocity dispersion \sigN, the mark correlation function \mcf, and counts-in-cells \cic.
For each of these clustering statistics, we must decide on the most accurate way to measure the model observables (see Section~\ref{sec:model}).
The ideal numerical measurement of the model observables is the mean measurement across many high-resolution mock galaxy catalogs constructed from the same HOD and differing only due to cosmic variance.
Obtaining such a large suite of high-resolution simulations and using them to explore a large number of HOD points in an MCMC both have a prohibitively high computational cost.
We instead have just two high-resolution boxes which we use to estimate the model observables.
Still, we have a choice to measure each clustering statistic using either full boxes or mocks carved out from the boxes.
Additionally we may use just one box or both boxes.
We explore this choice in this section.

While using more volume results in more precisely measured model observables, these measurements may also be more biased (i.e., further from the desired mean than the measurements made on a smaller volume).
Because we do not have a large number of high-resolution simulations, we cannot use our high-resolution boxes to test how close a given method of measurement is to the mean measurement across many mocks.
The closeness of a given box's measurements to the desired mean measurements is driven by the cosmic variance error of that box.
While the measurements themselves are affected by resolution, we do not expect that the cosmic variance of a box is affected by resolution.
Therefore, we can use our suite of 100 low-resolution simulation boxes (two of which have the same initial condition seeds and thus the same large-scale density modes as our high-resolution boxes) to determine the most appropriate method for measuring each clustering statistic.
We assume that the results of this test would be the same if we had a set of 100 high-resolution boxes with the same seeds as our low-resolution boxes.

To make this choice, we first select a fiducial HOD\footnote{Here, we show results using the HOD parameters in Table~\ref{tab:fid}. We have repeated this test using different HOD parameters, and the results do not change.} and populate each of our 100 low-resolution boxes with galaxies.
From these boxes we create 400 mock galaxy catalogs and measure each of our clustering statistics on these mocks.
We use the term ``true mean" to refer to the mean measurement of an observable across these 400 mocks.
We next consider different estimations of the model observables using the two boxes for which we have high-resolution counterparts (see Table~\ref{tab:simulations}).
For all clustering statistics, we consider (1) the mean measurement of four mocks from box 1, (2) the mean measurement of four mocks from box 2, and (3) the mean measurement of eight mocks from both boxes.
For \ngal, \wprp, \zxi, and \cic, we also consider (4) the measurement on box 1, (5) the measurement on box 2, and (6) the average of these two measurements.
We do not consider box measurements for \gmf, \sigN, and \mcf because these statistics all depend on galaxy group identification, a process which is affected by survey geometry \citep{Berlind2006,Campbell2015}.
When we measure statistics on a full box, we first distort galaxies along the z-direction using the distant-observer approximation.
Once we have each of our estimates of the model observables, we compare them to the ``true mean" in order to determine the most appropriate method of measurement.

\begin{figure}
    \centering
    \includegraphics[width=\columnwidth]{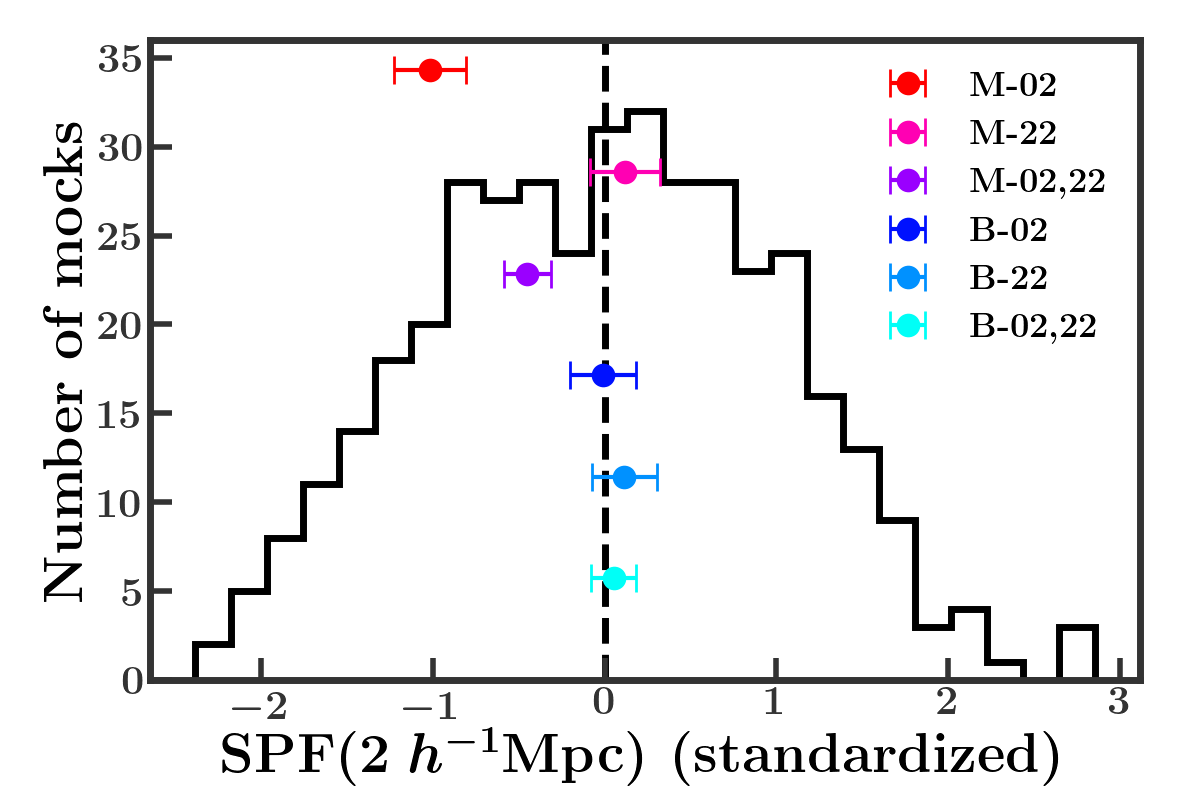}
    \caption{The accuracy of six different estimates of \spf on a scale of 2 \hmpc for the Consuelo boxes. The black histogram shows the distribution of measurements on 400 low-resolution mock galaxy catalogs all created with the same HOD, and the black dashed vertical line shows the mean of this distribution. The colored points show estimates of this mean when using fewer boxes/mocks, as is indicated in the legend. The error bar on these points comes from the stochasticity of the process of populating dark matter halos. All measurements are standardized using the mean and standard deviation of the 400 mocks.}
    \label{fig:one_model}
\end{figure}

This test is illustrated for one observable, \spf on a scale of 2 \hmpc, for the $-19$ sample in Figure~\ref{fig:one_model}.
The black histogram shows measurements of \spf on 400 mocks, while the vertical black line shows the ``true mean."
Each measurement $x_\mathrm{obs}$ in this plot is standardized via the equation $x_\mathrm{std} = \dfrac{x_\mathrm{obs}-\overline{x}_\mathrm{obs}}{\sigma_\mathrm{obs}}$, where $\overline{x}_\mathrm{obs}$ and $\sigma_\mathrm{obs}$ are the mean and standard deviation of measurements across 400 mocks, respectively.
The estimates using boxes and mocks are shown with the multi-colored dots, as is indicated in the legend.
Each time we populate a halo catalog with galaxies, the exact numbers of galaxies in each halo and the placement of these galaxies can vary (see Section~\ref{subsec:mocks}), resulting in a stochastic scatter associated with each estimate of the model observables.
In practice, this process is controlled by a random number generator and is seeded with a ``population seed."
To get a handle on this scatter, we repopulate each of our two low-resolution boxes 100 times, each time using a different population seed.
We then recalculate our estimates of the model 100 times.
Lastly, we standardize each of these 100 measurements by subtracting $\overline{x}_\mathrm{obs}$ and dividing by $\sigma_\mathrm{obs}$.
For each type of estimate, we show the mean and standard deviation of these 100 measurements in Figure~\ref{fig:one_model}, with vertical offset added for display purposes.

Standardizing allows us to easily compare and interpret the accuracy of model estimates across all of our observables.
For example, we can see in Figure~\ref{fig:one_model} that using the 4002 mocks to estimate this scale of \spf results in an estimate that is $\sim 1\ \sigma_\mathrm{obs}$ (i.e., the cosmic variance uncertainty) lower than we desire, while using the 4022 mocks produces a much more accurate estimate.
Naively, if given the choice of averaging measurements from eight mocks or from four mocks, we would choose to average eight mocks.
However, we can see from this plot that the 4002 mocks happen to be outliers for this particular observable.
Thus, including them as part of the eight-mock average results in a more biased model observable than just using the four mocks from box 4022.
On this standardized scale, we will refer to the difference between an estimate and zero as the ``model deviation" and the associated scatter as the ``model scatter."

\begin{figure*}
    \centering
    \includegraphics[width=6in]{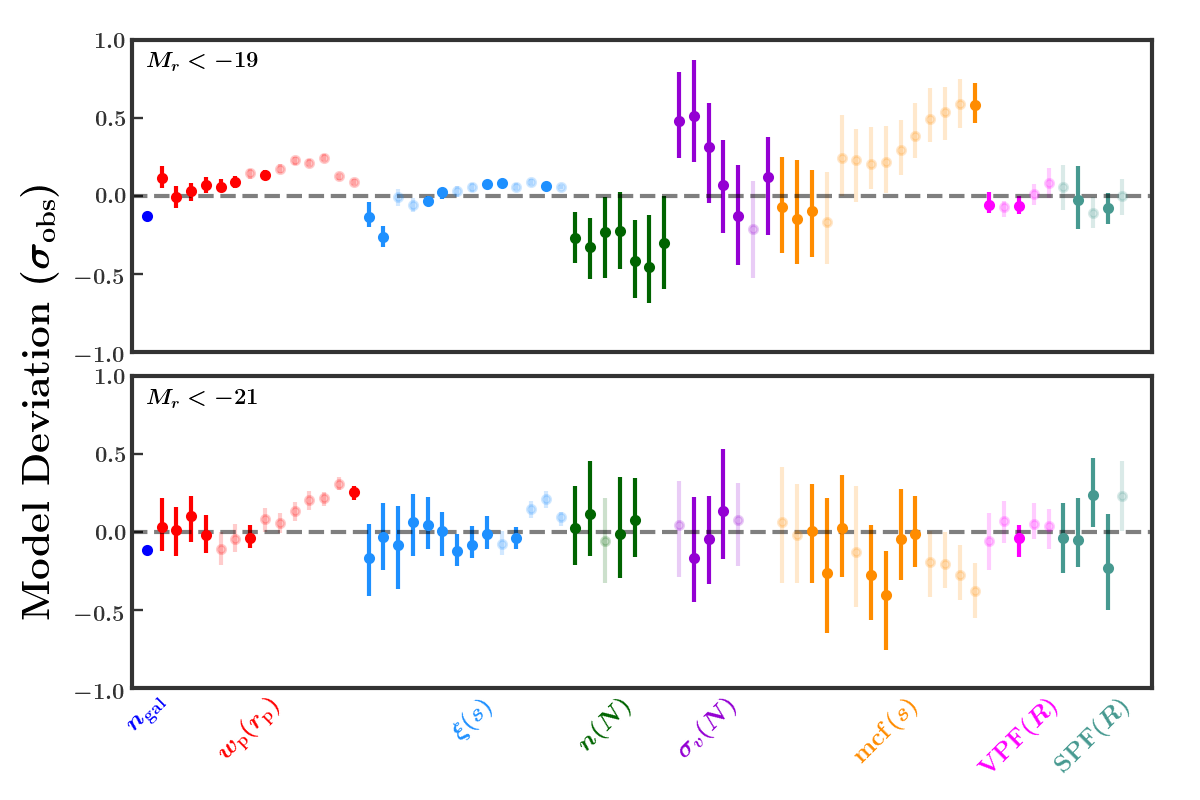}
    \caption{The standardized error in our estimates of the model, after choosing a measurement method for each clustering statistic. The top and bottom panels show the model deviation for the $-19$ and $-21$ samples, respectively. Each clustering statistic is given a different color, as is indicated on the horizontal axis. The error bar associated with each point comes from the stochastic population process and reflects the precision of the model estimate. The more opaque points are the observables we ultimately use in our analysis (see Section~\ref{subsec:choose_obs}).}
    \label{fig:model_error}
\end{figure*}

Looking at Figure~\ref{fig:one_model}, there are several good choices for how to measure this specific scale of \spf.
To maintain some simplicity in our methodology, however, we choose one form of measurement per clustering statistic (not per observable).
Thus, we must take into account how a given method performs for all scales of a clustering statistic.
We treat \vpf and \spf as one clustering statistic when making this choice.
While we certainly desire precise estimates of the model observables (small model scatter), our priority is to choose estimates that are accurate (small model deviation) across all scales.
We make this choice simply by visually comparing the different methods as opposed to establishing any sort of comparison metric.
We summarize the ultimate choices we make for how to measure each clustering statistic in Table~\ref{tab:method} (see Section~\ref{sec:model} for a description).

In Figure~\ref{fig:model_error} for each observable we show the model deviation, given our choice of how to measure said observable. 
We show results for the $-19$ sample in the top panel and for the $-21$ sample in the bottom.
The different clustering statistics are assigned different colors as indicated by the label on the x-axis.
Each point represents a clustering statistic measured at a different scale, ordered in increasing physical size. 
The more opaque points represent the observable we use in our analysis.
For example, the dark green points in both panels show the different scales of \gmf.
For the $-19$ sample we can see that the smallest scale has a model deviation of about -0.25 $\sigma_\mathrm{obs}$.
Thus, we expect our estimate of the model observable to be about 0.25 $\sigma_\mathrm{obs}$ lower than it should be on average.
The scatter in this estimate is about 0.2 $\sigma_\mathrm{obs}$, meaning that our estimate of this model observable is between 0.05 and 0.45 $\sigma_\mathrm{obs}$ lower than it should be for 68\% of the HOD points we evaluate in our MCMC\footnote{As we discuss in Section~\ref{sec:model}, for a given HOD point, we repeat the population process four times (each time using a different population seed) and average the measurements across these populations.
This process produces a scatter that is only half as large as what is shown in Figure~\ref{fig:model_error} for all observables.}.
We display the observables we use in our final SDSS analysis (Section~\ref{subsec:sdss}) in bold.

While we could cull observables by establishing thresholds for model deviation and scatter, we do not know what impact these errors have when they are included in a chain.
Establishing any sort of cutoff then seems rather arbitrary, especially when considering that the model deviation is generally small ($\lesssim 0.5\ \sigma_\mathrm{obs}$ for almost every observable).
We are content to quantify this error and choose methods of measurement that minimize its effects.
In any case, in Appendix~\ref{app:mock_chain} we show that we are able to recover the correct HOD when running a chain on a mock galaxy catalog using the same observables we use in our SDSS analysis (Section~\ref{subsec:sdss}).
This result suggests that the model deviations shown in Figure~\ref{fig:model_error} do not significantly affect the locations of our posterior results.

\section{Errors in Covariance Matrix} \label{app:cov_error}

While using two high resolution boxes is sufficient for estimating the model observables, the covariance matrix requires many mock galaxy catalogs.
Approximate methods, such as jackknifing and bootstrapping, have been shown to produce biased results \citep{Norberg2009}.
We follow the approach of S18 in using many low-resolution mock galaxy catalogs to create a covariance matrix.
While we have made efforts to improve the covariance matrices we use compared to S18 (see Section~\ref{sec:model}), our covariance matrices have four identifiable sources of error:
\begin{enumerate}[noitemsep]
    \item The limited number of mocks. 
    \item The stochastic variation due to population seed.
    \item The use of a fixed fiducial HOD.
    \item The resolution of the simulations.
\end{enumerate}
We discuss the relative importance of the first three of these points in Appendix~\ref{subapp:rel_cov_error} and the last in Appendix~\ref{subapp:resolution}.
We examine the noise only in the cosmic variance uncertainty $\sigma_\mathrm{obs}$ of each observable, ignoring off-diagonal elements of the matrix.

\subsection{Comparing relative errors}
\label{subapp:rel_cov_error}

Assuming Gaussian errors, for a sample of size $N$ and true standard deviation $\sigma$, the error in a calculation of the sample standard deviation is $\sigma/\sqrt{2N}$.
For $N=400$ mocks, this gives a fractional error of $\sim$0.035.
We have verified that the distribution of mock measurements for all observables we consider is roughly Gaussian and so this fractional error holds for all values of $\sigma_\mathrm{obs}$.

The elements of our covariance matrix have an additional error due to the stochastic process of populating dark matter halos, which is controlled by the ``population seed."
Using a different population seed will produce a different matrix.
To quantify this error, using our fiducial HOD we remake our 400 mocks 100 different times with 100 different population seeds.
This gives us 100 estimates of $\sigma_\mathrm{obs}$.
For a particular observable, the fractional error in $\sigma_\mathrm{obs}$ is given by
\begin{equation}
    \label{eq:frac_err}
    \delta(\sigma_\mathrm{obs}) = \sqrt{\frac{1}{100}\sum_{i=1}^{100}(\sigma_{\mathrm{obs},i}-\overline{\sigma}_{\mathrm{obs}})^2}\Bigg/\overline{\sigma}_{\mathrm{obs}} ,
\end{equation}
where $\sigma_{\mathrm{obs},i}$ is the $i^{th}$ estimate of the standard deviation of this observable and $\overline{\sigma}_{\mathrm{obs}}$ is the average of 100 estimates of the standard deviation of this observable.

Lastly, there is an error stemming from our decision to not vary the matrix when we consider new HOD parameters in our chain.
Ideally, each time we evaluate a new set of HOD parameters, we would reconstruct the covariance matrix from 400 mocks built using those HOD parameters.
Because building and measuring clustering statistics on 400 mocks for the $\sim$500,000 HOD points we evaluate in a chain is computationally infeasible, we instead keep our covariance matrix fixed.
To investigate the error we are making in using a fixed matrix, we calculate the fractional error $\delta(\sigma_\mathrm{obs})$ again.
This time, however, instead of varying the population seed while keeping HOD fixed, we build 100 different matrices, each with a different HOD.
To choose the 100 HODs, we randomly sample points from our OPT chains (see Section~\ref{subsec:sdss}).
A broader range of HODs would likely give us a broader error, but we only care about this error insofar as it affects our results.
Thus we only consider the range of HODs explored in our analysis.

\begin{figure*}
    \centering
    \includegraphics[width=6in]{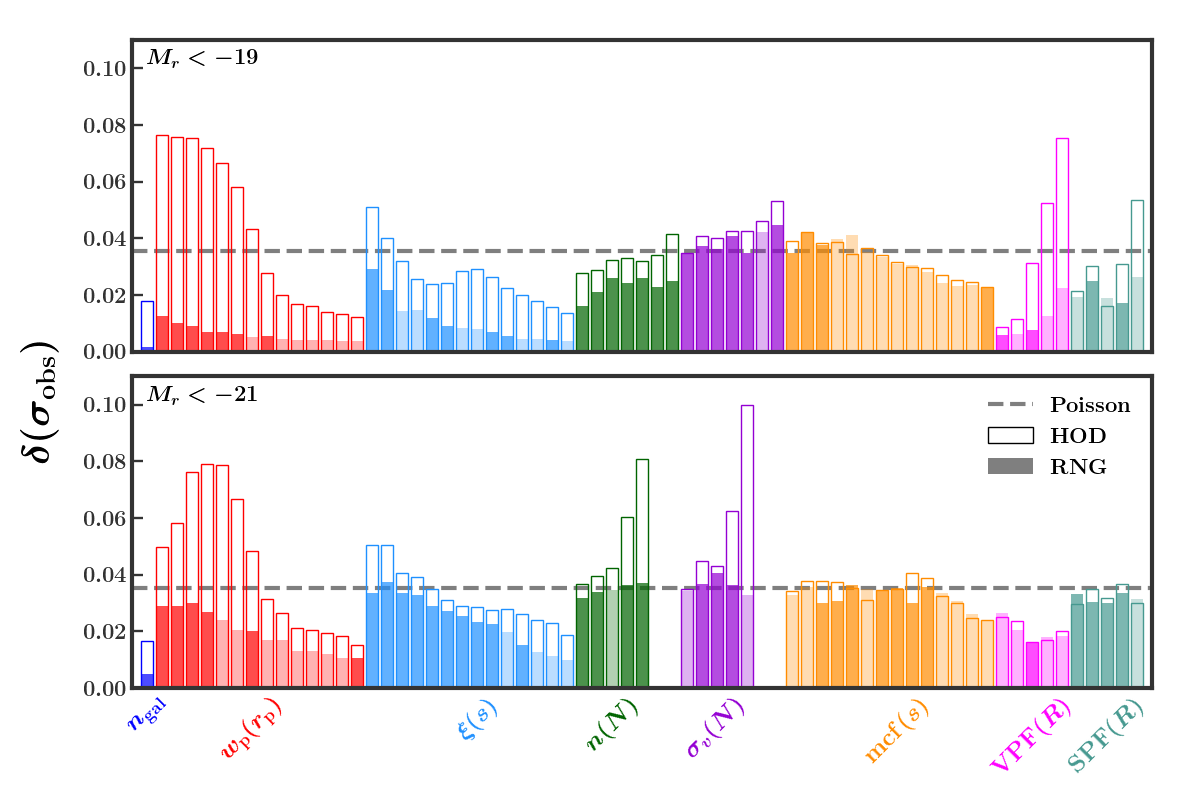}
    \caption{The fractional error in the cosmic variance uncertainty due to various processes. The horizontal line gives the shot noise due to using only 400 mocks to construct the matrix. The height of the hollow vertical bars shows the error from reconstructing the matrix from 100 different HODs, while the height of the filled vertical bars shows the error from varying the ``population seed" (i.e., the seed which controls the numbers of centrals/satellites assigned to each halo) 100 times. The clustering statistics are colored and ordered in the same way as in Figure~\ref{fig:model_error}. The more opaque bars show the observables we actually use in our analysis. For both samples shot noise is the dominant source of error for most observables, though the error from not varying HOD is dominant for many scales.}
    \label{fig:stoc}
\end{figure*}

In Figure~\ref{fig:stoc}, we show these three contributions to the fractional error in $\sigma_\mathrm{obs}$.
The format of this figure is similar to Figure~\ref{fig:model_error}, with clustering statistics assigned different colors and each bar representing a different scale, ordered in increasing physical size. 
The height of the open bar gives the fractional error in $\sigma_\mathrm{obs}$ due to not varying HOD, while the filled bar gives the error due to the stochastic population process.
The Poisson noise associated with the number of mocks we use is the same for all observables and is indicated with the horizontal dashed line.
The clustering statistics we use in our SDSS analysis (Section~\ref{subsec:sdss}) are indicated with greater opacity in the shaded bars.

The fractional errors in $\sigma_\mathrm{obs}$ are all smaller than 10\% for both samples.
For most observables, the dominant contribution to the noise is the number of mocks we use when constructing the matrix.
A few scales of observables (e.g., small scales of \wprp) show an exception to this trend, with the dominant error being due to not varying HOD.
For many observables (e.g., \mcf), there appears to be an ``equal" contribution from all sources of noise.
It is worth noting that the HOD error effectively contains the stochastic population error because we choose to supply unique population seeds for each HOD\footnote{The way our population process works, even if we supply the same seed, different HODs require the generation of different amounts of random numbers, which affects the random number sequencing. Thus, we simply choose to supply different random numbers.}.
Therefore, in instances where these two errors appear equal, the stochastic population error may in fact be dominant.

\subsection{Resolution effects}
\label{subapp:resolution}

An assumption in our modeling procedure is that our covariance matrix is not significantly affected by resolution.
In other words, we assume that, if we were to build the same covariance matrix using 400 high resolution mocks, the differences in matrices would be small.
We cannot accurately test this assumption without actually having the 100 high-resolution boxes, which would negate the need to test the assumption.
However, what we do have is a matching set of two low-resolution and high-resolution boxes.
These boxes are identical in their density fields and only differ in resolution.
We can therefore use these boxes to test the impact of resolution on our matrix.

To test the effect of resolution, we first divide each box into 27 equal-volume  sub-boxes.
In the cases of both samples, the volume of each of these sub-boxes is about half the volume of SDSS.
With two seeds, this gives us 54 low-resolution and 54 matching high-resolution sub-boxes.
Using a fiducial HOD (Table~\ref{tab:fid}), we populate each of these sub-boxes with galaxies and measure clustering statistics.
For each observable, we compute a high-resolution standard deviation ($\sigma_\mathrm{HD}$) and a low-resolution standard deviation ($\sigma_\mathrm{LR}$) across the 54 sub-boxes.

As discussed in Appendix~\ref{app:model_error}, the process of populating dark matter halos is stochastic and is controlled by a ``population seed."
This stochasticity introduces noise into any single calculation of the quantity $\sigma_\mathrm{LR}/\sigma_\mathrm{HD}$.
Therefore, we repeat the population process of both low-resolution and high-resolution sub-boxes with 100 different population seeds.
We measure clustering statistics for each of these 100 sets of 54 sub-boxes and get 100 measurements of $\sigma_\mathrm{LR}$ and $\sigma_\mathrm{HD}$ for each observable. We then average to eliminate the noise due to this stochasticity.

\begin{figure*}
    \centering
    \includegraphics[width=6in]{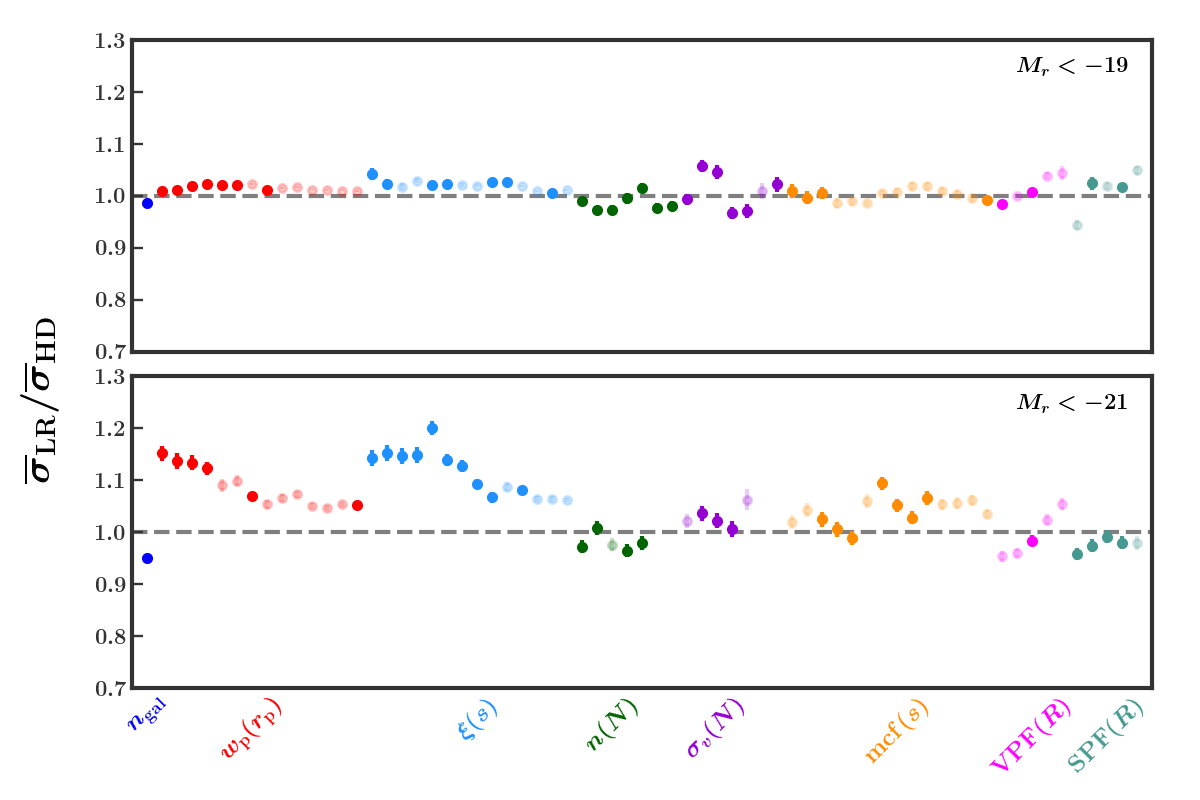}
    \caption{The error in the cosmic variance uncertainty due to resolution. For each observable we calculate the standard deviation across 54 matched high (``HD") and low (``LR") resolution ``sub-boxes." This process is repeated 100 times while varying the ``population seed" to create a distribution for high and low resolution values of $\sigma$. We show here the ratio of means of these distributions. Error bars (barely visible) show the propagation of the errors in the mean of these two distributions.}
    \label{fig:res}
\end{figure*}

In Figure~\ref{fig:res}, we show the results of this test for both $-19$ (top) and $-21$ (bottom).
As in previous figures, different clustering statistics are assigned different colors as indicated by the labels on the x-axis.
Each point represents a clustering statistic measured at a different scale, ordered in increasing physical size.
For each observable we plot the mean of the 100 values of $\sigma_\mathrm{LR}$ divided by the mean of the 100 values of $\sigma_\mathrm{HD}$.
We henceforth refer to this ratio as the ``resolution error."
The error bar (barely visible) associated with each point is found via propagation of the errors in the means of the two distributions.
The optimal observables we use in our analysis (Section~\ref{sec:results}) are indicated with greater opacity.

For the $-19$ sample, all observables have a resolution error smaller than 6\%.
For the $-21$ sample, the resolution errors are larger, but most errors are still less than 10\%.
The observables most affected by resolution are the small scales of \wprp and \zxi in the $-21$ sample, which show a $10-20$\% error.
The larger resolution error in the $-21$ sample is due to the fact that halos of mass \mmin ($10^{12.72}$ \hMsun) in Carmen have fewer particles than halos of mass \mmin ($10^{11.54}$ \hMsun) in Consuelo.
Furthermore, the larger value of \siglogm in Carmen leads us to place galaxies in even lower mass halos that are even less resolved.
While the resolution error appears in general to be more significant than the errors discussed in Appendix~\ref{subapp:rel_cov_error}, we emphasize that overall these errors are quite small and likely have very little impact on our analysis.

Given this result, we do not make any attempts to correct for the resolution error.
We are in general against adding complexity to our modeling procedure unless it seems necessary and well-motivated.
Ultimately our covariance matrix is built from mocks which have a different volume and geometry than the sub-boxes we use to test resolution.
We are not confident that, given these differences, the results of this test would be the same in the case of SDSS-like mocks.
Additionally, larger cosmic variance uncertainties result in lower $\chi^2$ measurements and in general make it more difficult to rule out incorrect models.
If, as this tests suggests, we are using larger cosmic variance uncertainties, then the results of our MCMC will allow a greater space of HOD parameters.
We are content to have artificially broad constraints, as opposed to making a correction that might result in artificially tight constraints.

\begin{figure*}
    \centering
    \includegraphics[width=5.9in]{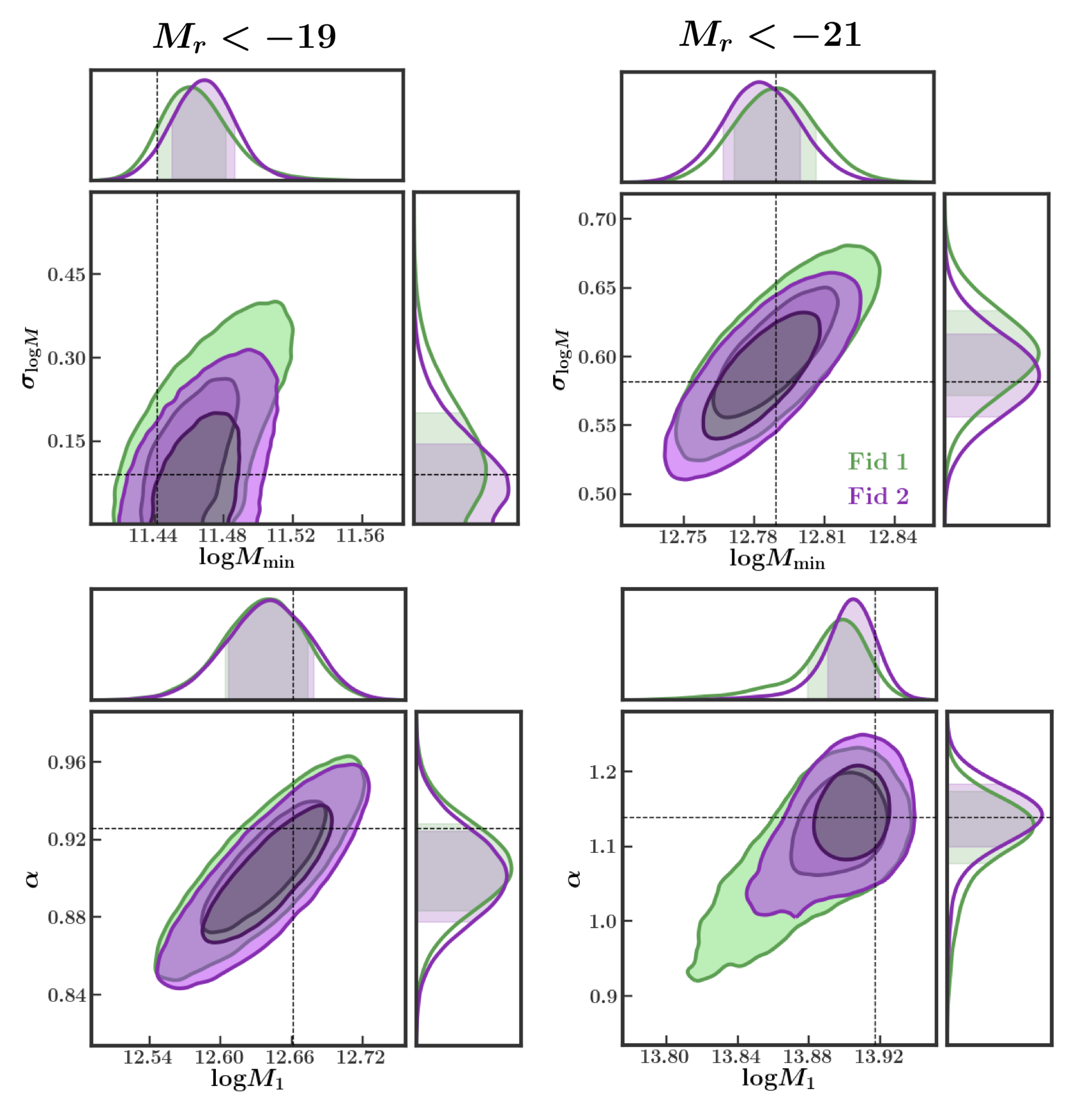}
    \caption{Similar to Fig.~\ref{fig:sdss_chain}, except that these chains are run on mock galaxy catalogs instead of the SDSS. The left and right panels show results for the $-19$ and $-21$ mocks, respectively. Both the green contours and the purple contours come from chains run on the same mock galaxy catalog using the same observables as in our OPT SDSS chains. For the ``Fid 1" chain (green), the HOD used to make the covariance matrix is not the same HOD used to generate the mock. For ``Fid 2" (purple), the matrix and mock are made using the same HOD. The cross-hairs mark the true HOD parameters of the mock galaxy catalog. We are able to recover the true HOD when using our OPT observables, and the difference in covariance matrix has little impact on the results. The HOD parameters for sets Fid 1 and Fid 2 are given in Table~\ref{tab:mock_fid}.}
    \label{fig:mock_chains}
\end{figure*}

Ultimately, we are concerned with what impact the resolution errors will have on the results of our chains.
In Appendix~\ref{app:mock_chain} we show that we are able to recover the correct HOD when running a chain on a mock galaxy catalog using the same observables we use in our SDSS analysis (Section~\ref{subsec:sdss}).
This result suggests that the errors in the matrix due to resolution and the processes discussed in Appendix~\ref{subapp:rel_cov_error} do not have a detectable impact on the locations of our posterior results.

\section{Mock chains} \label{app:mock_chain}

After choosing the set of observables for our SDSS chains, we also used the process of importance sampling to conclude that we could reasonably recover the true HOD parameters when running a chain on a mock galaxy catalog (see end of Section~\ref{subsec:choose_obs}).
Still, because all importance-sampled chains contain some noise, we desire a more robust test of the ability of a chain run using our chosen set of observables to recover the true HOD parameters.
In this section we explicitly test whether or not we can recover the true HOD parameters when running a chain on a mock galaxy catalog.
We also investigate the effect that changing the HOD used to build our fiducial covariance matrix has on our posterior results.

To perform this test, we run an MCMC on a mock galaxy catalog using the same set of observables we use in our SDSS OPT chains.
We construct both $-19$ and $-21$ mock galaxy catalogs from the best-fit HOD values of the SDSS OPT chains\footnote{For this test, we use the best-fit points from the chains run using unadjusted SDSS observables (see Appendix~\ref{app:fib}).}.
We use these HODs (Fid 2 in Table~\ref{tab:mock_fid}) so that our mock has clustering similar to that of SDSS.
When running this MCMC, we use the same covariance matrices we used in our SDSS OPT chains (Fid 1 in Table~\ref{tab:mock_fid}).
These choices are made so as to best replicate the OPT chains of Section~\ref{subsec:sdss} but in the case where we know the true HOD parameters.

\begin{deluxetable}{ccccccc}
\tablenum{8}
\tablecaption{Fiducial HOD Parameters for Mock Chains\label{tab:mock_fid}}
\tablewidth{0pt}
\tablehead{
\colhead{Name} & \colhead{$M_r^\mathrm{lim}$} & \colhead{\logmmin} & \colhead{\siglogm} & \colhead{\logmzero} & \colhead{\logmone} & \colhead{$\alpha$}
}
\startdata
Fid 1 & $-19$ & 11.54 & 0.22 & 12.01 & 12.74 & 0.92 \\
& $-21$ & 12.72 & 0.46 & 7.87 & 13.95 & 1.17 \\
Fid 2 & $-19$ & 11.44 & 0.09 & 11.91 & 12.66 & 0.93 \\
& $-21$ & 12.79 & 0.58 & 7.03 & 13.92 & 1.14 \\
\enddata
\tablecomments{For each sample, we run two chains on a mock galaxy catalog generated using the HOD parameters given by Fid 2. In one of these chains, we construct our covariance matrix using Fid 1, and in the other we construct our matrix using Fid 2.}
\vspace{-5mm}
\end{deluxetable}

In Figure~\ref{fig:mock_chains} we show the results of this test with the green contours (labeled Fid 1 in reference to the matrix).
This figure has the exact same general formatting as Figures~\ref{fig:sdss_chain} and \ref{fig:halomass}.
The cross-hairs indicate the true HOD of the mock galaxy catalog.
We can see that, for both -19 and -21, the true HOD lies inside the joint 68\% probability region for both central and satellite parameters.
Where exactly the truth \textit{should} lie is heavily dependent on the cosmic variance of the mock galaxy catalog.
Thus, an offset of the middle of the contours from the truth, such as we observe here, is completely within reason.
Therefore, we conclude that when using our set of optimal observables, we are able to recover the true HOD parameters.

In Section~\ref{subsec:sdss}, we noted that the HOD used to construct the fiducial covariance matrix was far from the location of our posterior constraints.
To investigate what impact this has, we also run (for both $-19$ and $-21$ mocks) a second chain in which we use a covariance matrix constructed from the same HOD as the mock galaxy catalog (Fid 2).
These results are shown in Figure~\ref{fig:mock_chains} with the purple contours (labeled Fid 2).
When comparing chains run on the same mock but with different matrices (i.e., purple and green contours), there are very minor differences in the posterior results.
Critically, the shifts seen for all parameters are not enough to affect the recovery of the true HOD parameters.
We considered performing an iterative procedure in which we would reconstruct the covariance matrix from the best-fit HOD parameters from the OPT chain and then rerun the OPT chain using this matrix.
In light of these results, we conclude that this iterative procedure is unnecessary and that our choice of fiducial HOD has little impact overall on our posterior results.

Additionally, we check whether our model is able to fit the clustering of the mocks and find that this is indeed the case.
In all of these chains, the p-value of the best-fit HOD point is between 0.74 and 0.94.
Compare this with the SDSS OPT results, where the p-values of the best-fit point indicate that our model is unable to jointly fit the clustering of SDSS.
These results suggest that the tension we find in Section~\ref{sec:results} is real and not an artifact of our modeling procedure.

\bibliography{clustering}{}

\begin{thebibliography}{}
\expandafter\ifx\csname natexlab\endcsname\relax\def\natexlab#1{#1}\fi
\providecommand{\url}[1]{\href{#1}{#1}}
\providecommand{\dodoi}[1]{doi:~\href{http://doi.org/#1}{\nolinkurl{#1}}}
\providecommand{\doeprint}[1]{\href{http://ascl.net/#1}{\nolinkurl{http://ascl.net/#1}}}
\providecommand{\doarXiv}[1]{\href{https://arxiv.org/abs/#1}{\nolinkurl{https://arxiv.org/abs/#1}}}

\bibitem[{{Abazajian} {et~al.}(2009){Abazajian}, {Adelman-McCarthy},
  {Ag{\"u}eros}, {Allam}, {Allende Prieto}, {An}, {Anderson}, {Anderson},
  {Annis}, {Bahcall}, \& et~al.}]{Abazajian2009}
{Abazajian}, K.~N., {Adelman-McCarthy}, J.~K., {Ag{\"u}eros}, M.~A., {et~al.}
  2009, \apjs, 182, 543, \dodoi{10.1088/0067-0049/182/2/543}

\bibitem[{{Artale} {et~al.}(2018){Artale}, {Zehavi}, {Contreras}, \&
  {Norberg}}]{Artale2018}
{Artale}, M.~C., {Zehavi}, I., {Contreras}, S., \& {Norberg}, P. 2018, \mnras,
  480, 3978, \dodoi{10.1093/mnras/sty2110}

\bibitem[{{Astropy Collaboration} {et~al.}(2013){Astropy Collaboration},
  {Robitaille}, {Tollerud}, {Greenfield}, {Droettboom}, {Bray}, {Aldcroft},
  {Davis}, {Ginsburg}, {Price-Whelan}, {Kerzendorf}, {Conley}, {Crighton},
  {Barbary}, {Muna}, {Ferguson}, {Grollier}, {Parikh}, {Nair}, {Unther},
  {Deil}, {Woillez}, {Conseil}, {Kramer}, {Turner}, {Singer}, {Fox}, {Weaver},
  {Zabalza}, {Edwards}, {Azalee Bostroem}, {Burke}, {Casey}, {Crawford},
  {Dencheva}, {Ely}, {Jenness}, {Labrie}, {Lim}, {Pierfederici}, {Pontzen},
  {Ptak}, {Refsdal}, {Servillat}, \& {Streicher}}]{2013A&A...558A..33A}
{Astropy Collaboration}, {Robitaille}, T.~P., {Tollerud}, E.~J., {et~al.} 2013,
  \aap, 558, A33, \dodoi{10.1051/0004-6361/201322068}

\bibitem[{{Astropy Collaboration} {et~al.}(2018){Astropy Collaboration},
  {Price-Whelan}, {Sip{\H o}cz}, {G{\"u}nther}, {Lim}, {Crawford}, {Conseil},
  {Shupe}, {Craig}, {Dencheva}, {Ginsburg}, {VanderPlas}, {Bradley},
  {P{\'e}rez-Su{\'a}rez}, {de Val-Borro}, {Aldcroft}, {Cruz}, {Robitaille},
  {Tollerud}, {Ardelean}, {Babej}, {Bach}, {Bachetti}, {Bakanov}, {Bamford},
  {Barentsen}, {Barmby}, {Baumbach}, {Berry}, {Biscani}, {Boquien}, {Bostroem},
  {Bouma}, {Brammer}, {Bray}, {Breytenbach}, {Buddelmeijer}, {Burke},
  {Calderone}, {Cano Rodr{\'{\i}}guez}, {Cara}, {Cardoso}, {Cheedella},
  {Copin}, {Corrales}, {Crichton}, {D'Avella}, {Deil}, {Depagne}, {Dietrich},
  {Donath}, {Droettboom}, {Earl}, {Erben}, {Fabbro}, {Ferreira}, {Finethy},
  {Fox}, {Garrison}, {Gibbons}, {Goldstein}, {Gommers}, {Greco}, {Greenfield},
  {Groener}, {Grollier}, {Hagen}, {Hirst}, {Homeier}, {Horton}, {Hosseinzadeh},
  {Hu}, {Hunkeler}, {Ivezi{\'c}}, {Jain}, {Jenness}, {Kanarek}, {Kendrew},
  {Kern}, {Kerzendorf}, {Khvalko}, {King}, {Kirkby}, {Kulkarni}, {Kumar},
  {Lee}, {Lenz}, {Littlefair}, {Ma}, {Macleod}, {Mastropietro}, {McCully},
  {Montagnac}, {Morris}, {Mueller}, {Mumford}, {Muna}, {Murphy}, {Nelson},
  {Nguyen}, {Ninan}, {N{\"o}the}, {Ogaz}, {Oh}, {Parejko}, {Parley}, {Pascual},
  {Patil}, {Patil}, {Plunkett}, {Prochaska}, {Rastogi}, {Reddy Janga},
  {Sabater}, {Sakurikar}, {Seifert}, {Sherbert}, {Sherwood-Taylor}, {Shih},
  {Sick}, {Silbiger}, {Singanamalla}, {Singer}, {Sladen}, {Sooley},
  {Sornarajah}, {Streicher}, {Teuben}, {Thomas}, {Tremblay}, {Turner},
  {Terr{\'o}n}, {van Kerkwijk}, {de la Vega}, {Watkins}, {Weaver}, {Whitmore},
  {Woillez}, {Zabalza}, \& {Astropy Contributors}}]{2018AJ....156..123A}
{Astropy Collaboration}, {Price-Whelan}, A.~M., {Sip{\H o}cz}, B.~M., {et~al.}
  2018, \aj, 156, 123, \dodoi{10.3847/1538-3881/aabc4f}

\bibitem[{{Behroozi} {et~al.}(2010){Behroozi}, {Conroy}, \&
  {Wechsler}}]{Behroozi2010}
{Behroozi}, P.~S., {Conroy}, C., \& {Wechsler}, R.~H. 2010, \apj, 717, 379,
  \dodoi{10.1088/0004-637X/717/1/379}

\bibitem[{{Behroozi} {et~al.}(2013){Behroozi}, {Wechsler}, \&
  {Wu}}]{Behroozi2013}
{Behroozi}, P.~S., {Wechsler}, R.~H., \& {Wu}, H.-Y. 2013, \apj, 762, 109,
  \dodoi{10.1088/0004-637X/762/2/109}

\bibitem[{{Beltz-Mohrmann} \& {Berlind}(2021)}]{Beltz-Mohrmann2021}
{Beltz-Mohrmann}, G.~D., \& {Berlind}, A.~A. 2021, arXiv e-prints,
  arXiv:2103.05076.
\newblock \doarXiv{2103.05076}

\bibitem[{{Beltz-Mohrmann} {et~al.}(2020){Beltz-Mohrmann}, {Berlind}, \&
  {Szewciw}}]{Beltz-Mohrmann2020}
{Beltz-Mohrmann}, G.~D., {Berlind}, A.~A., \& {Szewciw}, A.~O. 2020, \mnras,
  491, 5771, \dodoi{10.1093/mnras/stz3442}

\bibitem[{{Benson} {et~al.}(2000){Benson}, {Cole}, {Frenk}, {Baugh}, \&
  {Lacey}}]{Benson2000}
{Benson}, A.~J., {Cole}, S., {Frenk}, C.~S., {Baugh}, C.~M., \& {Lacey}, C.~G.
  2000, \mnras, 311, 793, \dodoi{10.1046/j.1365-8711.2000.03101.x}

\bibitem[{{Berlind} \& {Weinberg}(2002)}]{Berlind2002}
{Berlind}, A.~A., \& {Weinberg}, D.~H. 2002, \apj, 575, 587,
  \dodoi{10.1086/341469}

\bibitem[{{Berlind} {et~al.}(2003){Berlind}, {Weinberg}, {Benson}, {Baugh},
  {Cole}, {Dav{\'e}}, {Frenk}, {Jenkins}, {Katz}, \& {Lacey}}]{Berlind2003}
{Berlind}, A.~A., {Weinberg}, D.~H., {Benson}, A.~J., {et~al.} 2003, \apj, 593,
  1, \dodoi{10.1086/376517}

\bibitem[{{Berlind} {et~al.}(2006){Berlind}, {Frieman}, {Weinberg}, {Blanton},
  {Warren}, {Abazajian}, {Scranton}, {Hogg}, {Scoccimarro}, {Bahcall},
  {Brinkmann}, {Gott}, {Kleinman}, {Krzesinski}, {Lee}, {Miller}, {Nitta},
  {Schneider}, {Tucker}, {Zehavi}, \& {SDSS Collaboration}}]{Berlind2006}
{Berlind}, A.~A., {Frieman}, J., {Weinberg}, D.~H., {et~al.} 2006, \apjs, 167,
  1, \dodoi{10.1086/508170}

\bibitem[{Blanton \& Roweis(2007)}]{Blanton2007}
Blanton, M.~R., \& Roweis, S. 2007, The Astronomical Journal, 133, 734,
  \dodoi{10.1086/510127}

\bibitem[{{Blanton} {et~al.}(2003){Blanton}, {Brinkmann}, {Csabai}, {Doi},
  {Eisenstein}, {Fukugita}, {Gunn}, {Hogg}, \& {Schlegel}}]{Blanton2003b}
{Blanton}, M.~R., {Brinkmann}, J., {Csabai}, I., {et~al.} 2003, \aj, 125, 2348,
  \dodoi{10.1086/342935}

\bibitem[{{Blanton} {et~al.}(2005){Blanton}, {Schlegel}, {Strauss},
  {Brinkmann}, {Finkbeiner}, {Fukugita}, {Gunn}, {Hogg}, {Ivezi{\'c}}, {Knapp},
  {Lupton}, {Munn}, {Schneider}, {Tegmark}, \& {Zehavi}}]{Blanton2005}
{Blanton}, M.~R., {Schlegel}, D.~J., {Strauss}, M.~A., {et~al.} 2005, \aj, 129,
  2562, \dodoi{10.1086/429803}

\bibitem[{{Bose} {et~al.}(2019){Bose}, {Eisenstein}, {Hernquist}, {Pillepich},
  {Nelson}, {Marinacci}, {Springel}, \& {Vogelsberger}}]{Bose2019}
{Bose}, S., {Eisenstein}, D.~J., {Hernquist}, L., {et~al.} 2019, \mnras, 490,
  5693, \dodoi{10.1093/mnras/stz2546}

\bibitem[{{Boylan-Kolchin} {et~al.}(2010){Boylan-Kolchin}, {Springel}, {White},
  \& {Jenkins}}]{Boylan-Kolchin2010}
{Boylan-Kolchin}, M., {Springel}, V., {White}, S.~D.~M., \& {Jenkins}, A. 2010,
  \mnras, 406, 896, \dodoi{10.1111/j.1365-2966.2010.16774.x}

\bibitem[{{Bryan} \& {Norman}(1998)}]{Bryan1998}
{Bryan}, G.~L., \& {Norman}, M.~L. 1998, \apj, 495, 80, \dodoi{10.1086/305262}

\bibitem[{{Campbell} {et~al.}(2015){Campbell}, {van den Bosch}, {Hearin},
  {Padmanabhan}, {Berlind}, {Mo}, {Tinker}, \& {Yang}}]{Campbell2015}
{Campbell}, D., {van den Bosch}, F.~C., {Hearin}, A., {et~al.} 2015, \mnras,
  452, 444, \dodoi{10.1093/mnras/stv1091}

\bibitem[{{Contreras} {et~al.}(2019){Contreras}, {Zehavi}, {Padilla}, {Baugh},
  {Jim{\'e}nez}, \& {Lacerna}}]{Contreras2019}
{Contreras}, S., {Zehavi}, I., {Padilla}, N., {et~al.} 2019, \mnras, 484, 1133,
  \dodoi{10.1093/mnras/stz018}

\bibitem[{{Cooray} \& {Sheth}(2002)}]{Cooray2002}
{Cooray}, A., \& {Sheth}, R. 2002, \physrep, 372, 1,
  \dodoi{10.1016/S0370-1573(02)00276-4}

\bibitem[{{Coupon} {et~al.}(2015){Coupon}, {Arnouts}, {van Waerbeke},
  {Moutard}, {Ilbert}, {van Uitert}, {Erben}, {Garilli}, {Guzzo}, {Heymans},
  {Hildebrandt}, {Hoekstra}, {Kilbinger}, {Kitching}, {Mellier}, {Miller},
  {Scodeggio}, {Bonnett}, {Branchini}, {Davidzon}, {De Lucia}, {Fritz}, {Fu},
  {Hudelot}, {Hudson}, {Kuijken}, {Leauthaud}, {Le F{\`e}vre}, {McCracken},
  {Moscardini}, {Rowe}, {Schrabback}, {Semboloni}, \& {Velander}}]{Coupon2015}
{Coupon}, J., {Arnouts}, S., {van Waerbeke}, L., {et~al.} 2015, \mnras, 449,
  1352, \dodoi{10.1093/mnras/stv276}

\bibitem[{{Crocce} {et~al.}(2006){Crocce}, {Pueblas}, \&
  {Scoccimarro}}]{Crocce2006}
{Crocce}, M., {Pueblas}, S., \& {Scoccimarro}, R. 2006, \mnras, 373, 369,
  \dodoi{10.1111/j.1365-2966.2006.11040.x}

\bibitem[{{Croton} {et~al.}(2007){Croton}, {Gao}, \& {White}}]{Croton2007}
{Croton}, D.~J., {Gao}, L., \& {White}, S.~D.~M. 2007, \mnras, 374, 1303,
  \dodoi{10.1111/j.1365-2966.2006.11230.x}

\bibitem[{{Foreman-Mackey} {et~al.}(2013){Foreman-Mackey}, {Hogg}, {Lang}, \&
  {Goodman}}]{EMCEE2013}
{Foreman-Mackey}, D., {Hogg}, D.~W., {Lang}, D., \& {Goodman}, J. 2013, \pasp,
  125, 306, \dodoi{10.1086/670067}

\bibitem[{{Gao} {et~al.}(2005){Gao}, {Springel}, \& {White}}]{Gao2005}
{Gao}, L., {Springel}, V., \& {White}, S.~D.~M. 2005, \mnras, 363, L66,
  \dodoi{10.1111/j.1745-3933.2005.00084.x}

\bibitem[{{Guo} {et~al.}(2015{\natexlab{a}}){Guo}, {Zheng}, {Zehavi}, {Dawson},
  {Skibba}, {Tinker}, {Weinberg}, {White}, \& {Schneider}}]{Guo2015a}
{Guo}, H., {Zheng}, Z., {Zehavi}, I., {et~al.} 2015{\natexlab{a}}, \mnras, 446,
  578, \dodoi{10.1093/mnras/stu2120}

\bibitem[{{Guo} {et~al.}(2015{\natexlab{b}}){Guo}, {Zheng}, {Zehavi},
  {Behroozi}, {Chuang}, {Comparat}, {Favole}, {Gottloeber}, {Klypin}, {Prada},
  {Weinberg}, \& {Yepes}}]{Guo2015b}
---. 2015{\natexlab{b}}, \mnras, 453, 4368, \dodoi{10.1093/mnras/stv1966}

\bibitem[{{Hadzhiyska} {et~al.}(2021{\natexlab{a}}){Hadzhiyska}, {Bose},
  {Eisenstein}, \& {Hernquist}}]{Hadzhiyska2021a}
{Hadzhiyska}, B., {Bose}, S., {Eisenstein}, D., \& {Hernquist}, L.
  2021{\natexlab{a}}, \mnras, 501, 1603, \dodoi{10.1093/mnras/staa3776}

\bibitem[{{Hadzhiyska} {et~al.}(2020){Hadzhiyska}, {Bose}, {Eisenstein},
  {Hernquist}, \& {Spergel}}]{Hadzhiyska2020}
{Hadzhiyska}, B., {Bose}, S., {Eisenstein}, D., {Hernquist}, L., \& {Spergel},
  D.~N. 2020, \mnras, 493, 5506, \dodoi{10.1093/mnras/staa623}

\bibitem[{{Hadzhiyska} {et~al.}(2021{\natexlab{b}}){Hadzhiyska}, {Tacchella},
  {Bose}, \& {Eisenstein}}]{Hadzhiyska2021b}
{Hadzhiyska}, B., {Tacchella}, S., {Bose}, S., \& {Eisenstein}, D.~J.
  2021{\natexlab{b}}, \mnras, 502, 3599, \dodoi{10.1093/mnras/stab243}

\bibitem[{Harris {et~al.}(2020)Harris, Millman, van~der Walt, Gommers,
  Virtanen, Cournapeau, Wieser, Taylor, Berg, Smith, Kern, Picus, Hoyer, van
  Kerkwijk, Brett, Haldane, del R{'{\i}}o, Wiebe, Peterson,
  G{'{e}}rard-Marchant, Sheppard, Reddy, Weckesser, Abbasi, Gohlke, \&
  Oliphant}]{harris2020array}
Harris, C.~R., Millman, K.~J., van~der Walt, S.~J., {et~al.} 2020, Nature, 585,
  357, \dodoi{10.1038/s41586-020-2649-2}

\bibitem[{{Hearin} {et~al.}(2016){Hearin}, {Zentner}, {van den Bosch},
  {Campbell}, \& {Tollerud}}]{Hearin2016}
{Hearin}, A.~P., {Zentner}, A.~R., {van den Bosch}, F.~C., {Campbell}, D., \&
  {Tollerud}, E. 2016, \mnras, 460, 2552, \dodoi{10.1093/mnras/stw840}

\bibitem[{{Hinton}(2016)}]{Hinton2016}
{Hinton}, S.~R. 2016, The Journal of Open Source Software, 1, 00045,
  \dodoi{10.21105/joss.00045}

\bibitem[{Hunter(2007)}]{Hunter:2007}
Hunter, J.~D. 2007, Computing In Science \& Engineering, 9, 90

\bibitem[{{Jim{\'e}nez} {et~al.}(2019){Jim{\'e}nez}, {Contreras}, {Padilla},
  {Zehavi}, {Baugh}, \& {Gonzalez-Perez}}]{Jimenez2019}
{Jim{\'e}nez}, E., {Contreras}, S., {Padilla}, N., {et~al.} 2019, \mnras, 490,
  3532, \dodoi{10.1093/mnras/stz2790}

\bibitem[{{Jing} {et~al.}(1998){Jing}, {Mo}, \& {B{\"o}rner}}]{Jing1998}
{Jing}, Y.~P., {Mo}, H.~J., \& {B{\"o}rner}, G. 1998, \apj, 494, 1,
  \dodoi{10.1086/305209}

\bibitem[{{Kravtsov} {et~al.}(2004){Kravtsov}, {Berlind}, {Wechsler}, {Klypin},
  {Gottl{\"o}ber}, {Allgood}, \& {Primack}}]{Kravtsov2004}
{Kravtsov}, A.~V., {Berlind}, A.~A., {Wechsler}, R.~H., {et~al.} 2004, \apj,
  609, 35, \dodoi{10.1086/420959}

\bibitem[{{Lacey} \& {Cole}(1994)}]{Lacey1994}
{Lacey}, C., \& {Cole}, S. 1994, \mnras, 271, 676,
  \dodoi{10.1093/mnras/271.3.676}

\bibitem[{{Landy} \& {Szalay}(1993)}]{Landy1993}
{Landy}, S.~D., \& {Szalay}, A.~S. 1993, \apj, 412, 64, \dodoi{10.1086/172900}

\bibitem[{{Leauthaud} {et~al.}(2012){Leauthaud}, {Tinker}, {Bundy}, {Behroozi},
  {Massey}, {Rhodes}, {George}, {Kneib}, {Benson}, {Wechsler}, {Busha},
  {Capak}, {Cort{\^e}s}, {Ilbert}, {Koekemoer}, {Le F{\`e}vre}, {Lilly},
  {McCracken}, {Salvato}, {Schrabback}, {Scoville}, {Smith}, \&
  {Taylor}}]{Leauthaud2012}
{Leauthaud}, A., {Tinker}, J., {Bundy}, K., {et~al.} 2012, \apj, 744, 159,
  \dodoi{10.1088/0004-637X/744/2/159}

\bibitem[{{Ma} \& {Fry}(2000)}]{Ma2000}
{Ma}, C.-P., \& {Fry}, J.~N. 2000, \apj, 543, 503, \dodoi{10.1086/317146}

\bibitem[{{Mao} {et~al.}(2015){Mao}, {Williamson}, \& {Wechsler}}]{Mao2015}
{Mao}, Y.-Y., {Williamson}, M., \& {Wechsler}, R.~H. 2015, \apj, 810, 21,
  \dodoi{10.1088/0004-637X/810/1/21}

\bibitem[{{McBride} {et~al.}(2009){McBride}, {Berlind}, {Scoccimarro},
  {Wechsler}, {Busha}, {Gardner}, \& {van den Bosch}}]{McBride2009}
{McBride}, C., {Berlind}, A., {Scoccimarro}, R., {et~al.} 2009, in Bulletin of
  the American Astronomical Society, Vol.~41, American Astronomical Society
  Meeting Abstracts \#213, 253

\bibitem[{{McClelland} \& {Silk}(1977)}]{McClelland1977}
{McClelland}, J., \& {Silk}, J. 1977, \apj, 217, 331, \dodoi{10.1086/155583}

\bibitem[{{McCullagh} {et~al.}(2017){McCullagh}, {Norberg}, {Cole},
  {Gonzalez-Perez}, {Baugh}, \& {Helly}}]{McCullagh2017}
{McCullagh}, N., {Norberg}, P., {Cole}, S., {et~al.} 2017, arXiv e-prints.
\newblock \doarXiv{1705.01988}

\bibitem[{McKinney(2010)}]{McKinney_2010}
McKinney, W. 2010, in Proceedings of the 9th Python in Science Conference, Vol.
  445, Austin, TX, 51--56

\bibitem[{McKinney(2011)}]{McKinney_2011}
McKinney, W. 2011, Python for High Performance and Scientific Computing, 14

\bibitem[{{Montero-Dorta} {et~al.}(2021){Montero-Dorta}, {Chaves-Montero},
  {Artale}, \& {Favole}}]{Montero2021}
{Montero-Dorta}, A.~D., {Chaves-Montero}, J., {Artale}, M.~C., \& {Favole}, G.
  2021, arXiv e-prints, arXiv:2105.05274.
\newblock \doarXiv{2105.05274}

\bibitem[{{Neyman} \& {Scott}(1952)}]{Neyman1952}
{Neyman}, J., \& {Scott}, E.~L. 1952, \apj, 116, 144, \dodoi{10.1086/145599}

\bibitem[{{Norberg} {et~al.}(2009){Norberg}, {Baugh}, {Gazta{\~n}aga}, \&
  {Croton}}]{Norberg2009}
{Norberg}, P., {Baugh}, C.~M., {Gazta{\~n}aga}, E., \& {Croton}, D.~J. 2009,
  \mnras, 396, 19, \dodoi{10.1111/j.1365-2966.2009.14389.x}

\bibitem[{{Padilla} {et~al.}(2019){Padilla}, {Contreras}, {Zehavi}, {Baugh}, \&
  {Norberg}}]{Padilla2019}
{Padilla}, N., {Contreras}, S., {Zehavi}, I., {Baugh}, C.~M., \& {Norberg}, P.
  2019, \mnras, 486, 582, \dodoi{10.1093/mnras/stz824}

\bibitem[{{Parejko} {et~al.}(2013){Parejko}, {Sunayama}, {Padmanabhan}, {Wake},
  {Berlind}, {Bizyaev}, {Blanton}, {Bolton}, {van den Bosch}, {Brinkmann},
  {Brownstein}, {da Costa}, {Eisenstein}, {Guo}, {Kazin}, {Maia},
  {Malanushenko}, {Maraston}, {McBride}, {Nichol}, {Oravetz}, {Pan},
  {Percival}, {Prada}, {Ross}, {Ross}, {Schlegel}, {Schneider}, {Simmons},
  {Skibba}, {Tinker}, {Tojeiro}, {Weaver}, {Wetzel}, {White}, {Weinberg},
  {Thomas}, {Zehavi}, \& {Zheng}}]{Parejko2013}
{Parejko}, J.~K., {Sunayama}, T., {Padmanabhan}, N., {et~al.} 2013, \mnras,
  429, 98, \dodoi{10.1093/mnras/sts314}

\bibitem[{{Peacock} \& {Smith}(2000)}]{Peacock2000}
{Peacock}, J.~A., \& {Smith}, R.~E. 2000, \mnras, 318, 1144,
  \dodoi{10.1046/j.1365-8711.2000.03779.x}

\bibitem[{{Peebles}(1974)}]{Peebles1974}
{Peebles}, P.~J.~E. 1974, \aap, 32, 197

\bibitem[{P\'erez \& Granger(2007)}]{PER-GRA:2007}
P\'erez, F., \& Granger, B.~E. 2007, Computing in Science and Engineering, 9,
  21, \dodoi{10.1109/MCSE.2007.53}

\bibitem[{{Piscionere} {et~al.}(2015){Piscionere}, {Berlind}, {McBride}, \&
  {Scoccimarro}}]{Piscionere2015}
{Piscionere}, J.~A., {Berlind}, A.~A., {McBride}, C.~K., \& {Scoccimarro}, R.
  2015, \apj, 806, 125, \dodoi{10.1088/0004-637X/806/1/125}

\bibitem[{{Planck Collaboration} {et~al.}(2014){Planck Collaboration}, {Ade},
  {Aghanim}, {Armitage-Caplan}, {Arnaud}, {Ashdown}, {Atrio-Barand ela},
  {Aumont}, {Baccigalupi}, {Banday}, {Barreiro}, {Bartlett}, {Battaner},
  {Benabed}, {Beno{\^\i}t}, {Benoit-L{\'e}vy}, {Bernard}, {Bersanelli},
  {Bielewicz}, {Bobin}, {Bock}, {Bonaldi}, {Bond}, {Borrill}, {Bouchet},
  {Bridges}, {Bucher}, {Burigana}, {Butler}, {Calabrese}, {Cappellini},
  {Cardoso}, {Catalano}, {Challinor}, {Chamballu}, {Chary}, {Chen}, {Chiang},
  {Chiang}, {Christensen}, {Church}, {Clements}, {Colombi}, {Colombo},
  {Couchot}, {Coulais}, {Crill}, {Curto}, {Cuttaia}, {Danese}, {Davies},
  {Davis}, {de Bernardis}, {de Rosa}, {de Zotti}, {Delabrouille}, {Delouis},
  {D{\'e}sert}, {Dickinson}, {Diego}, {Dolag}, {Dole}, {Donzelli}, {Dor{\'e}},
  {Douspis}, {Dunkley}, {Dupac}, {Efstathiou}, {Elsner}, {En{\ss}lin},
  {Eriksen}, {Finelli}, {Forni}, {Frailis}, {Fraisse}, {Franceschi}, {Gaier},
  {Galeotta}, {Galli}, {Ganga}, {Giard}, {Giardino}, {Giraud-H{\'e}raud},
  {Gjerl{\o}w}, {Gonz{\'a}lez-Nuevo}, {G{\'o}rski}, {Gratton}, {Gregorio},
  {Gruppuso}, {Gudmundsson}, {Haissinski}, {Hamann}, {Hansen}, {Hanson},
  {Harrison}, {Henrot-Versill{\'e}}, {Hern{\'a}ndez-Monteagudo}, {Herranz},
  {Hildebrand t}, {Hivon}, {Hobson}, {Holmes}, {Hornstrup}, {Hou}, {Hovest},
  {Huffenberger}, {Jaffe}, {Jaffe}, {Jewell}, {Jones}, {Juvela},
  {Keih{\"a}nen}, {Keskitalo}, {Kisner}, {Kneissl}, {Knoche}, {Knox}, {Kunz},
  {Kurki-Suonio}, {Lagache}, {L{\"a}hteenm{\"a}ki}, {Lamarre}, {Lasenby},
  {Lattanzi}, {Laureijs}, {Lawrence}, {Leach}, {Leahy}, {Leonardi},
  {Le{\'o}n-Tavares}, {Lesgourgues}, {Lewis}, {Liguori}, {Lilje},
  {Linden-V{\o}rnle}, {L{\'o}pez-Caniego}, {Lubin}, {Mac{\'\i}as-P{\'e}rez},
  {Maffei}, {Maino}, {Mand olesi}, {Maris}, {Marshall}, {Martin},
  {Mart{\'\i}nez-Gonz{\'a}lez}, {Masi}, {Massardi}, {Matarrese}, {Matthai},
  {Mazzotta}, {Meinhold}, {Melchiorri}, {Melin}, {Mendes}, {Menegoni},
  {Mennella}, {Migliaccio}, {Millea}, {Mitra}, {Miville-Desch{\^e}nes},
  {Moneti}, {Montier}, {Morgante}, {Mortlock}, {Moss}, {Munshi}, {Murphy},
  {Naselsky}, {Nati}, {Natoli}, {Netterfield}, {N{\o}rgaard-Nielsen},
  {Noviello}, {Novikov}, {Novikov}, {O'Dwyer}, {Osborne}, {Oxborrow}, {Paci},
  {Pagano}, {Pajot}, {Paladini}, {Paoletti}, {Partridge}, {Pasian},
  {Patanchon}, {Pearson}, {Pearson}, {Peiris}, {Perdereau}, {Perotto},
  {Perrotta}, {Pettorino}, {Piacentini}, {Piat}, {Pierpaoli}, {Pietrobon},
  {Plaszczynski}, {Platania}, {Pointecouteau}, {Polenta}, {Ponthieu}, {Popa},
  {Poutanen}, {Pratt}, {Pr{\'e}zeau}, {Prunet}, {Puget}, {Rachen}, {Reach},
  {Rebolo}, {Reinecke}, {Remazeilles}, {Renault}, {Ricciardi}, {Riller},
  {Ristorcelli}, {Rocha}, {Rosset}, {Roudier}, {Rowan-Robinson},
  {Rubi{\~n}o-Mart{\'\i}n}, {Rusholme}, {Sandri}, {Santos}, {Savelainen},
  {Savini}, {Scott}, {Seiffert}, {Shellard}, {Spencer}, {Starck}, {Stolyarov},
  {Stompor}, {Sudiwala}, {Sunyaev}, {Sureau}, {Sutton}, {Suur-Uski}, {Sygnet},
  {Tauber}, {Tavagnacco}, {Terenzi}, {Toffolatti}, {Tomasi}, {Tristram},
  {Tucci}, {Tuovinen}, {T{\"u}rler}, {Umana}, {Valenziano}, {Valiviita}, {Van
  Tent}, {Vielva}, {Villa}, {Vittorio}, {Wade}, {Wandelt}, {Wehus}, {White},
  {White}, {Wilkinson}, {Yvon}, {Zacchei}, \& {Zonca}}]{Planck2014}
{Planck Collaboration}, {Ade}, P.~A.~R., {Aghanim}, N., {et~al.} 2014, \aap,
  571, A16, \dodoi{10.1051/0004-6361/201321591}

\bibitem[{{Pujol} \& {Gazta{\~n}aga}(2014)}]{Pujol2014}
{Pujol}, A., \& {Gazta{\~n}aga}, E. 2014, \mnras, 442, 1930,
  \dodoi{10.1093/mnras/stu1001}

\bibitem[{{Pujol} {et~al.}(2017){Pujol}, {Hoffmann}, {Jim{\'e}nez}, \&
  {Gazta{\~n}aga}}]{Pujol2017}
{Pujol}, A., {Hoffmann}, K., {Jim{\'e}nez}, N., \& {Gazta{\~n}aga}, E. 2017,
  \aap, 598, A103, \dodoi{10.1051/0004-6361/201629121}

\bibitem[{{Salcedo} {et~al.}(2018){Salcedo}, {Maller}, {Berlind}, {Sinha},
  {McBride}, {Behroozi}, {Wechsler}, \& {Weinberg}}]{Salcedo2018}
{Salcedo}, A.~N., {Maller}, A.~H., {Berlind}, A.~A., {et~al.} 2018, \mnras,
  475, 4411, \dodoi{10.1093/mnras/sty109}

\bibitem[{{Salcedo} {et~al.}(2020){Salcedo}, {Zu}, {Zhang}, {Wang}, {Yang},
  {Wu}, {Jing}, {Mo}, \& {Weinberg}}]{Salcedo2020}
{Salcedo}, A.~N., {Zu}, Y., {Zhang}, Y., {et~al.} 2020, arXiv e-prints,
  arXiv:2010.04176.
\newblock \doarXiv{2010.04176}

\bibitem[{{Schaller} {et~al.}(2015){Schaller}, {Frenk}, {Bower}, {Theuns},
  {Jenkins}, {Schaye}, {Crain}, {Furlong}, {Dalla Vecchia}, \&
  {McCarthy}}]{Schaller2015}
{Schaller}, M., {Frenk}, C.~S., {Bower}, R.~G., {et~al.} 2015, \mnras, 451,
  1247, \dodoi{10.1093/mnras/stv1067}

\bibitem[{{Scherrer} \& {Bertschinger}(1991)}]{Scherrer1991}
{Scherrer}, R.~J., \& {Bertschinger}, E. 1991, \apj, 381, 349,
  \dodoi{10.1086/170658}

\bibitem[{{Scoccimarro}(1998)}]{Scoccimarro1998}
{Scoccimarro}, R. 1998, \mnras, 299, 1097,
  \dodoi{10.1046/j.1365-8711.1998.01845.x}

\bibitem[{{Scoccimarro} {et~al.}(2001){Scoccimarro}, {Sheth}, {Hui}, \&
  {Jain}}]{Scoccimarro2001}
{Scoccimarro}, R., {Sheth}, R.~K., {Hui}, L., \& {Jain}, B. 2001, \apj, 546,
  20, \dodoi{10.1086/318261}

\bibitem[{{Seljak}(2000)}]{Seljak2000}
{Seljak}, U. 2000, \mnras, 318, 203, \dodoi{10.1046/j.1365-8711.2000.03715.x}

\bibitem[{{Seljak} \& {Zaldarriaga}(1996)}]{Seljak1996}
{Seljak}, U., \& {Zaldarriaga}, M. 1996, \apj, 469, 437, \dodoi{10.1086/177793}

\bibitem[{{Sheth} {et~al.}(2001){Sheth}, {Hui}, {Diaferio}, \&
  {Scoccimarro}}]{Sheth2001}
{Sheth}, R.~K., {Hui}, L., {Diaferio}, A., \& {Scoccimarro}, R. 2001, \mnras,
  325, 1288, \dodoi{10.1046/j.1365-8711.2001.04222.x}

\bibitem[{{Sinha} {et~al.}(2018){Sinha}, {Berlind}, {McBride}, {Scoccimarro},
  {Piscionere}, \& {Wibking}}]{Sinha2018}
{Sinha}, M., {Berlind}, A.~A., {McBride}, C.~K., {et~al.} 2018, \mnras, 478,
  1042, \dodoi{10.1093/mnras/sty967}

\bibitem[{Sinha \& Garrison(2019)}]{Corrfunc2019}
Sinha, M., \& Garrison, L. 2019, in Software Challenges to Exascale Computing,
  ed. A.~Majumdar \& R.~Arora (Singapore: Springer Singapore), 3--20.
\newblock \url{https://doi.org/10.1007/978-981-13-7729-7_1}

\bibitem[{{Sinha} \& {Garrison}(2020)}]{Corrfunc2020}
{Sinha}, M., \& {Garrison}, L.~H. 2020, \mnras, 491, 3022,
  \dodoi{10.1093/mnras/stz3157}

\bibitem[{{Sinha} \& {Holley-Bockelmann}(2012)}]{Sinha2012}
{Sinha}, M., \& {Holley-Bockelmann}, K. 2012, \apj, 751, 17,
  \dodoi{10.1088/0004-637X/751/1/17}

\bibitem[{{Springel}(2005)}]{Springel2005}
{Springel}, V. 2005, \mnras, 364, 1105,
  \dodoi{10.1111/j.1365-2966.2005.09655.x}

\bibitem[{{Springel} {et~al.}(2018){Springel}, {Pakmor}, {Pillepich},
  {Weinberger}, {Nelson}, {Hernquist}, {Vogelsberger}, {Genel}, {Torrey},
  {Marinacci}, \& {Naiman}}]{Springel2018}
{Springel}, V., {Pakmor}, R., {Pillepich}, A., {et~al.} 2018, \mnras, 475, 676,
  \dodoi{10.1093/mnras/stx3304}

\bibitem[{{Tinker} {et~al.}(2006{\natexlab{a}}){Tinker}, {Weinberg}, \&
  {Warren}}]{Tinker2006}
{Tinker}, J.~L., {Weinberg}, D.~H., \& {Warren}, M.~S. 2006{\natexlab{a}},
  \apj, 647, 737, \dodoi{10.1086/504795}

\bibitem[{{Tinker} {et~al.}(2006{\natexlab{b}}){Tinker}, {Weinberg}, \&
  {Zheng}}]{Tinker2006a}
{Tinker}, J.~L., {Weinberg}, D.~H., \& {Zheng}, Z. 2006{\natexlab{b}}, \mnras,
  368, 85, \dodoi{10.1111/j.1365-2966.2006.10114.x}

\bibitem[{{Vakili} \& {Hahn}(2019)}]{Vakili2019}
{Vakili}, M., \& {Hahn}, C. 2019, \apj, 872, 115,
  \dodoi{10.3847/1538-4357/aaf1a1}

\bibitem[{{Van den Bosch} {et~al.}(2005){Van den Bosch}, {Weinmann}, {Yang},
  {Mo}, {Li}, \& {Jing}}]{vandenBosch2005}
{Van den Bosch}, F.~C., {Weinmann}, S.~M., {Yang}, X., {et~al.} 2005, \mnras,
  361, 1203, \dodoi{10.1111/j.1365-2966.2005.09260.x}

\bibitem[{{Virtanen} {et~al.}(2020){Virtanen}, {Gommers}, {Oliphant},
  {Haberland}, {Reddy}, {Cournapeau}, {Burovski}, {Peterson}, {Weckesser},
  {Bright}, {van der Walt}, {Brett}, {Wilson}, {Jarrod Millman}, {Mayorov},
  {Nelson}, {Jones}, {Kern}, {Larson}, {Carey}, {Polat}, {Feng}, {Moore}, {Vand
  erPlas}, {Laxalde}, {Perktold}, {Cimrman}, {Henriksen}, {Quintero}, {Harris},
  {Archibald}, {Ribeiro}, {Pedregosa}, {van Mulbregt}, \&
  {Contributors}}]{Virtanen_2020}
{Virtanen}, P., {Gommers}, R., {Oliphant}, T.~E., {et~al.} 2020, Nature
  Methods, 17, 261, \dodoi{https://doi.org/10.1038/s41592-019-0686-2}

\bibitem[{{Vogelsberger} {et~al.}(2014){Vogelsberger}, {Genel}, {Springel},
  {Torrey}, {Sijacki}, {Xu}, {Snyder}, {Nelson}, \&
  {Hernquist}}]{IntroducingIllustris}
{Vogelsberger}, M., {Genel}, S., {Springel}, V., {et~al.} 2014, \mnras, 444,
  1518, \dodoi{10.1093/mnras/stu1536}

\bibitem[{{Walsh} \& {Tinker}(2019)}]{Walsh2019}
{Walsh}, K., \& {Tinker}, J. 2019, \mnras, \dodoi{10.1093/mnras/stz1351}

\bibitem[{{Wang} {et~al.}(2019){Wang}, {Mao}, {Zentner}, {van den Bosch},
  {Lange}, {Schafer}, {Villarreal}, {Hearin}, \& {Campbell}}]{Wang2019}
{Wang}, K., {Mao}, Y.-Y., {Zentner}, A.~R., {et~al.} 2019, \mnras,
  \dodoi{10.1093/mnras/stz1733}

\bibitem[{{Watson} {et~al.}(2012){Watson}, {Berlind}, {McBride}, {Hogg}, \&
  {Jiang}}]{Watson2012}
{Watson}, D.~F., {Berlind}, A.~A., {McBride}, C.~K., {Hogg}, D.~W., \& {Jiang},
  T. 2012, \apj, 749, 83, \dodoi{10.1088/0004-637X/749/1/83}

\bibitem[{{Wechsler} {et~al.}(2006){Wechsler}, {Zentner}, {Bullock},
  {Kravtsov}, \& {Allgood}}]{Wechsler2006}
{Wechsler}, R.~H., {Zentner}, A.~R., {Bullock}, J.~S., {Kravtsov}, A.~V., \&
  {Allgood}, B. 2006, \apj, 652, 71, \dodoi{10.1086/507120}

\bibitem[{{Wetzel} \& {White}(2010)}]{Wetzel2010}
{Wetzel}, A.~R., \& {White}, M. 2010, \mnras, 403, 1072,
  \dodoi{10.1111/j.1365-2966.2009.16191.x}

\bibitem[{{White} {et~al.}(2001){White}, {Hernquist}, \&
  {Springel}}]{White2001}
{White}, M., {Hernquist}, L., \& {Springel}, V. 2001, \apjl, 550, L129,
  \dodoi{10.1086/319644}

\bibitem[{{Xu} \& {Zheng}(2018)}]{Xu2018}
{Xu}, X., \& {Zheng}, Z. 2018, arXiv e-prints.
\newblock \doarXiv{1812.11210}

\bibitem[{{York} {et~al.}(2000){York}, {Adelman}, {Anderson}, {Anderson},
  {Annis}, {Bahcall}, {Bakken}, {Barkhouser}, {Bastian}, {Berman}, {Boroski},
  {Bracker}, {Briegel}, {Briggs}, {Brinkmann}, {Brunner}, {Burles}, {Carey},
  {Carr}, {Castander}, {Chen}, {Colestock}, {Connolly}, {Crocker}, {Csabai},
  {Czarapata}, {Davis}, {Doi}, {Dombeck}, {Eisenstein}, {Ellman}, {Elms},
  {Evans}, {Fan}, {Federwitz}, {Fiscelli}, {Friedman}, {Frieman}, {Fukugita},
  {Gillespie}, {Gunn}, {Gurbani}, {de Haas}, {Haldeman}, {Harris}, {Hayes},
  {Heckman}, {Hennessy}, {Hindsley}, {Holm}, {Holmgren}, {Huang}, {Hull},
  {Husby}, {Ichikawa}, {Ichikawa}, {Ivezi{\'c}}, {Kent}, {Kim}, {Kinney},
  {Klaene}, {Kleinman}, {Kleinman}, {Knapp}, {Korienek}, {Kron}, {Kunszt},
  {Lamb}, {Lee}, {Leger}, {Limmongkol}, {Lindenmeyer}, {Long}, {Loomis},
  {Loveday}, {Lucinio}, {Lupton}, {MacKinnon}, {Mannery}, {Mantsch}, {Margon},
  {McGehee}, {McKay}, {Meiksin}, {Merelli}, {Monet}, {Munn}, {Narayanan},
  {Nash}, {Neilsen}, {Neswold}, {Newberg}, {Nichol}, {Nicinski}, {Nonino},
  {Okada}, {Okamura}, {Ostriker}, {Owen}, {Pauls}, {Peoples}, {Peterson},
  {Petravick}, {Pier}, {Pope}, {Pordes}, {Prosapio}, {Rechenmacher}, {Quinn},
  {Richards}, {Richmond}, {Rivetta}, {Rockosi}, {Ruthmansdorfer}, {Sandford},
  {Schlegel}, {Schneider}, {Sekiguchi}, {Sergey}, {Shimasaku}, {Siegmund},
  {Smee}, {Smith}, {Snedden}, {Stone}, {Stoughton}, {Strauss}, {Stubbs},
  {SubbaRao}, {Szalay}, {Szapudi}, {Szokoly}, {Thakar}, {Tremonti}, {Tucker},
  {Uomoto}, {Vanden Berk}, {Vogeley}, {Waddell}, {Wang}, {Watanabe},
  {Weinberg}, {Yanny}, {Yasuda}, \& {SDSS Collaboration}}]{York2000}
{York}, D.~G., {Adelman}, J., {Anderson}, Jr., J.~E., {et~al.} 2000, \aj, 120,
  1579, \dodoi{10.1086/301513}

\bibitem[{{Zaldarriaga} \& {Seljak}(2000)}]{Zaldarriaga2000}
{Zaldarriaga}, M., \& {Seljak}, U. 2000, \apjs, 129, 431,
  \dodoi{10.1086/313423}

\bibitem[{{Zaldarriaga} {et~al.}(1998){Zaldarriaga}, {Seljak}, \&
  {Bertschinger}}]{Zaldarriaga1998}
{Zaldarriaga}, M., {Seljak}, U., \& {Bertschinger}, E. 1998, \apj, 494, 491,
  \dodoi{10.1086/305223}

\bibitem[{{Zehavi} {et~al.}(2018){Zehavi}, {Contreras}, {Padilla}, {Smith},
  {Baugh}, \& {Norberg}}]{Zehavi2018}
{Zehavi}, I., {Contreras}, S., {Padilla}, N., {et~al.} 2018, \apj, 853, 84,
  \dodoi{10.3847/1538-4357/aaa54a}

\bibitem[{{Zehavi} {et~al.}(2002){Zehavi}, {Blanton}, {Frieman}, {Weinberg},
  {Mo}, {Strauss}, {Anderson}, {Annis}, {Bahcall}, {Bernardi}, {Briggs},
  {Brinkmann}, {Burles}, {Carey}, {Castander}, {Connolly}, {Csabai},
  {Dalcanton}, {Dodelson}, {Doi}, {Eisenstein}, {Evans}, {Finkbeiner},
  {Friedman}, {Fukugita}, {Gunn}, {Hennessy}, {Hindsley}, {Ivezi{\'c}}, {Kent},
  {Knapp}, {Kron}, {Kunszt}, {Lamb}, {Leger}, {Long}, {Loveday}, {Lupton},
  {McKay}, {Meiksin}, {Merrelli}, {Munn}, {Narayanan}, {Newcomb}, {Nichol},
  {Owen}, {Peoples}, {Pope}, {Rockosi}, {Schlegel}, {Schneider}, {Scoccimarro},
  {Sheth}, {Siegmund}, {Smee}, {Snir}, {Stebbins}, {Stoughton}, {SubbaRao},
  {Szalay}, {Szapudi}, {Tegmark}, {Tucker}, {Uomoto}, {Vanden Berk}, {Vogeley},
  {Waddell}, {Yanny}, \& {York}}]{Zehavi2002}
{Zehavi}, I., {Blanton}, M.~R., {Frieman}, J.~A., {et~al.} 2002, \apj, 571,
  172, \dodoi{10.1086/339893}

\bibitem[{{Zehavi} {et~al.}(2011){Zehavi}, {Zheng}, {Weinberg}, {Blanton},
  {Bahcall}, {Berlind}, {Brinkmann}, {Frieman}, {Gunn}, {Lupton}, {Nichol},
  {Percival}, {Schneider}, {Skibba}, {Strauss}, {Tegmark}, \&
  {York}}]{Zehavi2011}
{Zehavi}, I., {Zheng}, Z., {Weinberg}, D.~H., {et~al.} 2011, \apj, 736, 59,
  \dodoi{10.1088/0004-637X/736/1/59}

\bibitem[{{Zentner} {et~al.}(2019){Zentner}, {Hearin}, {van den Bosch},
  {Lange}, \& {Villarreal}}]{Zentner2019}
{Zentner}, A.~R., {Hearin}, A., {van den Bosch}, F.~C., {Lange}, J.~U., \&
  {Villarreal}, A. 2019, \mnras, 485, 1196, \dodoi{10.1093/mnras/stz470}

\bibitem[{{Zentner} {et~al.}(2014){Zentner}, {Hearin}, \& {van den
  Bosch}}]{Zentner2014}
{Zentner}, A.~R., {Hearin}, A.~P., \& {van den Bosch}, F.~C. 2014, \mnras, 443,
  3044, \dodoi{10.1093/mnras/stu1383}

\bibitem[{{Zheng} {et~al.}(2007){Zheng}, {Coil}, \& {Zehavi}}]{Zheng2007}
{Zheng}, Z., {Coil}, A.~L., \& {Zehavi}, I. 2007, \apj, 667, 760,
  \dodoi{10.1086/521074}

\bibitem[{{Zheng} \& {Weinberg}(2007)}]{Zheng2007a}
{Zheng}, Z., \& {Weinberg}, D.~H. 2007, \apj, 659, 1, \dodoi{10.1086/512151}

\bibitem[{{Zheng} {et~al.}(2005){Zheng}, {Berlind}, {Weinberg}, {Benson},
  {Baugh}, {Cole}, {Dav{\'e}}, {Frenk}, {Katz}, \& {Lacey}}]{Zheng2005}
{Zheng}, Z., {Berlind}, A.~A., {Weinberg}, D.~H., {et~al.} 2005, \apj, 633,
  791, \dodoi{10.1086/466510}

\bibitem[{{Zu} \& {Mandelbaum}(2018)}]{Zu2018}
{Zu}, Y., \& {Mandelbaum}, R. 2018, \mnras, 476, 1637,
  \dodoi{10.1093/mnras/sty279}

\end{thebibliography}
\bibliographystyle{aasjournal}



\end{document}